\documentclass[a4paper,12pt]{article}
\usepackage{jheppub}
\usepackage{tcilatex}
\usepackage{amsfonts}
\usepackage{amssymb}
\usepackage{multicol}
\usepackage{graphicx}
\usepackage{float}
\usepackage{caption}
\usepackage{xcolor}
\usepackage[utf8]{inputenc}
\usepackage{amsmath}

\affiliation[1]{ LPHE-MS, Science Faculty, Mohammed V University in Rabat, Morocco}
\affiliation[2]{ Hassan II Academy of Science and Technology, Kingdom of Morocco.}
\affiliation[3]{Centre of Physics and Mathematics, CPM- Morocco}
\emailAdd{e.saidi@um5r.ac.ma}
\abstract{ Using properties of OSp(4\TEXTsymbol{\vert}2) and PSL(2\TEXTsymbol{\vert}2),
we investigate the super geometry of the parametric D($2,1;\zeta $)
labeled by variable $\zeta $ belonging to $\mathbb{C}\backslash \{-1,0\}$ and we give applications in the study of integrable superspin chains. This $9|8$ dimensional Lie
supergroup has three orthogonal isospins in its even part SL($2,\mathbb{C}$)$^{\otimes 3}$ assembled by
the tri-fundamental $2^{\otimes 3}$ with odd parity. It undergoes contractions
at $\zeta =-1,0$ where an SL($2,\mathbb{C}$) gets decompactified into
commutative $\mathbb{C}^{3}$ interpreted in terms of three central
extensions. By help of the obtained characteristic features of D($2,1;\zeta $) and their
local structures at the special points $\zeta =\pm 1$, we calculate the Lax
operator $\mathcal{L}_{\mathfrak{d}(2,1;\zeta )}^{(\mathfrak{\eta})}$ solving the
RLL equation describing the integrability of the superspin chain $\mathfrak{d}$($2,1;\zeta $).
We also complete missing results regarding the calculation of
 $\mathcal{L}_{psl(2|2)}^{(\mathfrak{\mu })}$ and $\mathcal{L}_{osp(4|2)}^{(\mathfrak{\mu })}$.
Other features of the four super Dynkin diagrams $S\mathfrak{DD}_{\mathfrak{d}(2,1;\zeta )}^{(\mathfrak{\eta})}$ and weight graphs of $\mathfrak{d}$($2,1;\zeta $)
 as well as discrete automorphisms are also given.}
\input{tcilatex}
\begin{document}

\title{Complex D($2,1;\zeta $) and spin chain solutions from Chern-Simons
theory}
\title{Complex D($2,1;\zeta $) and spin chain solutions from Chern-Simons
theory}
\author{\qquad El Hassan Saidi}
\maketitle

\notoc
\flushbottom
%\newpage

\section{Introduction}

\qquad \label{sec:intro} The exceptional 17 dimensional Lie superalgebra $%
\mathfrak{d}$(2,1;$\zeta $) has been subject to increasing interest during
last decades; it has three orthogonal isospins $sl(2)^{\oplus 3}$ glued
together by 8 fermionic charges making it somehow a special super symmetric
system \textrm{\cite{1A}-\cite{8A}}. It is a parametric superalgebra labeled
by a continuous parameter $\zeta $ belonging to the complex line $\mathbb{C}$
deprived of the points $\{-1,0\}$ where it contracts into two semi-simple
sub- superalgebras. Generally speaking, the exceptional $\mathfrak{d}$(2,1;$%
\zeta $) was introduced by V. G Kac \textrm{\cite{1A}} as a deformation of
the orthosymplectic 17- $\dim $\ Lie superalgebra osp(4\TEXTsymbol{\vert}2)
located at $\zeta =1$. This deviation is imagined in this paper as a
characteristic datum of the exotic $\mathfrak{d}$(2,1;$\zeta $) \textrm{and}
will be used as a fundamental algorithm in the present study.

The even part $\mathfrak{d}$(2,1;$\zeta $)$_{\bar{0}}$ of this superalgebra
has several real forms including $su(1,1)^{\oplus 3}$ and $su(1,1)\oplus
so(1,3)$ as well as the remarkable $sl(2,\mathbb{R})\oplus su(2,\mathbb{R}%
)^{\oplus 2}$ \textrm{\cite{relform}} which found interesting applications
in type II strings on AdS$_{3}\times \mathbb{S}^{3}\times \mathbb{S}^{3}$
\textrm{\cite{AdS1}-\cite{Rajae2}}. It has been also considered in the study
of exotic integrable superspin chains \textrm{\cite{9A}-\cite{10AA} }going
beyond the classical ABCD superchains\textrm{\ \cite%
{fr,fr1,fr2,fr3,ADE1,ADE2,ODE} }by using Bethe ansatz and algebraic methods
\textrm{\cite{IS1,SYM,GR,1C}}. Here, the superspin chain $\mathfrak{d}$(2,1;$%
\zeta $) will be revisited from the view of 4D Chern-Simons (CS) theory%
\textrm{\ \cite{1AA,2AA,CYW,S,2AB,W0} }where the solution of the RLL\
equation (\ref{1}) signing the integrability of the superspin chain is given
by the computation of the Lax operator\textrm{\ that we denote like }%
\begin{equation*}
\mathcal{L}_{\mathfrak{d}(2,1;\zeta )}
\end{equation*}%
\textrm{\ The CS investigation of the} exceptional integrable superspin
chain $\mathfrak{d}$(2,1;$\zeta $) and the calculation of the parametric
L-operator $\mathcal{L}_{\mathfrak{d}(2,1;\zeta )}$\textrm{\ }constitute
therefore one of the main results of the present work (see \autoref{sec5}
and \autoref{appD}).

\begin{figure}[tbph]
\begin{center}
\includegraphics[width=12cm]{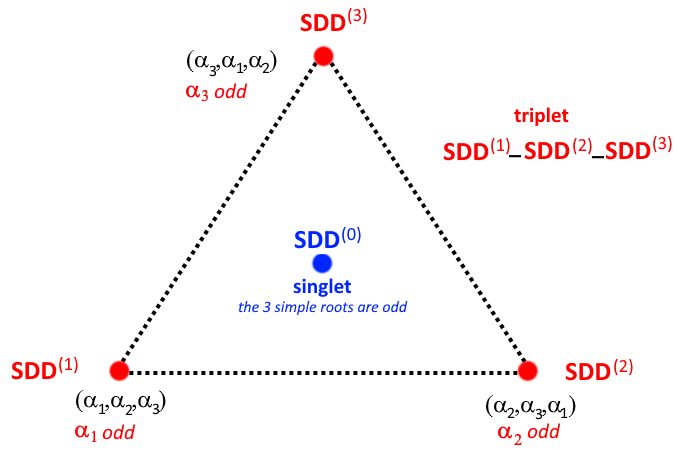}
\end{center}
\par
\vspace{-0.5cm}
\caption{A unified representation of the four super Dynkin diagrams of the
Lie superalgebra $\mathfrak{d}\left( 2,1,\protect\zeta \right) $ given by
the Figure \textbf{\protect\ref{3D}}. The four super Dynkin diagrams S$%
\mathfrak{DD}^{(\protect\eta )}$ organise into a singlet S$\mathfrak{DD}%
^{(0)}$ and a triplet S$\mathfrak{DD}^{(i)}$ of the permutation group $%
\mathcal{S}_{3}$. The closed S$\mathfrak{DD}^{(0)}$ has three odd simple
roots versus an odd simple root for the open S$\mathfrak{DD}^{(i)}$; it is
in the blue centre of the triangle.}
\label{4DD}
\end{figure}

\begin{figure}[tbph]
\begin{center}
\includegraphics[width=6cm]{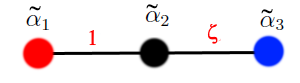}
\end{center}
\par
\vspace{-0.5cm}
\caption{Even simple roots: $\mathbf{\tilde{\protect\alpha}}_{1}=\left(
\mathbf{\protect\epsilon }_{1}-\mathbf{\protect\epsilon }_{2}\right) $ and $%
\mathbf{\tilde{\protect\alpha}}_{3}=\protect\sqrt{\protect\zeta }\left(
\mathbf{\protect\epsilon }_{1}+\mathbf{\protect\epsilon }_{2}\right) $. Odd
simple root $\mathbf{\tilde{\protect\alpha}}_{2}=\frac{1}{2}\protect\sqrt{%
2\left( 1+\protect\zeta \right) }\mathbf{\protect\delta }-\frac{1}{2}\left(
\mathbf{\protect\epsilon }_{1}-\mathbf{\protect\epsilon }_{2}\right) -\frac{1%
}{2}\protect\sqrt{\protect\zeta }\left( \mathbf{\protect\epsilon }_{1}+%
\mathbf{\protect\epsilon }_{2}\right) .$\ These roots are functions of $%
\protect\zeta .$}
\label{2}
\end{figure}

Algebraically, the complexified super $\mathfrak{d}_{\mathbb{C}}$(2,1;$\zeta
$) ---for short $\mathfrak{d}$(2,1;$\zeta $)--- is classified as a basic Lie
superalgebra in the dictionary of Frappat and Sorba \textrm{\cite{FS1,FS2}}.
I\textrm{t has four parametric super Dynkin diagrams with three nodes }$%
\left( \mathbf{\alpha }_{1},\mathbf{\alpha }_{2}\mathbf{,\alpha }_{3}\right)
$ \textrm{\cite{3AA,FS2,Y1}} that we denote collectively as
\begin{equation*}
S\mathfrak{DD}_{\zeta }^{(\mathtt{\eta })}
\end{equation*}%
with label $\mathtt{\eta }=0,1,2,3$. These super diagrams are effectively
given by the graphs of the \textbf{Figure} \textbf{\ref{3D}}; they can be
pictured into a unified form as depicted by the \textbf{Figure \ref{4DD}}.
However, the four super $\mathfrak{DD}_{\zeta }^{(\mathtt{\eta })}$ are not
completely independent; they split in two blocks like \textbf{1}$\oplus $%
\textbf{3} (blue singlet and red triplet in the \textbf{Figure \ref{4DD}}).
The three S$\mathfrak{DD}_{\zeta }^{(1)},$ S$\mathfrak{DD}_{\zeta }^{(2)},$ S%
$\mathfrak{DD}_{\zeta }^{(3)}$ are exchanged by an outer automorphism
symmetry given by the group $\mathcal{Z}_{3}$. Because of this symmetry, one
can then focus on one of them; say the S$\mathfrak{DD}_{\zeta }^{(2)}$ with
odd simple root $\mathbf{\alpha }_{2}$ as it will be done in this paper;
\textrm{see the }\textbf{Figure \ref{2} }for illustration.
Notice that the discrete outer automorphism $\mathcal{Z}_{3}$ is in fact a
subgroup of the non abelian symmetric group $\mathcal{S}_{3}$ that act by
permuting the so-called \emph{Kaplansky} parameters $\left( \text{\textsc{s}}%
_{1},\text{\textsc{s}}_{2},\text{\textsc{s}}_{3}\right) $ used for the
description of $\mathfrak{d}$(2,1;$\zeta $) by the isomorphic \emph{Scheunert%
} Lie superalgebra $\Gamma \left( \text{\textsc{s}}_{1},\text{\textsc{s}}%
_{2},\text{\textsc{s}}_{3}\right) $ \textrm{\cite{6,3A,GO}. Here, the
complex parameters }\textsc{s}$_{i}$\textrm{, which are constrained like }%
\textsc{s}$_{1}+$\textsc{s}$_{2}+$\textsc{s}$_{3}=0$\textrm{\ with }\textsc{s%
}$_{1}$\textsc{s}$_{2}$\textsc{s}$_{3}\neq 0$\textrm{, are realised as }$2$%
\textsc{s}$_{1}=1,$ $2$\textsc{s}$_{2}=-1-\zeta $ and $2$\textsc{s}$%
_{3}=\zeta $. This choice breaks $\mathcal{S}_{3}$ down to its subgroup $%
\mathcal{Z}_{2}$ which fixes \textsc{s}$_{1}$ and permutes \textsc{s}$_{2}$
and \textsc{s}$_{3}$; that is the $\mathcal{Z}_{2}$ generated by the
transposition $\tau _{23}:\left( \text{\textsc{s}}_{1},\text{\textsc{s}}_{2},%
\text{\textsc{s}}_{3}\right) \rightarrow \left( \text{\textsc{s}}_{1},\text{%
\textsc{s}}_{3},\text{\textsc{s}}_{2}\right) .$

The parametric super $\mathfrak{DD}_{\zeta }^{(\mathtt{\eta })}$ will play
an important role in this study; they share fundamental aspects with the
\textrm{super Dynkin }diagrams of psl(2\TEXTsymbol{\vert}2) and osp(4%
\TEXTsymbol{\vert}2) sitting at the special points $\zeta =-1$ and $+1;$
thus capturing important information on the exceptional super symmetry. For
the singular limit $\zeta \rightarrow -1,$ the exceptional $\mathfrak{d}$%
(2,1;$\zeta $) contracts into $psl(2|2)\oplus \mathbb{C}^{3}$ \textrm{\cite%
{1B,Le,2B}} where one of the three $sl(2,\mathbb{C})$s making the even part $%
\mathfrak{d}$(2,1;$\zeta $)$_{\bar{0}}$ gets replaced by the abelian $%
\mathbb{C}^{3}$ whose translation generators ($P_{1},$ $P_{2},$\ $P_{3}$)
have been interpreted like three commuting central extensions (\textrm{see %
\autoref{appA} for details)}. For the regular limit $\zeta \rightarrow +1,$
the $\mathfrak{d}$(2,1;$\zeta $) coincides with the 17 dimensional
orthosymplectic osp(4\TEXTsymbol{\vert}2) providing in turn an important
tool for engineering models with $\mathfrak{d}$(2,1;$\zeta $) symmetry
characterised by a finite set of deformation parameters \{$\kappa _{i}$\}
which are functions of the complex $\zeta $ as it will be shown throughout
this investigation.

In this paper, we aim for two main objectives motivated by $\left( \mathbf{1}%
\right) $ completing partial results on the super group D(2,1;$\zeta $) and $%
\left( \mathbf{2}\right) $ by the application of the construction in
integrable systems based on Chern-Simons potentials. For that, we first
revisit the study of the complexified $\mathfrak{d}$(2,1;$\zeta $) from
algebraic geometry view; then use this description to get more insight into
the parametric supergroup manifold D(2,1;$\zeta $) in order to develop the
study the\textrm{\ integrable }$\mathfrak{d}$(2,1;$\zeta $)\textrm{\
superspin chain }by using 4D CS theory. Because of the complexity in
directly solving the RLL integrability equation (\ref{lax}) of the $%
\mathfrak{d}$(2,1;$\zeta $)\textrm{\ superchain namely},
\begin{equation}
\boldsymbol{R}_{rs}^{ik}\left( z-w\right) \boldsymbol{L}_{j}^{r}\left(
z\right) \boldsymbol{L}_{l}^{s}\left( w\right) =\boldsymbol{L}_{r}^{i}\left(
w\right) \boldsymbol{L}_{s}^{k}\left( z\right) \boldsymbol{R}%
_{jl}^{rs}\left( z-w\right)  \label{1}
\end{equation}%
we get around this difficulty by taking advantage of its local structures at
the special points $\zeta =\pm 1$ where live psl(2\TEXTsymbol{\vert}2) and
osp(4\TEXTsymbol{\vert}2). Here, $\boldsymbol{R}_{rs}^{ik}$\ is the usual
R-matrix of Yang Baxter equation and $\boldsymbol{L}_{j}^{i}$ is the matrix
realisation of the Lax operator. In this way, we aim to reach the parametric
$\mathcal{L}_{\mathfrak{d}(2,1;\zeta )}$ by seeking to continuously join the
two superchain solutions at $\zeta =\pm 1$ while respecting the constraints
required by the exceptional symmetry.

However, an inspection of the literature on integrable superspin chains
reveals that there are no results on the derivation of the Lax operators for
the psl(2\TEXTsymbol{\vert}2) and osp(4\TEXTsymbol{\vert}2) superchains by
using the CS method. Results in this regard are only known for the
superalgebras psl($m|n$) with $m\neq n$ and osp($2k|2l$) with $k\geq 3$ and $%
l\geq 1.$ To overcome this, we start by filling in this gap by first
calculating the missing $\mathcal{L}_{p{\small sl(2|2)}}$ and $\mathcal{L}_{%
{\small osp(4|2)}}$ Lax operators; and then proceed to calculate $\mathcal{L}%
_{\mathfrak{d}(2,1;\zeta )}.$ As a result of this approach, we derive a
family of $\mathcal{L}_{\mathfrak{d}(2,1;\zeta )}$ operators labeled by
three complex parameters $\left( \kappa _{0},\kappa _{1},\kappa _{2}\right) $%
\ which are functions of $\zeta $. A particular matrix representation of the
obtained solutions is given by the following 6$\times $6 Lax matrix with $%
\kappa _{1}=1$ and $\kappa _{2}=-\kappa _{0},$%
\begin{equation*}
\end{equation*}%
\begin{equation}
\left(
\begin{array}{cccc}
(z^{\frac{1}{2}}+\kappa _{0}^{2}z^{-\frac{1}{2}}\mathrm{b}^{\prime }\mathrm{c%
}^{\prime })\delta _{j}^{i}+z^{-\frac{1}{2}}\mathrm{\beta }^{i}\mathrm{%
\gamma }_{j} & z^{-\frac{1}{2}}(\mathrm{c\beta }^{i}-\kappa _{0}^{2}\mathrm{b%
}^{\prime }\mathrm{\gamma }_{n}\varepsilon ^{ni}) & z^{-\frac{1}{2}}\mathrm{%
\beta }^{i} & \kappa _{0}z^{-\frac{1}{2}}\mathrm{b}^{\prime }\varepsilon
^{ji} \\
z^{-\frac{1}{2}}(\mathrm{b\gamma }_{j}-\kappa _{0}^{2}\mathrm{c}^{\prime
}\varepsilon _{jn}\mathrm{\beta }^{n}) & z^{\frac{1}{2}}+z^{-\frac{1}{2}}(%
\mathrm{bc}+\kappa _{0}^{2}\mathrm{\beta }^{n}\mathrm{\gamma }_{n}) & z^{-%
\frac{1}{2}}\mathrm{b} & -\kappa _{0}z^{-\frac{1}{2}}\mathrm{\beta }^{j} \\
z^{-\frac{1}{2}}\mathrm{\gamma }_{j} & z^{-\frac{1}{2}}\mathrm{c} & z^{-%
\frac{1}{2}} & 0 \\
\kappa _{0}z^{-\frac{1}{2}}\mathrm{c}^{\prime }\varepsilon _{ji} & -\kappa
_{0}z^{-\frac{1}{2}}\mathrm{\gamma }_{i} & 0 & z^{-\frac{1}{2}}\delta
_{i}^{j}%
\end{array}%
\right)  \label{L}
\end{equation}%
\begin{equation*}
\end{equation*}%
Here $z$ is the usual complex variable (rapidity) used in the 4D CS theory
whose field action S$_{4dCS}\left[ A\right] $ is reported in the \textrm{%
\autoref{appD}}; see eq\textrm{(\ref{act})}. The two pairs ($\mathrm{\beta }%
^{i},\mathrm{\gamma }_{i})$ are fermionic oscillators satisfying $\left\{
\mathrm{\beta }^{j},\mathrm{\gamma }_{i}\right\} =\delta _{i}^{j}$; while
the two singlets $\left( \mathrm{b},\mathrm{c}\right) $ and $\left( \mathrm{b%
}^{\prime },\mathrm{c}^{\prime }\right) $ are bosonic oscillators obeying $%
\left[ \mathrm{b},\mathrm{c}\right] =1=\left[ \mathrm{b}^{\prime },\mathrm{c}%
^{\prime }\right] $. Moreover, the parametric function $\kappa _{0}\left(
\zeta \right) $ in (\ref{L}) is equal to $\frac{1}{2}\zeta \left( \zeta
+1\right) $; it is given by the determinant $\det \left( \mathbf{\alpha }%
_{i}.\mathbf{\alpha }_{j}\right) $ of the intersection matrix of the simple
roots $\left( \mathbf{\alpha }_{1},\mathbf{\alpha }_{2},\mathbf{\alpha }%
_{3}\right) $ generating the distinguished root system of $\mathfrak{d}%
(2,1;\zeta )$ with odd $\mathbf{\alpha }_{2}$ and even $\left( \mathbf{%
\alpha }_{1},\mathbf{\alpha }_{3}\right) $ defining S$\mathfrak{DD}_{\zeta
}^{(2)}$. The values of the L-matrix (\ref{L}) at $\zeta =\pm 1$ recovers
precisely the expressions the matrix representations of obtained $\mathcal{L}%
_{p{\small sl(2|2)}}$ and $\mathcal{L}_{{\small osp(4|2)}}.$

The organisation of this paper is as follows: \textrm{In \autoref{sec2}}, we
study the algebraic geometry of the distinguished complex 14D supermanifold
of PSL(2\TEXTsymbol{\vert}2) with emphasis on its 6D even PSL(2\TEXTsymbol{%
\vert}2)$_{0}\simeq SL\left( 3\right) ^{\otimes 2}$ part and the 8D odd PSL(2%
\TEXTsymbol{\vert}2)$_{\bar{1}}$ module 2$\times $\textbf{2}$^{\otimes 2}$.
\textrm{The \autoref{sec3} and \autoref{sec4}} are deserved to the
exceptional D(2,1;$\zeta $). There, we investigate the geometry of the
complex 17D supermanifold of the supergroup D(2,1;$\zeta $), its algebraic
properties in link with the orthosymplectic OSp(4\TEXTsymbol{\vert}2) as
well as with its decompactification into $PSL(2|2)\times \mathbb{C}^{3}.$
\textrm{In \autoref{sec5}}, we study the integrable superspin chain $%
\mathfrak{d}$(2,1;$\zeta $) by using the 4D Chern-Simons theory. Because of
lack of results on integrable sl(n\TEXTsymbol{\vert}n) superchains in
Chern-Simons formulation, we fill in this gap by calculating $\mathcal{L}_{p%
{\small sl(2|2)}}$ in subsection 5.1. In subsection 5.2, we also determine $%
\mathcal{L}_{osp{\small (4|2)}}$ since it has not been calculated before;
and in subsection 5.3, we calculate the L-operator for\ the $\mathfrak{d}$%
(2,1;$\zeta $)\ superchain. \textrm{In \autoref{conc}, we give }conclusion
and additional comments. \textrm{Auxiliary sections are }devoted to \emph{%
four} appendices. \textrm{In \autoref{appA}}, we give useful properties of
the complex Lie superalgebras sl(2\TEXTsymbol{\vert}2) and its simple
subalgebra psl(2\TEXTsymbol{\vert}2). There, we use the 4$\times $4 super
matrix realisation of sl(2\TEXTsymbol{\vert}2) to describe the algebraic and
geometric properties of the super manifold SL(2\TEXTsymbol{\vert}2). \textrm{%
In \autoref{appB}}, we give some helpful ingredients of $osp(4|2)$ and study
its oscillator realisation as well as its link with the exceptional super D$%
(2,1;\zeta )$ at $\zeta =1$. \textrm{In \autoref{appC}, we give }complement
tools regarding D$(2,1;\zeta )$ and the realisation of its four possible
roots systems. \textrm{In \autoref{appD}, we recall some basic tools on the
4D Chern-Simons theory for integrable super systems.}

\section{Complex geometry of distinguished SL(2\TEXTsymbol{\vert}2)}

\qquad \label{sec2} In this section, we use the complex 4$\times $4 super
matrix realisation of SL(2\TEXTsymbol{\vert}2) with the 2$\times $2 block
matrix decomposition \textrm{\cite{SV},}%
\begin{equation}
\boldsymbol{M}_{{\small 4\times 4}}=\left(
\begin{array}{cc}
\boldsymbol{A}_{{\small 2\times 2}} & \boldsymbol{B}_{{\small 2\times 2}} \\
\boldsymbol{C}_{{\small 2\times 2}} & \boldsymbol{D}_{{\small 2\times 2}}%
\end{array}%
\right)  \label{M}
\end{equation}%
as well as the distinguished root system $\Phi _{{\small sl(2|2)}}^{\Pi
_{1}} $ to study useful algebraic and geometric properties of the SL(2%
\TEXTsymbol{\vert}2) supermanifold denoted here after like $\mathcal{X}_{%
{\small sl(2|2)}}^{\Pi _{1}}$. This 15D supermanifold is associated with the
distinguished simple root basis $\Pi _{1}=\left( \mathbf{\alpha }_{1},%
\mathbf{\alpha }_{2},\mathbf{\alpha }_{3}\right) $ having odd $\mathbf{%
\alpha }_{2}$ and even $\mathbf{\alpha }_{1},\mathbf{\alpha }_{3}$ \cite{FS1}%
; \textrm{see the \autoref{appA} for the four possible SL(2\TEXTsymbol{\vert}%
2) root systems }$\Phi _{{\small sl(2|2)}}^{\Pi _{\eta }}$\textrm{\textbf{\ }%
}labeled by $\eta =0,1,2,3$\textrm{\ and the associated graphs of the
\textbf{Figure} \textbf{\ref{Asl2}}}. Through this analysis, we also use the
two possible real forms of $SL(2,\mathbb{C})$ to study other properties $%
\mathcal{X}_{{\small sl(2|2)}}^{\Pi _{1}}$ such as critical limits given by
SU$\left( 2\right) $ singularity of complex surfaces and its{\small \ }%
SU(1,1) conical homologue thought of here in term of vanishing "gap energy"
ideas as described by eqs(\ref{vpm})-(\ref{VV}).

The investigation given in this section should be also viewed as a front
matter for addressing the (9\TEXTsymbol{\vert}8)-dimensional (17D)
supergeometry of the parametric $D\left( 2,1;\zeta \right) $ containing PSL(2%
\TEXTsymbol{\vert}2) as a complex (6\TEXTsymbol{\vert}8)-dimensional (14D)
submanifold. The results obtained here as well as the tools given below will
be used for the geometric interpretation of the contraction $D\left(
2,1;\zeta \right) \rightarrow PSL(2|2)$ in terms of the vanishing of the
(dual) 2-cycle
\begin{equation}
\mathcal{C}_{\mathbf{\mathbf{\tilde{\psi}}}}=\mathcal{C}_{\mathbf{\mathbf{%
\alpha }}_{1}+2\mathbf{\alpha }_{2}+\mathbf{\alpha }_{3}}  \label{cp}
\end{equation}%
sitting in the normal space $D\left( 2,1;\zeta \right) \backslash PSL(2|2)$.
In the above relation, the $\mathbf{\mathbf{\tilde{\psi}=\alpha }}_{1}+2%
\mathbf{\alpha }_{2}+\mathbf{\alpha }_{3}$ is the long root of $D\left(
2,1;\zeta \right) $ and $\mathcal{C}_{\mathbf{\mathbf{\tilde{\psi}}}}$ the
associated homological 2-cycle in the parametric $\mathcal{X}_{\mathfrak{d}%
\left( 2,1;\zeta \right) }^{\Pi _{1}}$; it area $\mathcal{A}(\mathcal{C}_{%
\mathbf{\mathbf{\tilde{\psi}}}})=\int_{\mathcal{C}_{\mathbf{\mathbf{\tilde{%
\psi}}}}}J_{Kahler}$ is a function of $\zeta $; it vanishes in the limit $%
\zeta \rightarrow -1$ as it will be shown later on.

\subsection{SL(2\TEXTsymbol{\vert}2) as a 15D hypersurface $\mathfrak{h}%
_{15} $ in $\mathbb{C}^{8|8}$}

\qquad We first study the algebraic relations describing the supermanifold $%
\mathcal{X}_{{\small sl(2|2)}}^{\Pi _{1}}$ of the distinguished Lie
supergroup SL(2\TEXTsymbol{\vert}2). Then, we turn to give useful properties
of this complex 15 dimensional- space $\mathfrak{h}_{15}$ and its
submanifolds; in particular PSL(2\TEXTsymbol{\vert}2), SL(2) and GL(1).

\subsubsection{Defining relations and properties}

\qquad The geometric description of the complex 15 dimensional SL(2%
\TEXTsymbol{\vert}2) is done in terms of graded complex 4$\times $4 matrices
(\ref{M}) that can be also presented like $\boldsymbol{M}_{{\small 4\times 4}%
}=\exp (\boldsymbol{H}_{{\small 4\times 4}})$ with vanishing supertrace $str(%
\boldsymbol{H}_{{\small 4\times 4}})=0.$ It is done by generalising usual
constructions of bosonic symmetries $SU(2)$ and $SL(2,\mathbb{R})\simeq
SU(1,1)$ describing the two real forms of SL(2,$\mathbb{C}$) \textrm{\cite%
{Rajae2, BY}}. To that purpose, we consider a generic complex super matrix%
\begin{equation}
\boldsymbol{M}_{{\small 4\times 4}}=\left(
\begin{array}{cc}
\boldsymbol{X}_{{\small 1}} & \boldsymbol{Y}_{{\small 2}} \\
\boldsymbol{Y}_{{\small 1}} & \boldsymbol{X}_{{\small 2}}%
\end{array}%
\right) \qquad \Leftrightarrow \qquad \left( \boldsymbol{M}\right) _{\text{%
\textsc{b}}}^{\text{\textsc{a}}}=\left(
\begin{array}{cc}
\left( \boldsymbol{X}_{{\small 1}}\right) _{b}^{a} & \left( \boldsymbol{Y}_{%
{\small 2}}\right) _{\mathrm{\beta }}^{a} \\
\left( \boldsymbol{Y}_{{\small 1}}\right) _{b}^{\mathrm{\alpha }} & \left(
\boldsymbol{X}_{{\small 2}}\right) _{\mathrm{\beta }}^{\mathrm{\alpha }}%
\end{array}%
\right)  \label{M44}
\end{equation}%
belonging to End($\mathbb{C}^{2|2}$) where the canonical vector basis \{$%
\mathbf{e}_{1},\mathbf{e}_{2};\mathbf{e}_{3},\mathbf{e}_{4}$\} of $\mathbb{C}%
^{2|2}$\ are ordered as (even, even, odd, odd) that is (bbff) in the
notation used in (\ref{pi}) termed as distinguished ordering; here the label
(b) refers to bosonic and the (f) to fermionic parities. In eq(\ref{M44}),
the complex $\boldsymbol{X}_{{\small 1,2}}$ and $\boldsymbol{Y}_{{\small 1,2}%
}$ are four 2$\times $2 block matrices with entry variables as%
\begin{equation}
\boldsymbol{X}_{i}=\left(
\begin{array}{cc}
x_{i} & u_{i} \\
y_{i} & v_{i}%
\end{array}%
\right) \qquad ,\qquad \boldsymbol{Y}_{i}=\left(
\begin{array}{cc}
\xi _{i} & \phi _{i} \\
\zeta _{i} & \chi _{i}%
\end{array}%
\right)  \label{xyx}
\end{equation}%
The 8+8\ complex entries of the super matrix (\ref{M44}) reads in a
condensed way like $\left( \boldsymbol{M}\right) _{\text{\textsc{b}}}^{\text{%
\textsc{a}}}$ with labels $a,b=1,2$ for the first SL(2)$_{1}$ and $\mathrm{%
\alpha },$ $\mathrm{\beta }=1,2$ for the second SL(2)$_{2}$ making the even
part PSL(2\TEXTsymbol{\vert}2)$_{\bar{0}}.$ This labeling shows that $%
\boldsymbol{Y}_{{\small 1}}$ and $\boldsymbol{Y}_{{\small 2}}$ sit in the
bi-fundamental of $SL(2)_{1}\times SL(2)_{2}$ while $\boldsymbol{X}_{{\small %
1}}$ and $\boldsymbol{X}_{{\small 2}}$ are in the adjoints of $GL(2)_{1}$
and $GL(2)_{2}$ respectively. By substituting (\ref{xyx}), we have
\begin{equation}
\boldsymbol{M}_{\text{\textsc{b}}}^{\text{\textsc{a}}}=\left(
\begin{array}{cccc}
x_{1} & u_{1} & \xi _{2} & \phi _{2} \\
y_{1} & v_{1} & \zeta _{2} & \chi _{2} \\
\xi _{1} & \phi _{1} & x_{2} & u_{2} \\
\zeta _{1} & \chi _{1} & y_{2} & v_{2}%
\end{array}%
\right)  \label{M45}
\end{equation}%
with entries belonging to $\mathbb{C}^{8|8}.$ In term of the super matrix $%
\boldsymbol{M}$, the defining equation of the complex SL(2\TEXTsymbol{\vert}%
2) manifold is given by the unimodular condition of the Berezinian (super
determinant: sdet) $Ber\left( \boldsymbol{M}_{{\small 4\times 4}}\right) =1$
turning into a constraint relation among the 16 complex variables in eq(\ref%
{M45}). This super determinant can be presented in a handy form using the 2$%
\times $2 sub-block matrices $\boldsymbol{X}_{{\small 1,2}}$ and $%
\boldsymbol{Y}_{{\small 1,2}}$ like $\det \left[ \boldsymbol{X}_{{\small 1}}-%
\boldsymbol{Y}_{{\small 1}}\boldsymbol{X}_{{\small 2}}^{-1}\boldsymbol{Y}_{%
{\small 2}}\right] \det \left( \boldsymbol{X}_{{\small 2}}\right) ^{-1}=1;$
thus defining a complex 15 dim hypersurface $\mathfrak{h}_{15}$ in $\mathbb{C%
}^{8|8}$. By rewriting this constraint relation like
\begin{equation}
\mathfrak{h}_{15}:\qquad \det \left( \boldsymbol{X}_{{\small 1}}\right)
\cdot \det \left[ \boldsymbol{I}-\boldsymbol{X}_{{\small 1}}^{-1}\boldsymbol{%
Y}_{{\small 1}}\boldsymbol{X}_{{\small 2}}^{-1}\boldsymbol{Y}_{{\small 2}}%
\right] \cdot \det \left( \boldsymbol{X}_{{\small 2}}^{-1}\right) =1
\label{detX}
\end{equation}%
one gets a correspondence between:

\begin{itemize}
\item On one hand, the triple factorisation of eq(\ref{detX}) given by the
three blocks $\det \left( \boldsymbol{X}_{{\small 1}}\right) $, $\det \left[
\boldsymbol{I}-\boldsymbol{X}_{{\small 1}}^{-1}\boldsymbol{Y}_{{\small 1}}%
\boldsymbol{X}_{{\small 2}}^{-1}\boldsymbol{Y}_{{\small 2}}\right] $ and $%
\det \left( \boldsymbol{X}_{{\small 2}}^{-1}\right) ;$ and

\item On the other hand, the subalgebras $sl(2,\mathbb{C})_{\text{\textsc{a}}%
}$, $sl(1|1)_{\text{\textsc{ab}}}$ and $sl(2,\mathbb{C})_{\text{\textsc{b}}}$
in sl(2\TEXTsymbol{\vert}2). These subalgebras are indexed by the simple
roots ($\mathbf{\mathbf{\alpha }}_{1}$)$,$ ($\mathbf{\alpha }_{2}$) and ($%
\mathbf{\alpha }_{3}$) where $\mathbf{\alpha }_{2}$ is fermionic and ($%
\mathbf{\mathbf{\alpha }}_{1},\mathbf{\alpha }_{3}$) bosonic.
\end{itemize}

One also derives other interesting properties on distinguished SL(2%
\TEXTsymbol{\vert}2) geometry; in particular the two useful ones:

$\left( \mathbf{A}\right) $ \textbf{projective symmetry}: Eq(\ref{detX}) has
a remarkable invariance given by the projective transformation
\begin{equation}
\mathbb{C}^{\ast }:\boldsymbol{X}_{{\small i}}\rightarrow \lambda
\boldsymbol{X}_{{\small i}}\qquad ,\qquad \boldsymbol{Y}_{{\small i}%
}\rightarrow \lambda \boldsymbol{Y}_{{\small i}}
\end{equation}%
where $\lambda $\ is a non vanishing $\mathbb{C}$-number. This projective
symmetry $\left( \mathbf{i}\right) $ allows to fix one of the complex
degrees of freedom in $\boldsymbol{M}_{{\small 4\times 4}}$; for example $%
\det \left( \boldsymbol{X}_{{\small 1}}\right) =1$ expanding like $%
x_{1}v_{1}-y_{1}u_{1}=1,$ and $\left( \mathbf{ii}\right) $ permits to
construct the defining relation of the 14 dimensional supermanifold PSL(2%
\TEXTsymbol{\vert}2) in terms of the coset
\begin{equation}
PSL(2|2)=SL(2|2)/\mathbb{C}^{\ast }\qquad ,\qquad \mathbb{C}^{\ast }\simeq
GL(1,\mathbb{C})
\end{equation}

$\left( \mathbf{B}\right) $ \textbf{sl(2) singularities}: For the case where
the two even submatrices $\boldsymbol{X}_{{\small i}}$ take the diagonal
values $z_{i}I_{2\times 2}$ where the $z_{i}$'s are complex numbers [i.e: $%
\boldsymbol{X}_{{\small i}}\in GL(1,\mathbb{C})_{i}$], we have $\det \left(
\boldsymbol{X}_{{\small i}}\right) =z_{i}^{2}.$ If moreover, the two complex
numbers are taken equal ($z_{1}=z_{2}=z$), that is
\begin{equation}
\boldsymbol{M}_{{\small 4\times 4}}=\left(
\begin{array}{cc}
z\boldsymbol{I}_{2\times 2} & \boldsymbol{Y}_{{\small 2}} \\
\boldsymbol{Y}_{{\small 1}} & z\boldsymbol{I}_{2\times 2}%
\end{array}%
\right)  \label{M4}
\end{equation}%
we have the equality $\det \left( \boldsymbol{X}_{{\small 1}}\right) =\det
\left( \boldsymbol{X}_{{\small 2}}\right) ;$ and then eq(\ref{detX}) becomes
\begin{equation}
\det \left[ \boldsymbol{I}_{2\times 2}-\frac{1}{z^{2}}\boldsymbol{Y}_{%
{\small 1}}\boldsymbol{Y}_{{\small 2}}\right] =1  \label{YY}
\end{equation}%
Natural solutions of this particular relation are obtained by taking $%
\boldsymbol{Y}_{{\small 1}}=w_{1}\boldsymbol{I}_{2\times 2}$ and $%
\boldsymbol{Y}_{{\small 2}}=w_{2}\boldsymbol{I}_{2\times 2}$ where $w_{1}$
and $w_{2}$\ are complex numbers; then putting into (\ref{YY}), we get
\begin{equation}
\left( 1-\frac{w_{1}w_{2}}{z^{2}}\right) ^{2}=1\qquad \Rightarrow \qquad 1-%
\frac{w_{1}w_{2}}{z^{2}}=\pm 1\qquad \Rightarrow \qquad w_{1}w_{2}=\left(
1\mp 1\right) z^{2}
\end{equation}%
thus leading to two remarkable solutions corresponding to the $\pm $ sectors:

\begin{description}
\item[$\left( \mathbf{1}\right) $] \textbf{First solution}: it is given by
the vanishing condition $w_{1}w_{2}=0$ corresponding to a triangular matrix $%
\boldsymbol{M}_{{\small 4\times 4}}=\boldsymbol{T}_{{\small 4\times 4}}$
where $w_{1}=0$ and $w_{2}\neq 0$ or $w_{1}\neq 0$ and $w_{2}=0$. For the
case $w_{1}=w_{2}=0$, it reduces simply to $\boldsymbol{D}_{{\small 4\times 4%
}}=z\boldsymbol{I}_{2\times 2}\oplus z\boldsymbol{I}_{2\times 2}.$ This
diagonal supermatrix describes a complex curve $\mathcal{C}$ within SL(2%
\TEXTsymbol{\vert}2). Also, the $\boldsymbol{D}_{{\small 4\times 4}}$\ is
the central element of the super group SL(2\TEXTsymbol{\vert}2) showing that
the hypersurface $\mathfrak{h}_{15}$ describing this complex 15 dimensional
supermanifold has within it a homological 2-cycle\textrm{\ which turns out
to be intimately related to the} $\mathcal{C}_{\mathbf{\mathbf{\tilde{\psi}}}%
}$ \textrm{given by (\ref{cp}).}

\item[$\left( \mathbf{2}\right) $] \textbf{Second solution}: it is given by
the interesting relation%
\begin{equation}
w_{1}w_{2}=2z^{2}\qquad \Rightarrow \qquad \boldsymbol{Y}_{{\small 1}}%
\boldsymbol{Y}_{{\small 2}}=2z^{2}I_{2\times 2}
\end{equation}%
leading to
\begin{equation}
\left( \boldsymbol{M}\right) _{\text{\textsc{b}}}^{\text{\textsc{a}}}=\left(
\begin{array}{cc}
z\delta _{b}^{a} & w_{2}\delta _{\mathrm{\beta }}^{a} \\
w_{1}\delta _{b}^{\mathrm{\alpha }} & z\delta _{\mathrm{\beta }}^{\mathrm{%
\alpha }}%
\end{array}%
\right)
\end{equation}%
It describes a complex surface $\Sigma $ inside SL(2\TEXTsymbol{\vert}2)
given by the complex equation $w_{1}w_{2}=2z^{2}$. This surface has \textrm{%
an sl(2) singularity} at z=0 \textrm{\cite{SE,SU2S}} corresponding to the
vanishing of the central 2-cycle $\mathcal{C}_{\mathbf{\mathbf{\tilde{\psi}}}%
}$. This feature shows that the complex 15 dimensional manifold $\mathfrak{h}%
_{15}$ describing SL(2\TEXTsymbol{\vert}2) factorises like
\begin{equation}
\mathfrak{h}_{15}=PSL(2|2)\times \mathcal{C}_{\mathbf{\mathbf{\tilde{\psi}}}}
\end{equation}
\end{description}

\subsubsection{GL(2) submanifolds inside SL(2\TEXTsymbol{\vert}2)}

\qquad The complex super SL(2\TEXTsymbol{\vert}2) has a 7-dim even part $%
SL(2|2)_{\bar{0}}$ and 8-dim odd one $SL(2|2)_{\bar{1}}$. The even part
factorises like $SL(2)_{1}\times SL(2)_{2}\otimes \mathbb{C}^{\ast };$ it
has two GL(2$,\mathbb{C}$) blocks reading in terms of the $\boldsymbol{X}%
_{i} $ 's in eq(\ref{xyx}) as follows
\begin{equation}
\left. \boldsymbol{M}_{{\small 4\times 4}}\right\vert _{SL(2|2)_{\bar{0}%
}}=\left(
\begin{array}{cc}
\boldsymbol{X}_{{\small 1}} & \boldsymbol{0} \\
\boldsymbol{0} & \boldsymbol{X}_{{\small 2}}%
\end{array}%
\right)
\end{equation}%
with $\left( \mathbf{i}\right) $ super determinant $Ber\left( \boldsymbol{M}%
_{{\small 4\times 4}}\right) =1$ requiring $\left( \det \boldsymbol{X}%
_{1}\right) \left( \det \boldsymbol{X}_{2}\right) ^{-1}=1;$ and $\left(
\mathbf{ii}\right) $ the projective symmetry $\mathbb{C}^{\ast }:\left(
\boldsymbol{X}_{{\small 1}},\boldsymbol{X}_{{\small 2}}\right) \rightarrow
\lambda \left( \boldsymbol{X}_{{\small 1}},\boldsymbol{X}_{{\small 2}%
}\right) $ permitting to set $\det \boldsymbol{X}_{1}=1$ and then $\det
\boldsymbol{X}_{2}=1;$ thus restricting the GL(2$,\mathbb{C}$)$_{i}$ blocks
to SL(2$,\mathbb{C}$)$_{i}.$ Below, we describe useful features of the
geometry of the GL(2$,\mathbb{C}$)s while focussing on one of them; say GL(2$%
,\mathbb{C}$)$_{1}$. To that purpose, we begin by the 2$\times $2 matrix
realisation of the Lie group $GL(2,\mathbb{C})_{1}$ given by
\begin{equation}
\boldsymbol{X}_{1}=\left(
\begin{array}{cc}
x_{1} & u_{1} \\
y_{1} & v_{1}%
\end{array}%
\right) \qquad ,\qquad \boldsymbol{X}_{1}^{-1}=\frac{1}{\mathrm{\mu }}\left(
\begin{array}{cc}
v_{1} & -u_{1} \\
-y_{1} & x_{1}%
\end{array}%
\right) :=\left(
\begin{array}{cc}
\tilde{v}_{1} & -\tilde{u}_{1} \\
-\tilde{y}_{1} & \tilde{x}_{1}%
\end{array}%
\right)  \label{1X}
\end{equation}%
with $\det \boldsymbol{X}_{1}=\mathrm{\mu }$ where $\mathrm{\mu }$ is a
fixed complex number ($\mathrm{\mu =cte}$) parameterising the abelian factor
GL(1$,\mathbb{C}$) in the group GL(2$,\mathbb{C}$)$_{1}$ which factorises
like $GL(2)_{1}=SL(2)_{1}\times GL(1)_{1}.$ The determinant of the above
even matrix describes a 3D complex space $\mathcal{V}_{1}$ (resp. $\mathcal{%
\tilde{V}}_{1}$) parameterised by the 4 coordinates ($%
x_{1},y_{1},u_{1},v_{1} $) [resp. ($\tilde{x}_{1},\tilde{y}_{1},\tilde{u}%
_{1},\tilde{v}_{1}$)] constrained as $x_{1}v_{1}-y_{1}u_{1}=\mathrm{\mu }$
(resp. $\tilde{x}_{1}\tilde{v}_{1}-\tilde{y}_{1}\tilde{u}_{1}=1/\mathrm{\mu }
$). For simplicity, we rewrite these constraints just like
\begin{equation}
\mathcal{V}:\quad xv-yu=\mathrm{\mu }\qquad ,\qquad \mathcal{\tilde{V}}%
:\quad \tilde{x}\tilde{v}-\tilde{y}\tilde{u}=\mathrm{\tilde{\mu}}=\frac{1}{%
\mathrm{\mu }}  \label{x1}
\end{equation}%
by dropping out the subscript 1. As we will see below, it turns out that it
is more interesting to work with the inverse matrix $\boldsymbol{X}_{1}^{-1}$
than the $\boldsymbol{X}_{1}.$ We refer to the use of $\boldsymbol{X}%
_{1}^{-1}$ as the dual parametrisation.

In the realisations (\ref{x1}), the complex threefold $\mathcal{V}_{1}$
(resp. $\mathcal{\tilde{V}}_{1}$) is given by a complex 3D hypersurface $%
\det \boldsymbol{X}_{1}=\mathrm{\mu }$ (resp. $\det \boldsymbol{X}%
_{1}^{-1}=1/\mathrm{\mu }$) in \textrm{the ambient complex space} $\mathbb{C}%
^{4|0}$ with coordinates ($x,y,u,v$) [resp. ($\tilde{x},\tilde{y},\tilde{u},%
\tilde{v}$)]. For later use, notice that when we study D($2,1;\zeta $), we
will think about the modulus $\mathrm{\mu }$ as a function of the $\zeta $\
parameter namely%
\begin{equation}
\frac{1}{\mathrm{\mu }}=\mathrm{\kappa }_{0}\left( \zeta \right) \qquad
,\qquad \mathrm{\kappa }_{0}=\frac{1}{2}\zeta \left( \zeta +1\right)
\end{equation}%
with $\mathrm{\kappa }_{0}\left( 1\right) =1$ and $\mathrm{\kappa }%
_{0}\left( -1\right) =0$ corresponding to OSp(4\TEXTsymbol{\vert}2) and PSL(2%
\TEXTsymbol{\vert}2) respectively. Notice moreover the following:

\begin{description}
\item[$\left( \mathbf{a}\right) $] the complex 3D hypersurfaces $\mathcal{V}$
and $\mathcal{\tilde{V}}$ are dual threefolds with duality transformation
given by the mapping $(\tilde{x},\tilde{y},\tilde{u},\tilde{v})=\frac{1}{%
\mathrm{\mu }}(v,-y,-u,x).$

\item[$\left( \mathbf{b}\right) $] the complex 3D hypersurface $xv-yu=%
\mathrm{\mu }$ contains two special real threefolds: \newline
$\left( \mathbf{i}\right) $ a parametric compact 3-sphere $\mathbb{S}%
^{3}\left( \mathrm{\mu }\right) $ that can roughly imagined in term of a
group coset as
\begin{equation}
\mathbb{S}^{3}\left( \mathrm{\mu }\right) \simeq SO(4)/SO(3)\simeq
SO(3)\simeq SU(2)
\end{equation}%
This 3-sphere sits in $\mathbb{C}^{4}$ at the $SU(2)$-locus given by
\begin{equation}
SU(2):\quad (u,v)=(-\bar{y},\bar{x})  \label{uvp}
\end{equation}%
and is described by the equation
\begin{equation}
\mathbb{S}^{3}\left( \mathrm{\mu }\right) :\quad x\bar{x}+y\bar{y}%
=\left\vert \mathrm{\mu }\right\vert \geq 0
\end{equation}%
$\left( \mathbf{ii}\right) $ a parametric non compact "pseudo 3-sphere" $%
\mathbb{S}^{1,2}(\mathrm{\mu })$ that can be also thought of in terms of
groups as follows
\begin{equation}
\mathbb{S}^{1,2}(\mathrm{\mu })\simeq SO(2,2)/SO(1,2)\simeq SO(1,2)\simeq
SU(1,1)
\end{equation}%
It sits in $\mathbb{C}^{4}$ at the $SU(1,1)$-locus
\begin{equation}
SU(1,1):\quad (u,v)=(\bar{y},\bar{x})  \label{uvm}
\end{equation}%
and is given by the algevraic equation
\begin{equation}
\mathbb{S}^{1,2}(\mathrm{\mu }):\quad x\bar{x}-y\bar{y}=\left\vert \mathrm{%
\mu }\right\vert
\end{equation}%
Comparing (\ref{uvp}) and (\ref{uvm}), we learn that these two loci can be
mapped into each other by a discrete Z$_{2}$ transformation acting on the
variables as follows
\begin{equation}
\begin{tabular}{ccccc}
SU(2) & : & $(u,v)$ & $=$ & $(-\bar{y},\bar{x})$ \\
$\downarrow $ &  &  & $\downarrow $ &  \\
SU(1,1) & : & $(u,v)$ & $=$ & $(+\bar{y},\bar{x})$%
\end{tabular}
\label{211}
\end{equation}%
In the limit $\left\vert \mathrm{\mu }\right\vert \rightarrow \infty $, we
have the "decompactifications" to the flat spaces
\begin{equation}
\lim_{\left\vert \mathrm{\mu }\right\vert \rightarrow \infty }\mathbb{S}%
^{3}\left( \mathrm{\mu }\right) \simeq \mathbb{R}^{3}\qquad ,\qquad
\lim_{\left\vert \mathrm{\mu }\right\vert \rightarrow \infty }\mathbb{S}%
^{1,2}(\mathrm{\mu })\simeq \mathbb{R}^{1,2}  \label{221}
\end{equation}%
By setting $x=\xi _{1}+i\xi _{2}$ and $y=\xi _{3}+i\xi _{4}$ with real $\xi
_{i}$, eqs(\ref{x1}) expand as%
\begin{equation}
\begin{tabular}{lllll}
$\mathbb{S}^{3}\left( \mathrm{\mu }\right) $ & $:$ & $\xi _{1}^{2}+\xi
_{2}^{2}+\xi _{3}^{2}+\xi _{4}^{2}$ & $=$ & $\left\vert \mathrm{\mu }%
\right\vert $ \\
$\mathbb{S}^{1,2}(\mathrm{\mu })$ & $:$ & $\xi _{1}^{2}+\xi _{2}^{2}-\xi
_{3}^{2}-\xi _{4}^{2}$ & $=$ & $\left\vert \mathrm{\mu }\right\vert $%
\end{tabular}
\label{v}
\end{equation}

\item[$\left( \mathbf{c}\right) $] the dual hypersurface $\mathcal{\tilde{V}}
$ described by the relation $\tilde{x}\tilde{v}-\tilde{y}\tilde{u}=1/\mathrm{%
\mu }$ contains also two special real threefolds given by the real forms of
SL(2,$\mathbb{C}$) namely: \newline
$\left( \mathbf{i}\right) $ a 3-sphere $\mathbb{\tilde{S}}^{3}\left( \mathrm{%
\mu }\right) \simeq SU(2)$ (dual to the real $\mathbb{S}^{3}\left( \mathrm{%
\mu }\right) $) sitting at the locus
\begin{equation}
SU(2):\quad (\tilde{u},\tilde{v})=(-\overline{\tilde{y}},\overline{\tilde{x}}%
)  \label{vup}
\end{equation}%
it is described by the equation $\left\vert \tilde{x}\right\vert
^{2}+\left\vert \tilde{y}\right\vert ^{2}=1/\left\vert \mathrm{\mu }%
\right\vert .$ This 3-sphere shrinks to a point in the limit $\left\vert
\mathrm{\mu }\right\vert \rightarrow \infty ,$ i.e:%
\begin{equation}
\lim_{\left\vert \mathrm{\mu }\right\vert \rightarrow \infty }\mathbb{\tilde{%
S}}^{3}\left( \mathrm{\mu }\right) \simeq \{\left( \tilde{x},\tilde{y}%
\right) =\left( 0,0\right) \}  \label{223}
\end{equation}%
We refer to this degenerate limit as an \emph{SU(2) singularity}. For later
use, notice that this singularity can be expressed in terms of vanishing "%
\emph{gap energy}" used in the study of topological insulators and
topological matter \textrm{\cite{W1,W2,FK,Z,TM1,TM2,TM3}}. Indeed,
expressing the equation of the 3-sphere in terms of $\left\vert \tilde{x}%
\right\vert _{\pm }$ and $\left\vert \tilde{y}\right\vert _{\pm }$ like
\begin{equation}
\left\vert \tilde{x}\right\vert _{\pm }=\pm \sqrt{\frac{1}{\left\vert
\mathrm{\mu }\right\vert }-\left\vert \tilde{y}\right\vert ^{2}}\qquad
\Leftrightarrow \qquad \left\vert \tilde{y}\right\vert _{\pm }=\pm \sqrt{%
\frac{1}{\left\vert \mathrm{\mu }\right\vert }-\left\vert \tilde{x}%
\right\vert ^{2}}
\end{equation}%
with $0\leq \left\vert \tilde{x}\right\vert ^{2}\leq \frac{1}{\left\vert
\mathrm{\mu }\right\vert }$ and $0\leq \left\vert \tilde{y}\right\vert
^{2}\leq \frac{1}{\left\vert \mathrm{\mu }\right\vert };$ it follows that
the "gap energy" defined as
\begin{equation}
\Delta \left\vert \tilde{x}\right\vert _{g}=\min (\left\vert \tilde{x}%
\right\vert _{+}-\left\vert \tilde{x}\right\vert _{-})\qquad ,\qquad \Delta
\left\vert \tilde{y}\right\vert _{g}=\min (\left\vert \tilde{y}\right\vert
_{+}-\left\vert \tilde{y}\right\vert _{-})
\end{equation}%
vanishes identically for $\left\vert \mathrm{\mu }\right\vert \rightarrow
\infty .$ \newline
$\left( \mathbf{ii}\right) $ a "pseudo 3-sphere" $\mathbb{\tilde{S}}^{1,2}(%
\mathrm{\mu })\simeq SU(1,1)$ (dual to the real $\mathbb{S}^{1,2}\left(
\mathrm{\mu }\right) $) sitting in $\mathbb{C}^{4}$ at the locus
\begin{equation}
SU(1,1):\quad (\tilde{u},\tilde{v})=(\overline{\tilde{y}},\overline{\tilde{x}%
})  \label{vpm}
\end{equation}%
it is described by the equation $\left\vert \tilde{x}\right\vert
^{2}-\left\vert \tilde{y}\right\vert ^{2}=1/\left\vert \mathrm{\mu }%
\right\vert .$ This relation can be also presented in terms of branches like%
\begin{equation}
\left\vert \tilde{x}\right\vert _{\pm }=\pm \sqrt{\left\vert \tilde{y}%
\right\vert ^{2}+\frac{1}{\left\vert \mathrm{\mu }\right\vert }}\qquad
\Leftrightarrow \qquad \left\vert \tilde{y}\right\vert _{\pm }=\pm \sqrt{%
\left\vert \tilde{x}\right\vert ^{2}-\frac{1}{\left\vert \mathrm{\mu }%
\right\vert }}
\end{equation}%
with $\left\vert \tilde{x}\right\vert ^{2}\geq \frac{1}{\left\vert \mathrm{%
\mu }\right\vert }$ and $\left\vert \tilde{y}\right\vert ^{2}\geq 0$. In
this case, the "gap energy" is as follows
\begin{equation}
\Delta \left\vert \tilde{x}\right\vert _{g}=2\sqrt{\frac{1}{\left\vert
\mathrm{\mu }\right\vert }}\qquad ,\qquad \Delta \left\vert \tilde{y}%
\right\vert _{g}=0\qquad ,\qquad \left\vert \tilde{x}\right\vert ^{2}=\frac{1%
}{\left\vert \mathrm{\mu }\right\vert }
\end{equation}%
In the limit $\left\vert \mathrm{\mu }\right\vert \rightarrow \infty $, this
gap energy vanishes and the two hyperbols $\left\vert \tilde{x}\right\vert
_{\pm }=\pm \sqrt{\left\vert \tilde{y}\right\vert ^{2}+1/\left\vert \mathrm{%
\mu }\right\vert }$ (resp. $\left\vert \tilde{y}\right\vert _{\pm }=\pm
\sqrt{\left\vert \tilde{x}\right\vert ^{2}-1/\left\vert \mathrm{\mu }%
\right\vert }$) are given by two cones $\left\vert \tilde{y}\right\vert
=+\left\vert \tilde{x}\right\vert $ and $\left\vert \tilde{y}\right\vert
=-\left\vert \tilde{x}\right\vert $ with opposite apex. In this case, the
pseudo 3-sphere shrinks to the cones
\begin{equation}
\lim_{\left\vert \mathrm{\mu }\right\vert \rightarrow \infty }\mathbb{\tilde{%
S}}^{1,2}(\mathrm{\mu })\simeq \{\left\vert \tilde{y}\right\vert =\pm
\left\vert \tilde{x}\right\vert \}  \label{VV}
\end{equation}%
We refer to this conical limit as an{\small \ "}\emph{SU(1,1) singularity}"
in analogy with the SU$\left( 2\right) $ singularity where the 3-sphere $%
\mathbb{\tilde{S}}^{3}(\mathrm{\mu })\rightarrow 0$.\newline
Similar relations to eqs(\ref{v}) can be also written down for the dual
manifold; we have%
\begin{equation}
\begin{tabular}{lllll}
$\mathbb{\tilde{S}}^{3}(\mathrm{\mu })$ & $:$ & $\tilde{\xi}_{1}^{2}+\tilde{%
\xi}_{2}^{2}+\tilde{\xi}_{3}^{2}+\tilde{\xi}_{4}^{2}$ & $=$ & $\frac{1}{%
\left\vert \mathrm{\mu }\right\vert }$ \\
$\mathbb{\tilde{S}}^{1,2}(\mathrm{\mu })$ & $:$ & $\tilde{\xi}_{1}^{2}+%
\tilde{\xi}_{2}^{2}-\tilde{\xi}_{3}^{2}-\tilde{\xi}_{4}^{2}$ & $=$ & $\frac{1%
}{\left\vert \mathrm{\mu }\right\vert }$%
\end{tabular}%
\end{equation}%
with%
\begin{equation}
\mathbb{\tilde{S}}^{1,2}(\mathrm{\mu }):\quad \tilde{\xi}_{1\pm }=\pm \sqrt{%
\frac{1}{\left\vert \mathrm{\mu }\right\vert }+\tilde{\xi}_{4}^{2}+\tilde{\xi%
}_{3}^{2}-\tilde{\xi}_{2}^{2}}
\end{equation}%
In the limit $\left\vert \mathrm{\mu }\right\vert \rightarrow \infty $, they
contract to%
\begin{equation}
\begin{tabular}{lllll}
$\mathbb{\tilde{S}}^{3}(\mathrm{\infty })$ & $:$ & $\tilde{\xi}_{1}^{2}+%
\tilde{\xi}_{2}^{2}+\tilde{\xi}_{3}^{2}+\tilde{\xi}_{4}^{2}$ & $=$ & $0$ \\
$\mathbb{\tilde{S}}^{1,2}(\mathrm{\infty })$ & $:$ & $\tilde{\xi}_{1}^{2}+%
\tilde{\xi}_{2}^{2}-\tilde{\xi}_{3}^{2}-\tilde{\xi}_{4}^{2}$ & $=$ & $0$%
\end{tabular}%
\end{equation}%
with the two hyperboloids
\begin{equation}
\mathbb{\tilde{S}}^{1,2}(\mathrm{\infty }):\quad \tilde{\xi}_{1\pm }=\pm
\sqrt{\tilde{\xi}_{4}^{2}+\tilde{\xi}_{3}^{2}-\tilde{\xi}_{2}^{2}}
\end{equation}%
colliding at $\tilde{\xi}_{4}^{2}+\tilde{\xi}_{3}^{2}=\tilde{\xi}_{2}^{2}.$
\end{description}

In what follows, we will use the dual $\mathcal{\tilde{V}}$ containing $%
\mathbb{\tilde{S}}^{3}$ and $\mathbb{\tilde{S}}^{1,2}$ instead of the $%
\mathcal{V}$ having $\mathbb{S}^{3}$ and the pseudo $\mathbb{S}^{1,2};$ this
is because $\mathbb{\tilde{S}}^{3}$ and $\mathbb{\tilde{S}}^{1,2}$ have
direct applications in our calculations; they are intimately linked with the
simple root $\mathbf{\alpha }$ of two real forms of SL(2,$\mathbb{C}$) and
the associated homological 2-cycle $\mathcal{C}_{\mathbf{\alpha }}$ of its
algebraic geometry; i.e:%
\begin{equation*}
\mathbf{\alpha \in \Phi }_{SL_{2}}\qquad \leftrightarrow \qquad \mathcal{C}_{%
\mathbf{\alpha }}\in H_{2}\left( SL_{2}\right)
\end{equation*}%
For convenience, we will refer to $\mathcal{\tilde{V}}$ simply like $%
\mathcal{X}$ and often drops out the twilda label on the $\mathbb{\tilde{S}}%
^{3}$ and $\mathbb{\tilde{S}}^{1,2}$ by referring to them just as $\mathbb{S}%
^{3}$ and $\mathbb{S}^{1,2}$.

Given (\ref{221}) as well as (\ref{223}) and (\ref{VV}), an interesting
question arises here: How much differs the non compact 3D hypersurface $%
\mathcal{X}$ from the flat $\mathbb{C}^{3}\simeq \mathbb{R}^{6}$?. We answer
this question by considering two local representations:

\paragraph{$\qquad $\textbf{A)} $\mathcal{X}$\ as cotangent T$^{\ast }%
\mathbb{\tilde{S}}^{3}(\mathrm{\protect\mu })$ or T$^{\ast }\mathbb{\tilde{S}%
}^{1,2}(\mathrm{\protect\mu })$}

\ \ \newline
Starting from (\ref{x1}) and taking the limit $\frac{1}{\mathrm{\mu }}%
\rightarrow \mathrm{0,}$ one learns that $\mathcal{X}$ is given by a complex
3D conifold \textrm{\cite{Conifold,cone2,cone3}}
\begin{equation}
\mathcal{X}:\quad vx-yu=\frac{1}{\mathrm{\mu }}\rightarrow 0
\end{equation}%
By setting setting $x=a+b$ and $v=\bar{a}-\bar{b}$ as well as $y=c+d$ and $u=%
\bar{c}-\bar{d}$ with complex (a,b,c,d), we can expand $vx-yu=1/\mathrm{\mu }
$ as follows (by using $\func{Im}\mathrm{\mu =0,}$ and $\func{Re}\mathrm{\mu
>>0}$)
\begin{equation}
\begin{tabular}{lllllll}
$\mathfrak{B}$ & $:$ & $\left\vert a\right\vert ^{2}+\left\vert d\right\vert
^{2}-\left\vert b\right\vert ^{2}-\left\vert c\right\vert ^{2}$ & $=$ & $+%
\frac{1}{\mathrm{\mu \bar{\mu}}}\func{Re}\mathrm{\mu }$ & $=$ & $\frac{1}{%
\func{Re}\mathrm{\mu }}$ \\
$\mathfrak{F}$ & $:$ & $b\bar{a}-a\bar{b}+d\bar{c}-c\bar{d}_{1}$ & $=$ & $-%
\frac{1}{\mathrm{\mu \bar{\mu}}}\func{Im}\mathrm{\mu }$ & $=$ & $0$%
\end{tabular}%
\end{equation}%
These two real equations can be solved in two interesting ways given by the
two real forms SU(2){\small \ and }SU(1,1){\small \ }as follows%
\begin{equation}
\begin{tabular}{|c|c|c|}
\hline
{\small SU(2)} & : & $b=c=0$ \\ \hline
$\mathfrak{B}=\mathbb{S}^{3}$ & : & $\left\vert a\right\vert ^{2}+\left\vert
d\right\vert ^{2}=\frac{1}{\func{Re}\mathrm{\mu }}$ \\ \hline
$\mathcal{X}$ & = & T$^{\ast }\mathbb{S}^{3}$ \\ \hline
\end{tabular}%
\qquad ,\qquad
\begin{tabular}{|c|c|c|}
\hline
{\small SU(1,1)} & : & $b=d=0$ \\ \hline
$\mathfrak{B}=\mathbb{S}^{1,2}$ & : & $\left\vert a\right\vert
^{2}-\left\vert c\right\vert ^{2}=\frac{1}{\func{Re}\mathrm{\mu }}$ \\ \hline
$\mathcal{X}$ & = & T$^{\ast }\mathbb{S}^{1,2}$ \\ \hline
\end{tabular}
\label{B3}
\end{equation}%
with $T^{\ast }E$ designating the cotangent space.

\begin{figure}[tbph]
\begin{center}
\includegraphics[width=15cm]{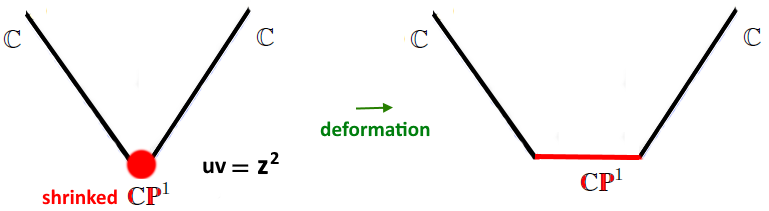}
\end{center}
\par
\vspace{-0.5cm}
\caption{A toric representation of the complex surface $\Sigma _{2}\sim
\mathbb{C}\times \mathbb{CP}^{1}$. The compact line is in red color. The non
compact directions is given by the line $\mathbb{C}$. In real variables, we
have $\mathcal{X}\simeq \mathbb{R}^{2}\times \mathbb{S}_{\protect\alpha %
}^{2} $; the 2-sphere (2-cycle $\mathcal{C}_{\mathbf{\protect\alpha }}$) is
associated with the simple root $\mathbf{\protect\alpha }$ of su(2).}
\label{0TR}
\end{figure}

\paragraph{$\qquad $\textbf{B)} $\mathcal{X}$\ as complex $\mathbb{C}\times
T^{\ast }\mathbb{\tilde{S}}^{2}(\mathrm{\protect\mu })$ or $\mathbb{C}\times
T^{\ast }\mathbb{\tilde{S}}^{1,1}(\mathrm{\protect\mu })$:}

\ \ \ \newline
These are local complex fibrations of $\mathcal{X}$ that may be derived by
using: $\left( \mathbf{i}\right) $ the cotangent $T^{\ast }\mathbb{S}^{3}$
of and the T$^{\ast }\mathbb{S}^{1,2}$ of eq(\ref{B3}); and $\left( \mathbf{%
ii}\right) $ substituting the Hopf fibration $\mathbb{S}^{3}\simeq \mathbb{S}%
^{2}\times \mathbb{S}^{1}$ and $\mathbb{S}^{1,2}$ $\simeq \mathbb{S}%
^{1,1}\times \mathbb{S}^{1}$. We end up with%
\begin{equation}
\begin{tabular}{|c|c|cccc|}
\hline
{\small SU(2)/U(1)} & : & $b=c$ & $=$ & $0$ &  \\ \hline
$\mathfrak{B}=\mathbb{S}^{2}$ & : & $\left\vert a\right\vert ^{2}+\left\vert
d\right\vert ^{2}$ & $=$ & $\frac{1}{\func{Re}\mathrm{\mu }}$ & $,\qquad
a\equiv e^{i\vartheta }a,\qquad d\equiv e^{-i\vartheta }d$ \\ \hline
$\mathcal{X}$ & = & $\mathbb{C}\times \Sigma _{2}$ & $=$ & $\mathbb{C}\times
T^{\ast }\mathbb{S}^{2}$ &  \\ \hline
$\Sigma _{2}$ & = & $T^{\ast }\mathbb{S}^{2}$ &  &  &  \\ \hline
\end{tabular}%
\end{equation}%
\begin{equation*}
\end{equation*}%
where we have used $\mathbb{C\simeq }T^{\ast }\mathbb{S}^{1}$\ and%
\begin{equation}
\begin{tabular}{|c|c|cccc|}
\hline
{\small SU(1,1)/U(1)} & : & $b=d$ & $=$ & $0$ &  \\ \hline
$\mathfrak{B}=\mathbb{S}^{1,1}$ & : & $\left\vert a\right\vert
^{2}-\left\vert d\right\vert ^{2}$ & $=$ & $\frac{1}{\func{Re}\mathrm{\mu }}$
& $,\qquad a\equiv e^{i\vartheta }a,\qquad c\equiv e^{-i\vartheta }d$ \\
\hline
$\mathcal{X}$ & = & $\mathbb{C}\times \Sigma _{1,1}$ & $=$ & $\mathbb{C}%
\times T^{\ast }\mathbb{S}^{1,1}$ &  \\ \hline
$\Sigma _{1,1}$ & = & $T^{\ast }\mathbb{S}^{1,1}$ &  &  &  \\ \hline
\end{tabular}%
\end{equation}%
\begin{equation*}
\end{equation*}%
To establish these local fibrations explicitly, we $\left( \mathbf{i}\right)
$ denote by $\left( z_{1},z_{2},z_{3}\right) $ the three free complex
variables parameterising the flat $\mathbb{C}^{3}$, and $\left( \mathbf{ii}%
\right) $ look for expressing the relation $xv-yv=1/\mathrm{\mu }$ as a non
linear function $v=f\left( z_{1},z_{2},z_{3}\right) $ with variable $v$ as
in (\ref{x1}). To that purpose, we set $\left( z_{1},z_{2},z_{3}\right)
=\left( 1/x,y,-u\right) ;$ this allows to factorise the constraint relation (%
\ref{x1}) like%
\begin{equation}
\begin{tabular}{ccccc}
$v$ & $=$ & z$_{1}\mathrm{G}\left( z\right) $ & $=$ & $z_{1}\left[ \frac{1}{%
\mathrm{\mu }}-\mathrm{\sigma }\left( z\right) \right] $ \\
$\mathrm{\sigma }$ & $=$ & z$_{2}z_{3}$ &  &
\end{tabular}
\label{fac}
\end{equation}%
where $z_{1}$ parameterises a complex line $\mathbb{C}$ and where $\mathrm{G}%
\left( z\right) $ describes a complex surface $\Sigma $ given by the
equation $\mathrm{G}\left( z\right) =1/\mathrm{\mu }-\mathrm{\sigma }\left(
z\right) $ with zeros at $z_{2}z_{3}=1/\mathrm{\mu }$ describing a
projective line. With the factorisation (\ref{fac}), we then have $\mathcal{X%
}=\mathbb{C}\times \Sigma $. Notice that by substituting $z_{2}=y$ and $%
z_{3}=-u$, the zeros equation $z_{2}z_{3}=1/\mathrm{\mu }$ reads as $yu=-1/%
\mathrm{\mu .}$ Moreover, substituting $u=\mp \bar{y}$ giving the loci of
the $\mathbb{S}^{3}$ and $\mathbb{S}^{1,2}$ as given by (\ref{211}), we end
up with
\begin{equation}
SU(2)\quad :\bar{y}y=+\frac{1}{\func{Re}\mathrm{\mu }}\qquad ,\qquad
SU(1,1)\quad :\bar{y}y=-\frac{1}{\func{Re}\mathrm{\mu }}
\end{equation}%
putting a constraint on the sign of $\func{Re}\mathrm{\mu ;}$ which is
positive for SU(2) and negative for SU(1,1)$\mathrm{.}$ Recall that for the
case where $z_{2}z_{3}=z^{2},$ we have an ALE space with SU(2) singularity
describing a complex surface $\Sigma _{2}\sim \mathbb{C}\times \mathbb{S}%
^{2} $ with shrinking 2-sphere to a point. This surface has a graphic
representation in terms of toric diagrams as depicted by the \textbf{Figure}
\textbf{\ref{0TR}.}

\subsubsection{Even and odd parts of SL(2\TEXTsymbol{\vert}2)}

\begin{figure}[tbph]
\begin{center}
\includegraphics[width=14cm]{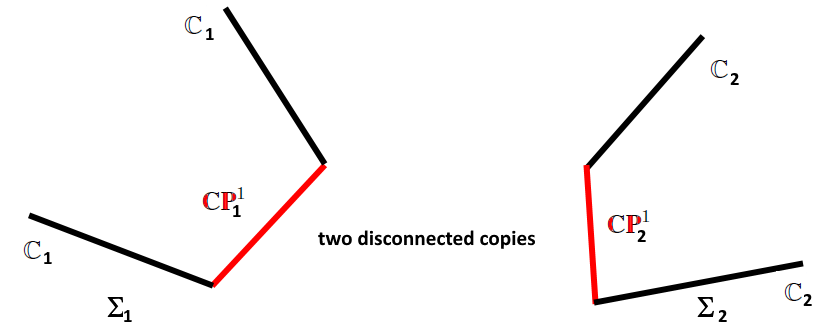}
\end{center}
\par
\vspace{-0.5cm}
\caption{Toric graph of the fibration $\mathfrak{B}_{12}\times \mathfrak{F}%
_{12}$ describing the $\left( \Sigma _{2}\right) _{1}\times \left( \Sigma
_{2}\right) _{2}$ geometry of the $SU(2)_{1}\times SU(2)_{2}$ symmetry
without bi-fundamentals. The red segments represent $\mathbb{CP}^{1}$
curves. }
\label{CP1}
\end{figure}

\qquad The even part of SL(2\TEXTsymbol{\vert}2)$_{\bar{0}}$ is given by $%
SL(2)_{1}\times SL(2)_{2}\times \mathbb{C}^{\ast }$; it has two isospin
copies; and as such it has a discrete $\mathcal{Z}_{2}$ outer automorphism
acting by the transposition $SL(2)_{1}\leftrightarrow SL(2)_{2}.$ To
implement this property, the geometry of SL(2\TEXTsymbol{\vert}2)$_{\bar{0}}$
denoted like $\mathcal{X}_{120}$ factorises as $\mathcal{X}_{12}\times
\mathbb{C}^{\ast }$ with $\mathcal{X}_{12}$ given by the product $\mathcal{X}%
_{1}\times \mathcal{X}_{2}.$ The complex manifold $\mathcal{X}_{12}$ is a
complex 6D manifold sitting in the $\mathbb{C}_{12}^{4|4}\sim \mathbb{C}%
_{1}^{4}\times \mathbb{C}_{2}^{4}$ parameterised by ($%
x_{1},y_{1},u_{1},v_{1} $) and ($x_{2},y_{2},u_{2},v_{2}$). Because $SL(2)$
has two real forms $SU(2) $ and $SU(1,1)$, one distinguishes the two
particular real forms of PSL(2\TEXTsymbol{\vert}2)$_{\bar{0}}$ given by%
\begin{equation}
\begin{tabular}{lllll}
$\left( \mathbf{i}\right) $ & : & $PSU(2|2)_{\bar{0}}$ & $=$ & $%
SU(2)_{1}\times SU(2)_{2}$ \\
$\left( \mathbf{ii}\right) $ & : & $PSU(1,1|2)_{\bar{0}}$ & $=$ & $%
SU(1,1)_{1}\times SU(2)_{2}$%
\end{tabular}%
\end{equation}%
For local graphic representations, we use the $SU(2|2)$ having compact
2-spheres $\mathbb{S}_{1}^{2}$ and $\mathbb{S}_{2}^{2}$. Using the \textrm{%
toric graph }of the \textbf{Figure} \textbf{\ref{0TR}}, the $\mathcal{X}%
_{12} $ can be represented by two disconnected toric diagrams \textrm{\cite%
{LEUNG,SEBB,AbouN}} as depicted by the \textbf{Figure} \textbf{\ref{CP1}}.
Notice that the two (red) 2-spheres in $\mathbb{S}_{1}^{2}\times \mathbb{S}%
_{2}^{2}$ have an interpretation in terms of the even\footnote{%
\ Sometimes these simple roots are denoted like $\mathbf{\beta }_{1}=\mathbf{%
\alpha }_{1}$\ and $\mathbf{\beta }_{2}=\mathbf{\alpha }_{3}$.} roots $%
\mathbf{\beta }_{1}$ and $\mathbf{\beta }_{2}$ of the even $psu(2|2)_{\bar{0}%
}=su(2)\oplus su(2)$.

The complex 6D fibration $\mathcal{X}_{12}$ is defined by the algebraic
equations%
\begin{equation}
\begin{tabular}{lllll}
$\det \boldsymbol{X}_{1}$ & : & $x_{1}y_{1}-u_{1}v_{1}$ & $=$ & $\frac{1}{%
\mathrm{\mu }}$ \\
$\det \boldsymbol{X}_{2}$ & : & $x_{2}y_{2}-u_{2}v_{2}$ & $=$ & $\frac{1}{%
\mathrm{\mu }}$%
\end{tabular}
\label{x12}
\end{equation}%
they can be merged into one constraint relation given by Berezinian $%
Ber\left( \boldsymbol{M}\right) =1$ of the super matrix $\boldsymbol{M}$
combining $\boldsymbol{X}_{1}$ and $\boldsymbol{X}_{2}$ as follows%
\begin{equation}
\boldsymbol{M}=\left(
\begin{array}{cc}
\boldsymbol{X}_{1} & \mathbf{0} \\
\mathbf{0} & \boldsymbol{X}_{2}%
\end{array}%
\right) ,\qquad \boldsymbol{X}_{i}=\left(
\begin{array}{cc}
x_{i} & u_{i} \\
y_{i} & v_{i}%
\end{array}%
\right)
\end{equation}%
Recall that the $Ber\left( \boldsymbol{M}\right) $ of the above super matrix
is given by $\left( \det \boldsymbol{X}_{1}\right) \left( \det \boldsymbol{X}%
_{2}\right) ^{-1}$; it reads explicitly as follows%
\begin{equation}
\frac{x_{1}y_{1}-u_{1}v_{1}}{x_{2}y_{2}-u_{2}v_{2}}=1  \label{sdet}
\end{equation}%
it describes a 7D hypersurface in $\mathbb{C}^{4|4}$ and gives the complex
7D geometry of SL(2\TEXTsymbol{\vert}2)$_{\bar{0}}$. In this regards, notice
the two following:

\begin{description}
\item[$\left( \mathbf{i}\right) $] the super determinant $Ber\left(
\boldsymbol{M}\right) $ is invariant under the scaling $\det \boldsymbol{X}%
_{i}\rightarrow \lambda ^{2}\det \boldsymbol{X}_{i}$ induced from the
projective transformation $\boldsymbol{X}_{i}\rightarrow \lambda \boldsymbol{%
X}_{i}.$

\item[$\left( \mathbf{ii}\right) $] this scaling describes precisely the
central element $H_{0}$ generating the GL(1,$\mathbb{C}$) subgroup of SL(2%
\TEXTsymbol{\vert}2)$_{\bar{0}};$ thus permitting to define the 14D geometry
of PSL(2\TEXTsymbol{\vert}2)$_{\bar{0}}.$
\end{description}

\begin{figure}[h]
\begin{center}
\includegraphics[width=16cm]{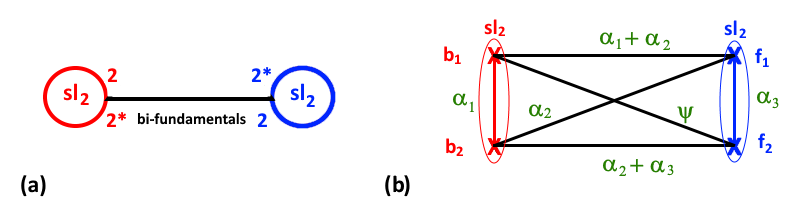}
\end{center}
\par
\vspace{-0.5cm}
\caption{On left, the nodes in red and blue represent the two sl(2,$\mathbb{C%
}$)s making the even psl(2\TEXTsymbol{\vert}2)$_{\bar{0}}$. The bridge
between the two nodes given by the bi-fundamentals represent the odd
generators of psl(2\TEXTsymbol{\vert}2)$_{\bar{1}}$ relating the two sl(2,$%
\mathbb{C}$)s. A dual description of this graph is given by the super Dynkin
diagrams of the Figure \textbf{\protect\ref{Asl2}}. On the right, the
graphic representation in terms of the roots.}
\label{23F}
\end{figure}

\begin{figure}[tbph]
\begin{center}
\includegraphics[width=14cm]{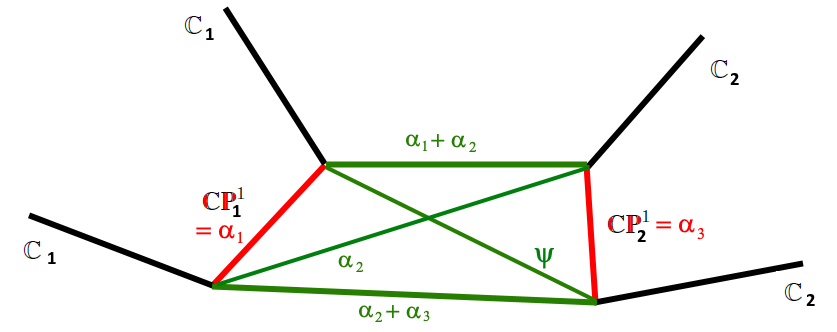}
\end{center}
\par
\vspace{-0.5cm}
\caption{Toric representation of the gluing of the surfaces $\Sigma _{1}$
and $\Sigma _{2}$ in the supergeometry of $PSL(2|2)$. The red segments refer
to the even cycles; the four green ones designate the odd cycles associated
with the odd roots.}
\label{CP2}
\end{figure}

For the eight dimensional odd part sl(2\TEXTsymbol{\vert}2)$_{\bar{1}}$; it
is given by the bi-fundamental representations $\mathbf{2}_{1}\otimes
\mathbf{2}_{2}^{c}$ and $\mathbf{2}_{1}^{c}\otimes \mathbf{2}_{2}$ of $SL(2,%
\mathbb{C})_{1}\times SL(2,\mathbb{C})_{2}$. These bi-fundamentals are
sketched by the \textbf{Figure} \textbf{\ref{23F}}.
The two nodes in this \textbf{Figure} corresponds to $\mathbf{2}_{1}\otimes
\mathbf{2}_{1}^{c}$ and $\mathbf{2}_{2}\otimes \mathbf{2}_{2}^{c}$; the
links to $\mathbf{2}_{i}\otimes \mathbf{2}_{j}^{c}$ and $\mathbf{2}%
_{i}^{c}\otimes \mathbf{2}_{j}.$ The bridge between the two (red) 2-spheres
is given by (green)\ fermionic 2-cycles generated by $\mathbf{\alpha }_{2}$
and $\mathbf{\mathbf{\alpha }}_{1}+\mathbf{\alpha }_{2}+\mathbf{\alpha }_{3}$%
\ as well as $\mathbf{\mathbf{\alpha }_{1}+\mathbf{\alpha }_{2}}$ and $%
\mathbf{\alpha }_{2}+\mathbf{\alpha }_{3}$. They are represented by the
green lines shown on the \textbf{Figure} \textbf{\ref{CP2}}. They permit to
glue together the two toric copies of the \textbf{Figure} \textbf{\ref{CP1}}%
.

In summary, the geometric description of the 15 dimensional sl(2\TEXTsymbol{%
\vert}2) is given by the complex super matrix (\ref{M44}-\ref{M45}) with
super determinant $Ber\left( \boldsymbol{M}\right) =1.$ This constraint
defines a complex 15 dim hypersurface in $\mathbb{C}^{8|8}$ reading
explicitly as follows%
\begin{equation}
\frac{\det \left( \boldsymbol{X}_{{\small 1}}\right) }{\det \left(
\boldsymbol{X}_{{\small 2}}\right) }\det \left[ \boldsymbol{I}-\boldsymbol{X}%
_{{\small 1}}^{-1}\boldsymbol{Y}_{{\small 1}}\boldsymbol{X}_{{\small 2}}^{-1}%
\boldsymbol{Y}_{{\small 2}}\right] =1
\end{equation}%
It shows that $\left( \mathbf{i}\right) $ the $\boldsymbol{X}_{{\small 1}}$
and $\boldsymbol{X}_{{\small 2}}$ are coupled through the bi-fundamentals $%
\boldsymbol{Y}_{{\small 1}}$ and $\boldsymbol{Y}_{{\small 2}}$ like $%
\boldsymbol{X}_{{\small 1}}^{-1}\boldsymbol{Y}_{{\small 1}}\boldsymbol{X}_{%
{\small 2}}^{-1}\boldsymbol{Y}_{{\small 2}}$ and $\left( \mathbf{ii}\right) $
the $Ber\left( \boldsymbol{M}\right) =1$ is invariant under the scaling $%
\boldsymbol{M}\rightarrow \lambda \boldsymbol{M}$ generated by the
projective transformations $\boldsymbol{X}_{{\small i}}\rightarrow \lambda
\boldsymbol{X}_{{\small i}}$ and $\boldsymbol{Y}_{{\small i}}\rightarrow
\lambda \boldsymbol{Y}_{{\small i}}$. By taking the coset with respect to
this projective scaling, we get the complex 14D manifold describing PSL(2%
\TEXTsymbol{\vert}2).

\subsection{Decompactifying the $\mathbb{S}^{2}$ inside SL($2,\mathbb{C}$)}

\begin{figure}[h]
\begin{center}
\includegraphics[width=16cm]{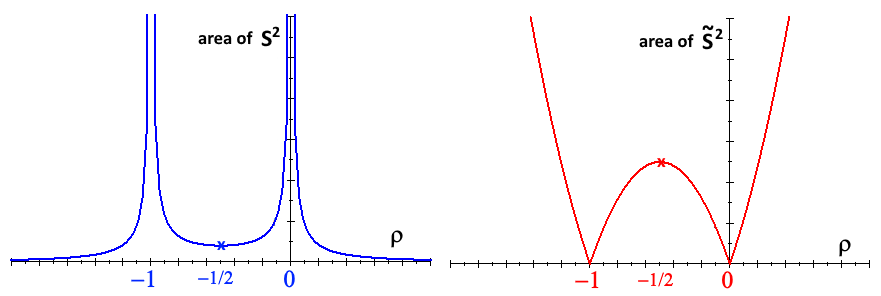}
\end{center}
\caption{Variation of the area of the real 2-sphere $\mathbb{S}^{2}$ (in
blue) and its dual $\mathbb{\tilde{S}}^{2}$ (in red). On the left, the
sphere is given by $\left\vert x^{2}\right\vert +\left\vert y^{2}\right\vert
=\left\vert \mathrm{\protect\mu }\right\vert $ and on the right it is given
by $\left\vert \tilde{x}^{2}\right\vert +\left\vert \tilde{y}^{2}\right\vert
=\left\vert \mathrm{\tilde{\protect\mu}}\right\vert $. They are functions of
the Khaler parameter $\protect\varrho $. The area of the real $\mathbb{S}%
^{2} $ explodes at the singular $\protect\zeta =-1,0$ in eq(\protect\ref{a})
while the dual $\mathbb{\tilde{S}}^{2}$\ shrinks.}
\label{area}
\end{figure}

\qquad Here, \textrm{we consider the} representation $\mathcal{X}\simeq
\mathbb{C}^{2}\times \mathbb{CP}^{1}$ to investigate the decompactification
of projective $\mathbb{CP}^{1}$ inside SL($2,\mathbb{C}$) towards the non
compact line $\mathbb{C}.$ This decompactification is done here with the
idea of applying it when studying the contraction of the exceptional super
group $D(2,1;\zeta )$ down to PSL(2\TEXTsymbol{\vert}2). It is achieved by
taking the infinite limit $\left\vert \mathrm{\mu }\right\vert \rightarrow
\infty $ of the area of the real 2-sphere (\ref{VV}); or equivalently the
shrinking of its dual given by the limit $\left\vert \mathrm{\tilde{\mu}}%
\right\vert \rightarrow 0$ as shown by the \textbf{Figure} \textbf{\ref{area}%
}.

\subsubsection{SL$(2,\mathbb{C})$ inside D$(2,1;\protect\zeta )_{\bar{0}}$}

\qquad We start by recalling that the even part of the exceptional group D$%
(2,1;\zeta )_{\bar{0}}$ is given by the product of three isospin groups as
follows%
\begin{equation}
\begin{tabular}{lll}
$D(2,1;\zeta )_{\bar{0}}$ & $=$ & $SL(2,\mathbb{C})_{1}\times SL(2,\mathbb{C}%
)_{2}\times SL(2,\mathbb{C})_{3}$ \\
& $\simeq $ & $SO(4,\mathbb{C})_{12}\times SL(2,\mathbb{C})_{3}$%
\end{tabular}%
\end{equation}%
The isospin group we want to decompactify is given by the third factor $SL(2,%
\mathbb{C})_{3}$ by using the analysis considered in deriving eq(\ref{VV}).
Instead of using the complex 2$\times $2 matrix $\boldsymbol{X}_{3}$ like%
\begin{equation}
\boldsymbol{X}_{3}=\left(
\begin{array}{cc}
x_{3} & u_{3} \\
y_{3} & v_{3}%
\end{array}%
\right) \qquad \rightarrow \qquad \boldsymbol{X}_{3}=\left(
\begin{array}{cc}
x & u \\
y & v%
\end{array}%
\right)  \label{a1}
\end{equation}%
with determinant $xv-yu=\mathrm{\mu }$, we work with the dual parametrisation%
\begin{equation}
\boldsymbol{X}_{3}^{-1}=\frac{1}{\mathrm{\mu }}\left(
\begin{array}{cc}
v_{3} & -u_{3} \\
-y_{3} & x_{3}%
\end{array}%
\right) \qquad \rightarrow \qquad \boldsymbol{X}_{3}^{-1}=\frac{1}{\mathrm{%
\mu }}\left(
\begin{array}{cc}
v & -u \\
-y & x%
\end{array}%
\right)
\end{equation}%
with determinant as%
\begin{equation}
\mathcal{X}_{3}:xv-yu=\frac{1}{\mathrm{\mu }}\qquad ,\qquad \mathcal{X}_{3}:=%
\mathcal{\tilde{V}}_{3}  \label{xv}
\end{equation}%
defining a complex 3D hypersurface $\mathcal{X}_{3}$ in $\mathbb{C}^{4}.$ To
study the decompactification of the $SL(2,\mathbb{C})_{3},$ we\textrm{\
focus on the locus }$\left( v,u\right) =\left( \bar{x},-\bar{y}\right) $%
\textrm{\ where lives the (dual) 3-sphere }$\mathbb{\tilde{S}}^{3},$\textrm{%
\ }
\begin{equation}
\left\vert x\right\vert ^{2}+\left\vert y\right\vert ^{2}=\frac{1}{\varrho }%
\qquad ,\qquad \varrho =\left\vert \mathrm{\mu }\right\vert  \label{3sp}
\end{equation}%
On this locus, it lives also a 2-sphere $\mathbb{\tilde{S}}^{2}$ descending
from the Hopf fibration of $\mathbb{\tilde{S}}^{3}\simeq \mathbb{\tilde{S}}%
^{2}\times \mathbb{\tilde{S}}^{1}.$ Under the identification $\left(
x,y\right) \simeq \left( e^{i\theta }x,e^{-i\theta }y\right) ,$ one gets a
2-sphere with area given by
\begin{equation}
\mathcal{A}(\mathbb{S}^{2})=4\pi \varrho \qquad ,\qquad \mathcal{A}(\mathbb{%
\tilde{S}}^{2})=\frac{4\pi }{\varrho }  \label{air}
\end{equation}

\subsubsection{Decompactified\emph{\ }$\mathbb{S}^{2}$\emph{\ }and shrinking
\emph{\ }$\mathbb{\tilde{S}}^{2}$}

\qquad From eq(\ref{air}), we see that the decompactification of the
2-sphere towards $\mathbb{C}$ can be realised by taking the limit $\varrho
\rightarrow \infty $ in the sphere; because $\frac{1}{\varrho }\rightarrow
0, $ the dual $\mathbb{\tilde{S}}^{2}$ shrinks then to a point.\ Below, we
think about the real parameter $\varrho $ as the absolute value of a
complexified Kahler parameter $t\left( \zeta \right) .$ This Kahler modulus
is a function of the parameter $\zeta $ of the exceptional super group D$%
(2,1;\zeta )$ as follows
\begin{equation}
t=\zeta \left( \zeta +1\right) \qquad ,\qquad t=\frac{1}{\varrho }e^{i%
\mathrm{\chi }}
\end{equation}%
It vanishes for the values $\zeta =0,-1$ where the area of the real 2-sphere
\begin{equation}
\mathcal{A}\left( \mathbb{S}^{2}\right) =\frac{4\pi }{\left\vert \zeta
\left( \zeta +1\right) \right\vert }  \label{a}
\end{equation}%
diverges towards the volume of $\mathbb{R}^{2}.$ For these values, the dual
area $\mathcal{A}(\mathbb{\tilde{S}}^{2})=4\pi \left\vert \zeta \left( \zeta
+1\right) \right\vert $ shrinks. The variation of this area is given by the
\textbf{Figure} \textbf{\ref{area}}.

\section{Graded 2-cycles in D$(2,1;\protect\zeta )$}

\qquad \label{sec3} In this section, we first describe the four parametric
super Dynkin graphs S$\mathfrak{DD}_{\mathfrak{d}(2,1;\zeta )}^{(\mathfrak{%
\eta })}$ of complex $\mathfrak{d}(2,1;\zeta )$ and their discrete outer
automorphisms by using $\left( \mathbf{i}\right) $ the complex parameter $%
\zeta $ and $\left( \mathbf{ii}\right) $ the \emph{Kaplansky} variables (%
\textsc{s}$_{1}$\textsc{,s}$_{2}$\textsc{,s}$_{3})$ constrained as $\sum
\text{\textsc{s}}_{i}=0$ and $\dprod $\textsc{s}$_{i}\neq 0$ \textrm{\cite%
{6,1B}.} Then, we use the root/2-cycle correspondence%
\begin{equation}
\begin{tabular}{ccc}
D$(2,1;\zeta )$ & $\qquad :\qquad $ & H$_{2}\left[ D(2,1;\zeta )\right] $ \\
$\text{algebraic roots }\mathbf{\alpha }$ & $\qquad \leftrightarrow \qquad $
& $\text{ \ geometric 2-cycles }\mathcal{\tilde{C}}_{\mathbf{\tilde{\alpha}}%
} $%
\end{tabular}
\label{cor}
\end{equation}%
and the ansatz \textrm{\cite{1A,SUP}}
\begin{equation}
\text{ D(}2,1;\zeta \text{)\ \ }\qquad \leftrightarrow \text{\ }\qquad \text{%
continuous deformation of OSp}(4|2)=\text{D(}2,1\text{)}  \label{az}
\end{equation}%
to extract useful information on the parametric super 2-cycles $\mathcal{%
\tilde{C}}_{\mathbf{\tilde{\alpha}}}\left[ \zeta \right] $ in the geometry
of complex D$(2,1;\zeta ).$ These parametric 2-cycles are functions of $%
\zeta $; they are thought of as a continuous deformation of the
orthosymplectic 2-cycles $\mathcal{C}_{\mathbf{\alpha }}^{osp}$ of the super
OSp(4\TEXTsymbol{\vert}2) further described in the \textrm{\autoref{appB}}.
Also, they contract in the super 2-cycles of PSL(2\TEXTsymbol{\vert}2) for $%
\zeta =-1.$\

Recall that $\mathfrak{d}(2,1;\zeta )$ has four possible basis sets $\Pi _{%
\mathfrak{d}(2,1;\zeta )}^{{\small (\mathfrak{\eta })}}$ of simple roots
that we denote like $\{\mathbf{\tilde{\alpha}}_{1}^{{\small (\mathfrak{\eta }%
)}},\mathbf{\tilde{\alpha}}_{2}^{{\small (\mathfrak{\eta })}},\mathbf{\tilde{%
\alpha}}_{3}^{{\small (\mathfrak{\eta })}}\}$ with index $\mathfrak{\eta }%
=0,1,2,3$; the twild label is used in order to distinguish the roots of $%
\mathfrak{d}(2,1;\zeta )$ from their homologue\footnote{%
\ To avoid confusion between the roots systems of\ sl(2\TEXTsymbol{\vert}2)
and osp(4\TEXTsymbol{\vert}2), we sometimes use the convention notation $%
\mathbf{\alpha }_{i}^{{\small sl(2|2)}}$ and $\mathbf{\alpha }_{i}^{{\small %
osp}}.$} $\left\{ \mathbf{\alpha }_{1},\mathbf{\alpha }_{2},\mathbf{\alpha }%
_{3}\right\} $ of psl(2\TEXTsymbol{\vert}2) and osp(4\TEXTsymbol{\vert}2).
These simple roots are functions of the deformation parameters; i.e:
\begin{equation}
\mathbf{\tilde{\alpha}}_{i}^{{\small (\mathfrak{\eta })}}\left( \mathbf{%
\alpha }^{{\small osp}};\text{\textsc{s}}\right) \qquad \Leftrightarrow
\qquad \mathbf{\tilde{\alpha}}_{i}^{{\small (\mathfrak{\eta })}}\left(
\mathbf{\alpha }^{{\small osp}};\zeta \right)
\end{equation}%
they are constrained by consistency conditions including the osp(4%
\TEXTsymbol{\vert}2) and the sl(2\TEXTsymbol{\vert}2) symmetries sitting at $%
\zeta =\pm 1;$ thus requiring $\left. \mathbf{\tilde{\alpha}}_{i}\right\vert
_{\zeta =1}=\mathbf{\alpha }_{i}^{{\small osp(4|2)}}$ and $\left. \mathbf{%
\tilde{\alpha}}_{i}\right\vert _{\zeta =-1}=\mathbf{\alpha }_{i}^{{\small %
sl(2|2)}}.$ Other useful properties on $\mathfrak{d}(2,1;\zeta )$ such as
the grading of the roots (homological 2-cycles) and outer automorphisms are
reported in the \textrm{\autoref{appC}}.

\subsection{Basis sets $\tilde{\Pi}_{\mathfrak{d}(2,1;\protect\zeta )}^{%
{\protect\small (\mathfrak{\protect\eta })}}$ and outer automorphisms}

\qquad \label{subsec31} The Lie superalgebra $\mathfrak{d}(2,1;\zeta )$ of
the super group D($2,1;\zeta $) is a parametric graded algebra with complex
parameter $\zeta $; it has rank 3 and 17 dimensions and was interpreted as a
deformation of the orthosymplectic osp(4\TEXTsymbol{\vert}2) sitting at $%
\zeta =1$ \textrm{\cite{1A,IV,IVA,bakas,DS,DSS}}. Also, the $\mathfrak{d}%
(2,1;\zeta )$ has the same rank 3 as for sl(2\TEXTsymbol{\vert}2); but two
dimensions bigger.

Its even part $\mathfrak{d}(2,1;\zeta )_{\bar{0}}$ has 9 dimensions; it is
given by
\begin{equation}
sl(2,\mathbb{C})_{1}\oplus sl(2,\mathbb{C})_{2}\oplus sl(2,\mathbb{C})_{3}
\end{equation}%
exhibiting a manifest $\mathcal{S}_{3}$ automorphism symmetry that permutes
the three isospins. Denoting the simple roots of these $sl(2)_{i}$s by $%
\mathbf{\beta }_{i}\ $(with diagonal intersection matrix $\mathbf{\beta }%
_{i}.\mathbf{\beta }_{j}=4\text{\textsc{s}}_{i}\delta _{ij}$)$,$ the group
elements $\mathrm{\sigma }$ of $\mathcal{S}_{3}$ acts like $\mathbf{\beta }%
_{i}\rightarrow \mathbf{\beta }_{\mathrm{\sigma }\left( i\right) }.$

\begin{figure}[tbph]
\begin{center}
\includegraphics[width=6cm]{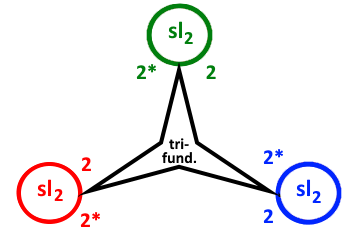}
\end{center}
\par
\vspace{-0.5cm}
\caption{The nodes in red, blue and green represent the three sl(2,$\mathbb{C%
}$)s making the even D(2,1;$\protect\zeta $)$_{\bar{0}}$. The bridges
between the three nodes given by the tri-fundamentals represent the odd
generators of D(2,1;$\protect\zeta $)$_{\bar{1}}$ relating the three sl(2,$%
\mathbb{C}$)s.}
\label{23}
\end{figure}

\begin{figure}[h]
\begin{center}
\includegraphics[width=12cm]{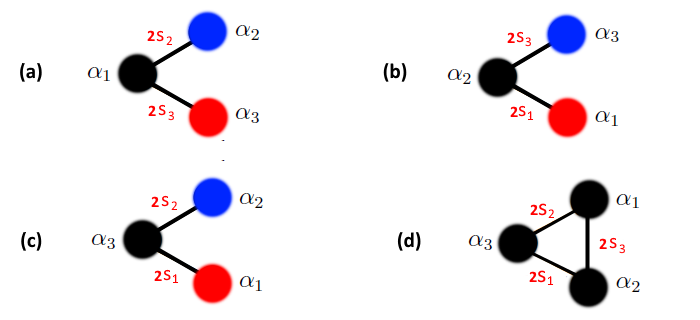}
\end{center}
\par
\vspace{-0.5cm}
\caption{The four super Dynkin diagrams of the Lie superalgebra D(2,1;$%
\protect\zeta $). They are as in \protect\cite{FS1} where here we have used
the \textsc{s}$_{i}$ parameters constrained like \textsc{s}$_{1}$\textsc{+s}$%
_{2}$\textsc{+s}$_{3}=0$ and \textsc{s}$_{1}$\textsc{s}$_{2}$\textsc{s}$%
_{3}\neq 0$. These diagrams will be imagined in terms of the intersection
matrix $\mathcal{I}_{ij}$ instead of $\mathcal{K}_{ij}$.}
\label{3D}
\end{figure}

The odd part $\mathfrak{d}$(2,1;$\zeta $)$_{\bar{1}}$ is eight dimensional;
it is given by the tri-fundamental representation $\mathbf{2}_{1}\otimes
\mathbf{2}_{2}\otimes \mathbf{2}_{3}$ as depicted by the \textbf{Figure}
\textbf{\ref{23}} extending the \textbf{Figure} \textbf{\ref{23F}}.
From this description, we see that the 14 dimensional PSL$(2|2)_{\bar{0}}$
sitting at $\zeta =-1$ descends from D$(2,1;\zeta )_{\bar{0}}$ by
decompactifying $SL(2,\mathbb{C})_{3}$ into $\mathbb{C}^{3}$ by taking the
area (\ref{a}) to infinity. In Lie algebra language we have for $\zeta
\rightarrow -1,$
\begin{equation}
\mathfrak{d}(2,1;\zeta )_{\bar{0}}\qquad \rightarrow \qquad \mathfrak{d}%
(2,1;-1)_{\bar{0}}=psl(2|2)_{\bar{0}}\oplus \mathbb{C}\oplus \mathbb{C}%
\oplus \mathbb{C}
\end{equation}%
The exceptional $\mathfrak{d}(2,1;\zeta )$ has four super Dynkin diagrams S$%
\mathfrak{DD}_{\mathfrak{d}(2,1;\zeta )}^{{\small (\mathfrak{\eta })}}$
given by the pictures (\textbf{a}), (\textbf{b}), (\textbf{c}) and (\textbf{d%
}) of the \textbf{Figure} \textbf{\ref{3D}}.\
The second picture \textbf{\ref{3D}}-(\textbf{b}) with odd $\mathbf{\tilde{%
\alpha}}_{2}$ will be further discussed in \textbf{subsection}\textrm{\
\textbf{3.2} in connection with the correspondence (\ref{cor})}. As for sl(2%
\TEXTsymbol{\vert}2), here also the sets $\Pi _{\mathfrak{d}(2,1;\zeta )}^{%
{\small (\mathfrak{\eta })}}$ (below $\tilde{\Pi}^{{\small (}\mathfrak{\mu }%
{\small )}}$) labeled by $\mathfrak{\eta }=0,1,2,3$ giving the four possible
bases of simple roots spanning the root system $\Phi _{\mathfrak{d}%
(2,1;\zeta )}^{{\small (\mathfrak{\eta })}}$ (for short $\tilde{\Phi}^{%
{\small (\mathfrak{\eta })}}$). Each basis set $\tilde{\Pi}^{{\small (%
\mathfrak{\eta })}}$ is generated by three simple roots $\left\{ \mathbf{%
\tilde{\alpha}}_{1},\mathbf{\tilde{\alpha}}_{2},\mathbf{\tilde{\alpha}}%
_{3}\right\} $ characterised by the Z$_{2}$- parity of the $\mathbf{\tilde{%
\alpha}}_{i}$'s.

\subsubsection{Gradation of super Dynkin S$\mathfrak{DD}_{\mathfrak{d}(2,1;%
\protect\zeta )}^{{\protect\small (\mathfrak{\protect\eta })}}$}

\qquad Depending on the parity of the simple roots, we distinguish four
graded roots systems%
\begin{equation}
\begin{tabular}{|c||c|c|c||c|}
\hline
basis & $\tilde{\Pi}^{{\small (1)}}$ & $\tilde{\Pi}^{{\small (2)}}$ & $%
\tilde{\Pi}^{{\small (3)}}$ & $\tilde{\Pi}^{{\small (0)}}$ \\ \hline
root system & $\tilde{\Phi}^{{\small (1)}}$ & $\tilde{\Phi}^{{\small (2)}}$
& $\tilde{\Phi}^{{\small (3)}}$ & $\tilde{\Phi}^{{\small (0)}}$ \\ \hline
$\mathcal{S}_{3}$ repres. & \multicolumn{3}{||c||}{triplet} & singlet \\
\hline
\end{tabular}%
\end{equation}%
The three $\tilde{\Pi}^{{\small (i)}}$ (resp. $\tilde{\Phi}^{{\small (i)}}$)
form a triplet of the discrete permutation group $\mathcal{S}_{3}$ while $%
\tilde{\Phi}^{{\small (0)}}$ is a singlet. Their roots contents are as
follows:

\begin{description}
\item[(\textbf{1)}] \textbf{the root system} $\tilde{\Phi}^{{\small (1)}}:$
(simple $\mathbf{\tilde{\alpha}}_{1}$ odd)\newline
It is generated by the simple roots $\mathbf{\tilde{\alpha}}_{1},$ $\mathbf{%
\tilde{\alpha}}_{2},$ $\mathbf{\tilde{\alpha}}_{3}$ with $\mathbf{\tilde{%
\alpha}}_{1}$ odd and $\mathbf{\tilde{\alpha}}_{2},$ $\mathbf{\tilde{\alpha}}%
_{3}$ even. The seven positive roots of $\tilde{\Phi}^{{\small (1)}}$ are
collected in the following Table%
\begin{equation}
\begin{tabular}{|c|c|c|}
\hline
{\small root systems} & {\small fermionic subset} & \ \ {\small bosonic
subset} \ \  \\ \hline\hline
$\tilde{\Phi}^{{\small (1)}}$ & $\left.
\begin{array}{c}
\mathbf{\tilde{\alpha}}_{1},\quad \\
\mathbf{\tilde{\alpha}_{1}+\tilde{\alpha}_{2},\quad \tilde{\alpha}}_{1}+%
\mathbf{\tilde{\alpha}}_{3},%
\end{array}%
\right. $ & $\mathbf{\tilde{\alpha}}_{2},\quad \mathbf{\tilde{\alpha}}_{3},$
\\
& $\mathbf{\tilde{\alpha}_{1}}+\mathbf{\tilde{\alpha}}_{2}+\mathbf{\tilde{%
\alpha}}_{3}$ & $2\mathbf{\tilde{\alpha}_{1}}+\mathbf{\tilde{\alpha}}_{2}+%
\mathbf{\tilde{\alpha}}_{3}$ \\ \hline\hline
\end{tabular}
\label{TD2}
\end{equation}%
The 6 even step operators E$_{\pm \mathbf{\beta }_{i}}$ labeled by the even
roots
\begin{equation}
\begin{tabular}{lllll}
$\mathbf{\beta }_{1}$ & $=$ & $2\mathbf{\tilde{\alpha}_{1}}+\mathbf{\tilde{%
\alpha}}_{2}+\mathbf{\tilde{\alpha}}_{3}$ & $\quad \leftrightarrow \quad $ &
$\mathcal{C}_{\mathbf{\beta }_{1}}:=\mathbb{S}_{\varrho }^{2}$ or $\mathbb{S}%
_{\varrho }^{1,1}$ \\
$\mathbf{\beta }_{2}$ & $=$ & $\mathbf{\tilde{\alpha}}_{2}$ & $\quad
\leftrightarrow \quad $ & $\mathcal{C}_{\mathbf{\beta }_{2}}$ \\
$\mathbf{\beta }_{3}$ & $=$ & $\mathbf{\tilde{\alpha}}_{3}$ & $\quad
\leftrightarrow \quad $ & $\mathcal{C}_{\mathbf{\beta }_{3}}$%
\end{tabular}
\label{3b}
\end{equation}%
generate $\oplus _{i}sl(2,\mathbb{C})_{i}.$ In (\ref{3b}), we have also
given the 2-sphere $\mathbb{S}_{i}^{2}$ or the pseudo $\mathbb{S}_{\varrho
}^{1,1}$ associated with the even roots $\mathbf{\beta }_{i}$ characterising
the isospin $sl(2,\mathbb{C})_{\mathbf{\beta }_{i}}$. Here $\mathbf{\beta }%
_{1}$ is the long root.

\item[(\textbf{2)}] \textbf{the root systems} $\tilde{\Phi}^{{\small (2)}}$
(with odd $\mathbf{\tilde{\alpha}}_{2}$) \textbf{and} $\tilde{\Phi}^{{\small %
(3)}}$ (with odd $\mathbf{\tilde{\alpha}}_{3}$)\newline
The root contents of these systems are obtained from $\tilde{\Phi}^{{\small %
(1)}}$ by performing cyclic permutations of the pictures (\textbf{a}), (%
\textbf{b}), (\textbf{c}) of the \textbf{Figure} \textbf{\ref{3D}}. This
discrete symmetry acts as
\begin{equation}
\left( \mathbf{\tilde{\alpha}}_{1},\mathbf{\tilde{\alpha}}_{2},\mathbf{%
\tilde{\alpha}}_{3}\right) \qquad \rightarrow \qquad \left( \mathbf{\tilde{%
\alpha}}_{2},\mathbf{\tilde{\alpha}}_{3},\mathbf{\tilde{\alpha}}_{1}\right)
\qquad \rightarrow \qquad \left( \mathbf{\tilde{\alpha}}_{3},\mathbf{\tilde{%
\alpha}}_{1},\mathbf{\tilde{\alpha}}_{2}\right)
\end{equation}%
The three basis sets $(\tilde{\Pi}_{1},\tilde{\Pi}_{2},\tilde{\Pi}_{3})$
form then a triplet under the $\mathcal{Z}_{3}$ automorphism symmetry.
Recall that this discrete symmetry $\mathcal{Z}_{3}$ is a subgroup of the
symmetric group $\mathcal{S}_{3}$ permuting the three canonical weight
vectors $\left( \mathbf{\varepsilon }_{1},\mathbf{\varepsilon }_{2},\mathbf{%
\varepsilon }_{3}\right) $ given by eq(\ref{3e}) \textrm{reported in \autoref%
{appC}}.\newline
The contents of the sets $\tilde{\Phi}^{{\small (2)}}$ and $\tilde{\Phi}^{%
{\small (3)}}$ are collected in the following Table%
\begin{equation}
\begin{tabular}{|c|c|c|}
\hline
{\small root systems} & {\small fermionic subset} & \ \ \ {\small bosonic
subset} $\ \ \ \ \ \ $ \\ \hline\hline
$\tilde{\Phi}^{{\small (2)}}$ & $\left.
\begin{array}{c}
\mathbf{\tilde{\alpha}_{2}},\quad \\
\mathbf{\tilde{\alpha}_{2}+\tilde{\alpha}_{1},\quad \tilde{\alpha}_{2}+%
\tilde{\alpha}}_{3},%
\end{array}%
\right. $ & $\mathbf{\tilde{\alpha}}_{1},\quad \mathbf{\tilde{\alpha}}_{3},$
\\
& $\mathbf{\tilde{\alpha}}_{2}+\mathbf{\tilde{\alpha}_{1}}+\mathbf{\tilde{%
\alpha}}_{3}$ & $\mathbf{\tilde{\alpha}_{1}}+2\mathbf{\tilde{\alpha}}_{2}+%
\mathbf{\tilde{\alpha}}_{3}$ \\ \hline\hline
$\tilde{\Phi}^{{\small (3)}}$ & $\left.
\begin{array}{c}
\mathbf{\tilde{\alpha}}_{3},\quad \\
\mathbf{\mathbf{\alpha }_{3}+\tilde{\alpha}_{2},\quad \tilde{\alpha}}_{3}+%
\mathbf{\tilde{\alpha}}_{1},%
\end{array}%
\right. $ & $\mathbf{\tilde{\alpha}}_{1},\quad \mathbf{\tilde{\alpha}}_{2},$
\\
& $\mathbf{\tilde{\alpha}_{1}}+\mathbf{\tilde{\alpha}}_{2}+\mathbf{\tilde{%
\alpha}}_{3}$ & $\mathbf{\tilde{\alpha}_{1}}+\mathbf{\tilde{\alpha}}_{2}+2%
\mathbf{\tilde{\alpha}}_{3}$ \\ \hline\hline
\end{tabular}
\label{RRD}
\end{equation}%
Here the 6 even step operators E$_{\pm \mathbf{\beta }_{i}}$ generating $%
\oplus _{i}sl(2,\mathbb{C})_{i}$ are labeled for the case of $\tilde{\Phi}_{%
\mathfrak{d}{\small (2,1;\zeta )}}^{{\small (2)}}$ by the even roots
\begin{equation}
\tilde{\Phi}^{{\small (2)}}:\quad
\begin{tabular}{lllll}
$\mathbf{\beta }_{1}^{\prime }$ & $=$ & $\mathbf{\tilde{\alpha}}_{1}$ & $%
\quad \leftrightarrow \quad $ & $\mathcal{C}_{\mathbf{\beta }_{1}^{\prime }}$
\\
$\mathbf{\beta }_{2}^{\prime }$ & $=$ & $\mathbf{\tilde{\alpha}_{1}}+2%
\mathbf{\tilde{\alpha}}_{2}+\mathbf{\tilde{\alpha}}_{3}$ & $\quad
\leftrightarrow \quad $ & $\mathcal{C}_{\mathbf{\beta }_{2}^{\prime }}:=%
\mathbb{S}_{\varrho }^{2}$ or $\mathbb{S}_{\varrho }^{1,1}$ \\
$\mathbf{\beta }_{3}^{\prime }$ & $=$ & $\mathbf{\tilde{\alpha}}_{3}$ & $%
\quad \leftrightarrow \quad $ & $\mathcal{C}_{\mathbf{\beta }_{3}^{\prime }}$%
\end{tabular}%
\end{equation}%
Because $\mathbf{\beta }_{i}^{\prime }.\mathbf{\beta }_{j}^{\prime }=0$ for $%
i<j,$ these 3-spheres have no direct intersections; they couple through the
odd roots. For the case of $\tilde{\Phi}^{{\small (3)}},$ we have%
\begin{equation}
\tilde{\Phi}^{{\small (3)}}:\quad
\begin{tabular}{lllll}
$\mathbf{\beta }_{1}^{\prime \prime }$ & $=$ & $\mathbf{\tilde{\alpha}}_{1}$
& $\quad \leftrightarrow \quad $ & $\mathcal{C}_{\mathbf{\beta }_{1}^{\prime
\prime }}$ \\
$\mathbf{\beta }_{2}^{\prime \prime }$ & $=$ & $\mathbf{\tilde{\alpha}}_{2}$
& $\quad \leftrightarrow \quad $ & $\mathcal{C}_{\mathbf{\beta }_{2}^{\prime
\prime }}$ \\
$\mathbf{\beta }_{3}^{\prime \prime }$ & $=$ & $\mathbf{\tilde{\alpha}_{1}}+%
\mathbf{\tilde{\alpha}}_{2}+2\mathbf{\tilde{\alpha}}_{3}$ & $\quad
\leftrightarrow \quad $ & $\mathcal{C}_{\mathbf{\beta }_{3}^{\prime \prime
}}:=\mathbb{S}_{\varrho }^{2}$ or $\mathbb{S}_{\varrho }^{1,1}$%
\end{tabular}%
\end{equation}

\item[(\textbf{3)}] \textbf{\ fourth root system} $\tilde{\Phi}^{{\small (0)}%
}:$ \newline
It is generated by the simple roots $\mathbf{\tilde{\alpha}}_{1}$, $\mathbf{%
\tilde{\alpha}}_{2},$ $\mathbf{\tilde{\alpha}}_{3}$ which all of them are
odd roots. Here, the seven positive roots are given by%
\begin{equation}
\begin{tabular}{|c|c|c|}
\hline
{\small root systems} & {\small fermionic subset} & \ \ \ \ \ \ {\small %
bosonic subset} \ \ \ \ \ \ \ \ \  \\ \hline\hline
$\tilde{\Phi}^{{\small (0)}}$ & $\mathbf{\tilde{\alpha}_{1},\quad \tilde{%
\alpha}_{2},\quad \tilde{\alpha}}_{3},$ & $\mathbf{\tilde{\alpha}_{1}+\tilde{%
\alpha}_{2},\quad \tilde{\alpha}}_{2}+\mathbf{\tilde{\alpha}}_{3}$ \\
& $\mathbf{\tilde{\alpha}}_{3}+\mathbf{\tilde{\alpha}_{1}}+\mathbf{\tilde{%
\alpha}}_{2}$ & $\mathbf{\tilde{\alpha}_{1}}+2\mathbf{\tilde{\alpha}}_{2}+%
\mathbf{\tilde{\alpha}}_{3}$ \\ \hline\hline
\end{tabular}
\label{TDZ}
\end{equation}%
For this special case, the 6 even step operators E$_{\pm \mathbf{\beta }%
_{i}} $ generating $\oplus _{i}sl(2,\mathbb{C})_{i}$ are labeled by the even
roots
\begin{equation}
case\text{ }\tilde{\Phi}^{{\small (0)}}:\quad
\begin{tabular}{lllll}
$\mathbf{\beta }_{1}$ & $=$ & $\mathbf{\tilde{\alpha}_{1}+\tilde{\alpha}_{2}}
$ & $\quad \leftrightarrow \quad $ & $\mathcal{C}_{\mathbf{\beta }_{1}}$ \\
$\mathbf{\beta }_{2}$ & $=$ & $\mathbf{\tilde{\alpha}_{1}}+2\mathbf{\tilde{%
\alpha}}_{2}+\mathbf{\tilde{\alpha}}_{3}$ & $\quad \leftrightarrow \quad $ &
$\mathcal{C}_{\mathbf{\beta }_{2}}$ \\
$\mathbf{\beta }_{3}$ & $=$ & $\mathbf{\tilde{\alpha}}_{2}+\mathbf{\tilde{%
\alpha}}_{3}$ & $\quad \leftrightarrow \quad $ & $\mathcal{C}_{\mathbf{\beta
}_{3}}$%
\end{tabular}%
\end{equation}
\end{description}

\subsubsection{Super 2-cycles in S$\mathfrak{DD}_{\mathfrak{d}(2,1;\protect%
\zeta )}^{{\protect\small (2)}}$}

\qquad Using $\left( \mathbf{i}\right) $ the correspondence (\ref{cor})
associating a homological 2-cycle $\mathcal{\tilde{C}}_{\mathbf{\tilde{\alpha%
}}}$ to each root $\mathbf{\tilde{\alpha}}$ in the set $\tilde{\Phi}_{%
\mathfrak{d}(2,1;\zeta )}^{{\small (\mathfrak{\eta })}}$, and $\left(
\mathbf{ii}\right) $ knowing the root systems content (\ref{TD2}-\ref{TDZ}),
one ends up with four parametric systems of homological 2-cycles that we
denote like%
\begin{equation*}
\mathcal{\tilde{C}}_{\mathfrak{d}(2,1;\zeta )}^{{\small (\mathfrak{\eta })}%
}\qquad ,\qquad \eta =0,1,2,3
\end{equation*}%
These 2-cycle systems are in 1:1 with the four $\tilde{\Phi}^{{\small (\eta )%
}}$ and then with the four parametric super Dynkin diagrams
\begin{equation*}
S\mathfrak{DD}_{\mathfrak{d}(2,1;\zeta )}^{{\small (\eta )}}\qquad ,\qquad
\eta =0,1,2,3
\end{equation*}%
Below, we consider the 2-cycles in the triplet $\tilde{\Phi}^{{\small (i)}%
}\simeq \mathcal{\tilde{C}}_{\mathfrak{d}(2,1;\zeta )}^{{\small (i)}}$ while
focussing on the system $\tilde{\Phi}^{{\small (2)}}$ due to the $\mathcal{S}%
_{3}$ symmetry. So, we have for the set $\mathcal{\tilde{C}}_{\mathfrak{d}%
(2,1;\zeta )}^{{\small (2)}}$ the following 2-cycle content%
\begin{equation}
\text{the set }\mathcal{\tilde{C}}_{\mathfrak{d}(2,1;\zeta )}^{{\small (2)}}%
\mathbf{:\quad }%
\begin{tabular}{|c|c|}
\hline
{\small odd 2-cycles} & \ \ \ {\small even 2-cycles} $\ \ \ \ \ \ $ \\
\hline\hline
$\left.
\begin{array}{c}
\mathcal{\tilde{C}}_{\mathbf{\tilde{\alpha}_{2}}} \\
\mathcal{\tilde{C}}_{\mathbf{\tilde{\alpha}_{2}+\tilde{\alpha}_{1}}}\mathbf{%
\quad ,\quad \mathcal{\tilde{C}}}_{\mathbf{\tilde{\alpha}_{2}+\tilde{\alpha}}%
_{3}} \\
\mathcal{\tilde{C}}_{\mathbf{\tilde{\alpha}}_{2}+\mathbf{\tilde{\alpha}_{1}}+%
\mathbf{\tilde{\alpha}}_{3}}%
\end{array}%
\right. $ & $\left.
\begin{array}{c}
\mathcal{\tilde{C}}_{\mathbf{\tilde{\alpha}}_{1}} \\
\mathcal{\tilde{C}}_{\mathbf{\tilde{\psi}}} \\
\mathcal{\tilde{C}}_{\mathbf{\tilde{\alpha}}_{3}}%
\end{array}%
\right. $ \\ \hline\hline
\end{tabular}
\label{s2}
\end{equation}%
where $\mathbf{\tilde{\psi}}$ stands for the long root given by $\mathbf{%
\tilde{\alpha}_{1}}+2\mathbf{\tilde{\alpha}}_{2}+\mathbf{\tilde{\alpha}}_{3}$%
. This \ corresponds to the super Dynkin diagram S$\mathfrak{DD}_{\mathfrak{d%
}(2,1;\zeta )}^{{\small (2)}}$ given by the \textbf{Figure} \textbf{\ref{3D}}%
-(b). In this regard, notice the following: $\left( \mathbf{i}\right) $ the
parametric S$\mathfrak{DD}_{\mathfrak{d}(2,1;\zeta )}^{{\small (2)}}$
contains the distinguished super Dynkin diagram S$\mathfrak{DD}_{psl(2|2)}^{%
{\small (1)}}$ given by the \textbf{Figure} \textbf{\ref{Asl2}}-(a) \textrm{%
in appendix A};
\begin{equation*}
S\mathfrak{DD}_{psl(2|2)}^{{\small (1)}}\subset S\mathfrak{DD}_{\mathfrak{d}%
(2,1;\zeta )}^{{\small (2)}}
\end{equation*}%
$\left( \mathbf{ii}\right) $\ The parametric super 2-cycle system (\ref{s2})
contains the 2-cycles $\mathcal{C}_{psl(2|2)}^{{\small (1)}}$ illustrated by
the \textbf{Figure} \textbf{\ref{CP2}} (green segments) namely%
\begin{equation}
\text{the set }\mathcal{C}_{psl(2|2)}^{{\small (1)}}\mathbf{:\quad }%
\begin{tabular}{|c|c|}
\hline
{\small odd 2-cycles} & \ \ \ {\small even 2-cycles} $\ \ \ \ \ \ $ \\
\hline\hline
$\left.
\begin{array}{c}
\mathcal{C}_{\mathbf{\alpha _{2}}} \\
\mathcal{C}_{\mathbf{\alpha _{1}}+\mathbf{\alpha }_{2}}\mathbf{\quad ,\quad
\mathcal{C}}_{\mathbf{\alpha _{2}+\alpha }_{3}} \\
\mathcal{C}_{\mathbf{\alpha _{1}}+\mathbf{\alpha }_{2}+\mathbf{\alpha }_{3}}%
\end{array}%
\right. $ & $\left.
\begin{array}{c}
\mathcal{C}_{\mathbf{\alpha }_{1}} \\
\mathcal{-} \\
\mathcal{C}_{\mathbf{\alpha }_{3}}%
\end{array}%
\right. $ \\ \hline\hline
\end{tabular}
\label{ss}
\end{equation}%
where the 2-cycle $\mathcal{\tilde{C}}_{\mathbf{\tilde{\psi}}}$ in (\ref{s2}%
) associated with the even root $\mathbf{\tilde{\psi}=\tilde{\alpha}_{1}}+2%
\mathbf{\tilde{\alpha}}_{2}+\mathbf{\tilde{\alpha}}_{3}$ gets decoupled; i.e
contracted to zero ($\mathcal{C}_{\mathbf{\tilde{\psi}}}\simeq 0$). The
other six 2-cycles $\mathcal{\tilde{C}}_{\mathbf{\tilde{\alpha}}}$ in (\ref%
{s2}) get mapped to the six $\mathcal{C}_{\mathbf{\alpha }}$ in the above (%
\ref{ss}).

\subsection{Realising the distinguished $\tilde{\Phi}_{\mathfrak{d}(2,1;%
\protect\zeta )}^{{\protect\small (2)}}$ and $\mathcal{\tilde{C}}_{\mathfrak{%
d}(2,1;\protect\zeta )}^{{\protect\small (2)}}$ systems}

\qquad \label{subsec32}To realise the positive roots of the system $\tilde{%
\Phi}_{\mathfrak{d}(2,1;\zeta )}^{{\small (2+)}}$ of the super $\mathfrak{d}$%
(2,1;$\zeta $) [resp. the associated $\mathcal{\tilde{C}}_{\mathfrak{d}%
(2,1;\zeta )}^{{\small (2)}}$], we use the three following quantities: $%
\left( \mathbf{1}\right) $ the three orthogonal canonical weight vectors \{$%
\mathbf{\epsilon }_{1},\mathbf{\epsilon }_{2},\mathbf{\delta }$\} with $%
\mathbf{\epsilon _{i}.\epsilon }_{j}=\delta _{ij}$ and $\mathbf{\delta
.\delta }=-1;$ $\left( \mathbf{2}\right) $ the \emph{Kaplansky} variables (%
\textsc{s}$_{1},$\textsc{s}$_{2},$\textsc{s}$_{3}$), and $\left( \mathbf{3}%
\right) $ the roots (\ref{RRD}) which for convenience we recall them here
after%
\begin{equation}
odd:\quad \left.
\begin{array}{c}
\mathbf{\tilde{\alpha}_{2}},\quad \mathbf{\tilde{\alpha}}_{2}+\mathbf{\tilde{%
\alpha}_{1}}+\mathbf{\tilde{\alpha}}_{3} \\
\mathbf{\tilde{\alpha}_{2}+\tilde{\alpha}_{1},\quad \tilde{\alpha}_{2}+%
\tilde{\alpha}}_{3}%
\end{array}%
\right. \qquad ;\qquad even:\quad \left.
\begin{array}{c}
\mathbf{\tilde{\alpha}}_{1},\quad \mathbf{\tilde{\alpha}}_{3}, \\
\mathbf{\tilde{\alpha}_{1}}+2\mathbf{\tilde{\alpha}}_{2}+\mathbf{\tilde{%
\alpha}}_{3}%
\end{array}%
\right. ,  \label{DR}
\end{equation}%
We represent the three simple roots $(\mathbf{\tilde{\alpha}}_{1},\mathbf{%
\tilde{\alpha}}_{2}\mathbf{,\tilde{\alpha}}_{3})$ generating $\tilde{\Phi}_{%
\mathfrak{d}(2,1;\zeta )}^{{\small (2+)}}$ as follows
\begin{eqnarray}
\mathbf{\tilde{\alpha}}_{1} &=&\sqrt{2\text{\textsc{s}}_{1}}\left( \mathbf{%
\epsilon }_{1}-\mathbf{\epsilon }_{2}\right)  \notag \\
\mathbf{\tilde{\alpha}}_{2} &=&\sqrt{\text{\textsc{s}}_{1}+\text{\textsc{s}}%
_{3}}\mathbf{\delta }-\sqrt{\frac{\text{\textsc{s}}_{1}}{2}}\left( \mathbf{%
\epsilon }_{1}-\mathbf{\epsilon }_{2}\right) -\sqrt{\frac{\text{\textsc{s}}%
_{3}}{2}}\left( \mathbf{\epsilon }_{1}+\mathbf{\epsilon }_{2}\right)
\label{se} \\
\mathbf{\tilde{\alpha}}_{3} &=&\sqrt{2\text{\textsc{s}}_{3}}\left( \mathbf{%
\epsilon }_{1}+\mathbf{\epsilon }_{2}\right)  \notag
\end{eqnarray}%
showing that the roots (\ref{DR}) are indeed function of the parameters
\textsc{s}$_{i}$. Notice that for $($\textsc{s}$_{1},$\textsc{s}$_{2},$%
\textsc{s}$_{3})=(1/2,-1,1/2),$ the above relations reduce to%
\begin{equation}
\begin{tabular}{lllll}
$\mathbf{\tilde{\alpha}}_{1}$ & $=$ & $\mathbf{\epsilon }_{1}-\mathbf{%
\epsilon }_{2}$ & $\equiv $ & $\mathbf{\alpha }_{1}^{{\small osp}}$ \\
$\mathbf{\tilde{\alpha}}_{2}$ & $=$ & $\mathbf{\delta }-\mathbf{\epsilon }%
_{1}$ & $\equiv $ & $\mathbf{\alpha }_{2}^{{\small osp}}$ \\
$\mathbf{\tilde{\alpha}}_{3}$ & $=$ & $\mathbf{\epsilon }_{1}+\mathbf{%
\epsilon }_{2}$ & $\equiv $ & $\mathbf{\alpha }_{3}^{{\small osp}}$%
\end{tabular}%
\end{equation}%
giving precisely the simple roots of the orthosymplectic Lie superalgebra
osp(4\TEXTsymbol{\vert}2) further commented in the\textrm{\ \autoref{appB}}.
These $\mathbf{\alpha }_{i}^{{\small osp}}$'s have well defined parities; $%
\mathbf{\alpha }_{1}^{{\small osp}}$ and $\mathbf{\alpha }_{3}^{{\small osp}%
} $ are even while $\mathbf{\alpha }_{2}^{{\small osp}}$\ is odd. From the
parametric realisation (\ref{se}), we also learn the following features:

\begin{description}
\item[$\left( \mathbf{i}\right) $] the realisation of the non simple roots (%
\ref{DR}) are given by
\begin{equation}
\begin{array}{lll}
\mathbf{\tilde{\alpha}}_{1}+\mathbf{\tilde{\alpha}_{2}} & = & \sqrt{\text{%
\textsc{s}}_{1}+\text{\textsc{s}}_{3}}\text{ }\mathbf{\delta }+\frac{1}{2}%
\sqrt{2\text{\textsc{s}}_{1}}\left( \mathbf{\epsilon }_{1}-\mathbf{\epsilon }%
_{2}\right) -\frac{1}{2}\sqrt{2\text{\textsc{s}}_{3}}\left( \mathbf{\epsilon
}_{1}+\mathbf{\epsilon }_{2}\right) \\
\mathbf{\tilde{\alpha}_{2}+\tilde{\alpha}}_{3} & = & \sqrt{\text{\textsc{s}}%
_{1}+\text{\textsc{s}}_{3}}\text{ }\mathbf{\delta }-\frac{1}{2}\sqrt{2\text{%
\textsc{s}}_{1}}\left( \mathbf{\epsilon }_{1}-\mathbf{\epsilon }_{2}\right) +%
\frac{1}{2}\sqrt{2\text{\textsc{s}}_{3}}\left( \mathbf{\epsilon }_{1}+%
\mathbf{\epsilon }_{2}\right) \\
\mathbf{\tilde{\alpha}_{1}}+\mathbf{\tilde{\alpha}}_{2}+\mathbf{\tilde{\alpha%
}}_{3} & = & \sqrt{\text{\textsc{s}}_{1}+\text{\textsc{s}}_{3}}\text{ }%
\mathbf{\delta }+\frac{1}{2}\sqrt{2\text{\textsc{s}}_{1}}\left( \mathbf{%
\epsilon }_{1}-\mathbf{\epsilon }_{2}\right) +\frac{1}{2}\sqrt{2\text{%
\textsc{s}}_{3}}\left( \mathbf{\epsilon }_{1}+\mathbf{\epsilon }_{2}\right)%
\end{array}%
\end{equation}%
together with the even long root $\mathbf{\tilde{\psi}}=\mathbf{\tilde{\alpha%
}}_{1}+2\mathbf{\tilde{\alpha}}_{2}+\mathbf{\tilde{\alpha}}_{3}$ pointing in
the $\mathbf{\delta }$-direction as follows\footnote{%
\ It would be interesting to explore the situation where the Kaplansky
parameters $\left( \text{\textsc{s}}_{1},\text{\textsc{s}}_{2},\text{\textsc{%
s}}_{3}\right) $ are solved like \textsc{s}$_{1}+$\textsc{s}$_{3}=n^{2}$
with integer n$.$ For this case we have \textsc{s}$_{2}=-n^{2}$ and the long
root is even and is given by $\mathbf{\tilde{\psi}}=2n$ $\mathbf{\delta .}$}:%
\begin{equation}
\mathbf{\tilde{\psi}}=2\sqrt{\text{\textsc{s}}_{1}+\text{\textsc{s}}_{3}}%
\text{ }\mathbf{\delta }
\end{equation}%
For \textsc{s}$_{1}+$\textsc{s}$_{3}=1,$ this long root reduce to $\mathbf{%
\psi }^{{\small osp}}=2\mathbf{\delta }$; this is the case of the
orthosymplectic point $($\textsc{s}$_{1},$\textsc{s}$_{2},$\textsc{s}$%
_{3})=(1/2,-1,1/2)$. For the case \textsc{s}$_{1}+$\textsc{s}$_{3}=0$ as for
the situation $($\textsc{s}$_{1},$\textsc{s}$_{2},$\textsc{s}$%
_{3})=(1/2,0,-1/2),$ the long root vanishes identically; this corresponds to
$\mathbf{\psi }^{{\small sl(2|2)}}=0.$

\item[$\left( \mathbf{ii}\right) $] Using the property \textsc{s}$_{1}+$%
\textsc{s}$_{2}+$\textsc{s}$_{3}=0,$ the intersection matrix of the simple
roots like $\mathcal{\tilde{J}}_{ij}\left( \text{\textsc{s}}\right) =\mathbf{%
\tilde{\alpha}}_{i}.\mathbf{\tilde{\alpha}}_{j}$ and the corresponding super
Cartan matrix $\mathcal{\tilde{K}}_{ij}\left( \text{\textsc{s}}\right) $
read in the patch \textsc{s}$_{1}\neq 0$ as follows
\begin{equation}
\mathcal{\tilde{J}}_{ij}\left( \text{\textsc{s}}\right) =\left(
\begin{array}{ccc}
4\text{\textsc{s}}_{1} & -2\text{\textsc{s}}_{1} & 0 \\
-2\text{\textsc{s}}_{1} & 0 & -2\text{\textsc{s}}_{3} \\
0 & -2\text{\textsc{s}}_{3} & 4\text{\textsc{s}}_{3}%
\end{array}%
\right) \quad ,\qquad \mathcal{\tilde{K}}_{ij}\left( \text{\textsc{s}}%
\right) =\left(
\begin{array}{ccc}
2 & -1 & 0 \\
1 & 0 & \text{\textsc{s}}_{3}/\text{\textsc{s}}_{1} \\
0 & -1 & 2%
\end{array}%
\right)  \label{JJ}
\end{equation}%
\begin{equation*}
\end{equation*}%
The determinant $\det \mathcal{\tilde{J}}_{ij}\left( \text{\textsc{s}}%
\right) =16$\textsc{s}$_{1}$\textsc{s}$_{2}$\textsc{s}$_{3};$ it is singular
for $\dprod \text{\textsc{s}}_{i}=0;$ it the case of the Lie superalgebra
psl(2\TEXTsymbol{\vert}2) sitting at (\textsc{s}$_{1},$\textsc{s}$_{2},$%
\textsc{s}$_{3})=(1/2,0,-1/2);$ i.e: $\det \mathcal{J}_{ij}^{{\small %
(sl(2|2))}}=0.$ For the Lie superalgebra osp(4\TEXTsymbol{\vert}2) sitting
at $($\textsc{s}$_{1},$\textsc{s}$_{2},$\textsc{s}$_{3})=(1/2,-1,1/2)$, we
have $\det \mathcal{J}_{ij}^{{\small (osp)}}=-4.$ The determinant of the
super Cartan%
\begin{equation}
\mathcal{\tilde{K}}_{ij}\left( \text{\textsc{s}}\right) =-\frac{2\text{%
\textsc{s}}_{2}}{\text{\textsc{s}}_{1}}{\Large \delta }_{{\small Dirac}%
}\left( \text{\textsc{s}}_{1}+\text{\textsc{s}}_{2}+\text{\textsc{s}}%
_{3}\right)
\end{equation}

\item[$\left( \mathbf{iii}\right) $] In terms of the even roots ($\mathbf{%
\tilde{\alpha}}_{1},\mathbf{\tilde{\psi}},\mathbf{\tilde{\alpha}}_{3}$), the
fermionic roots read as follows%
\begin{equation}
\begin{array}{lll}
\mathbf{\tilde{\alpha}}_{2} & = & +\frac{1}{2}\mathbf{\tilde{\psi}}-\frac{1}{%
2}\mathbf{\tilde{\alpha}}_{1}-\frac{1}{2}\mathbf{\tilde{\alpha}}_{3} \\
\mathbf{\tilde{\alpha}}_{1}+\mathbf{\tilde{\alpha}_{2}} & = & +\frac{1}{2}%
\mathbf{\tilde{\psi}}+\frac{1}{2}\mathbf{\tilde{\alpha}}_{1}-\frac{1}{2}%
\mathbf{\tilde{\alpha}}_{3} \\
\mathbf{\tilde{\alpha}_{2}+\tilde{\alpha}}_{3} & = & +\frac{1}{2}\mathbf{%
\tilde{\psi}}-\frac{1}{2}\mathbf{\tilde{\alpha}}_{1}+\frac{1}{2}\mathbf{%
\tilde{\alpha}}_{3} \\
\mathbf{\tilde{\alpha}_{1}}+\mathbf{\tilde{\alpha}}_{2}+\mathbf{\tilde{\alpha%
}}_{3} & = & +\frac{1}{2}\mathbf{\tilde{\psi}}+\frac{1}{2}\mathbf{\tilde{%
\alpha}}_{1}+\frac{1}{2}\mathbf{\tilde{\alpha}}_{3}%
\end{array}%
\end{equation}%
By using the correspondence $\mathbf{\tilde{\alpha}}$ $\leftrightarrow $ $%
\mathcal{\tilde{C}}_{\mathbf{\tilde{\alpha}}}$ of eq(\ref{cor}), we get the
system of 2-cycles (\ref{s2}) in the supermanifold describing D($2|1;\zeta $%
).
\end{description}

\subsubsection{Roots $\mathbf{\tilde{\protect\alpha}}$ as function of $%
\protect\zeta $}

\qquad Here, we solve the conditions \textsc{s}$_{1}+$\textsc{s}$_{2}+$%
\textsc{s}$_{3}=0$ and \textsc{s}$_{1}$\textsc{s}$_{2}$\textsc{s}$_{3}\neq 0$%
\ on the \emph{Kaplansky} variables in terms of the complex parameter $\zeta
$ labeling D(2\TEXTsymbol{\vert}1;$\zeta $) like
\begin{equation}
2\text{\textsc{s}}_{1}=1\quad ,\quad 2\text{\textsc{s}}_{2}=-1-\zeta \quad
,\quad 2\text{\textsc{s}}_{3}=\zeta  \label{ze}
\end{equation}%
This solution breaks the permutation symmetry $\mathcal{S}_{3}$ down to its
subgroup $\mathcal{Z}_{2}$. This discrete $\mathcal{Z}_{2}$ fixes the axis
\textsc{s}$_{1}$ and permutes \textsc{s}$_{2}$ and \textsc{s}$_{3}$; it has
a fix point at $\zeta =1/2.$ Putting (\ref{ze}) into (\ref{se}), we get%
\begin{eqnarray}
\mathbf{\tilde{\alpha}}_{1} &=&\left( \mathbf{\epsilon }_{1}-\mathbf{%
\epsilon }_{2}\right)  \notag \\
\mathbf{\tilde{\alpha}}_{2} &=&\frac{1}{2}\sqrt{2\left( 1+\zeta \right) }%
\text{ }\mathbf{\delta }-\frac{1}{2}\left( \mathbf{\epsilon }_{1}-\mathbf{%
\epsilon }_{2}\right) -\frac{1}{2}\sqrt{\zeta }\left( \mathbf{\epsilon }_{1}+%
\mathbf{\epsilon }_{2}\right)  \label{in} \\
\mathbf{\tilde{\alpha}}_{3} &=&\sqrt{\zeta }\left( \mathbf{\epsilon }_{1}+%
\mathbf{\epsilon }_{2}\right)  \notag
\end{eqnarray}%
and%
\begin{eqnarray}
\mathbf{\tilde{\alpha}}_{1}+\mathbf{\tilde{\alpha}_{2}} &\mathbf{=}&\sqrt{%
\frac{1+\zeta }{2}}\text{ }\mathbf{\delta }+\frac{1}{2}\left( \mathbf{%
\epsilon }_{1}-\mathbf{\epsilon }_{2}\right) -\frac{1}{2}\sqrt{\zeta }\left(
\mathbf{\epsilon }_{1}+\mathbf{\epsilon }_{2}\right)  \notag \\
\mathbf{\tilde{\alpha}_{2}+\tilde{\alpha}}_{3} &=&\sqrt{\frac{1+\zeta }{2}}%
\text{ }\mathbf{\delta }-\frac{1}{2}\left( \mathbf{\epsilon }_{1}-\mathbf{%
\epsilon }_{2}\right) +\frac{1}{2}\sqrt{\zeta }\left( \mathbf{\epsilon }_{1}+%
\mathbf{\epsilon }_{2}\right) \\
\mathbf{\tilde{\alpha}_{1}}+\mathbf{\tilde{\alpha}}_{2}+\mathbf{\tilde{\alpha%
}}_{3} &=&\sqrt{\frac{1+\zeta }{2}}\text{ }\mathbf{\delta }+\frac{1}{2}%
\left( \mathbf{\epsilon }_{1}-\mathbf{\epsilon }_{2}\right) +\frac{1}{2}%
\sqrt{\zeta }\left( \mathbf{\epsilon }_{1}+\mathbf{\epsilon }_{2}\right)
\notag
\end{eqnarray}%
as well as the long root%
\begin{equation}
\mathbf{\tilde{\psi}}=\sqrt{2\left( 1+\zeta \right) }\text{ }\mathbf{\delta }%
\qquad ,\qquad \mathbf{\tilde{\psi}}^{2}=-2\left( 1+\zeta \right)
\end{equation}%
The intersection matrix $\mathcal{\tilde{J}}_{ij}=\mathbf{\tilde{\alpha}}%
_{i}.\mathbf{\tilde{\alpha}}_{j}$ and the corresponding super Cartan matrix $%
\mathcal{\tilde{K}}_{ij}\left( \zeta \right) $ read in terms of $\zeta $ as
follows
\begin{equation}
\mathcal{\tilde{J}}_{ij}\left( \zeta \right) =\left(
\begin{array}{ccc}
2 & -1 & 0 \\
-1 & 0 & -\zeta \\
0 & -\zeta & 2\zeta%
\end{array}%
\right) \qquad ,\qquad \mathcal{\tilde{K}}_{ij}\left( \zeta \right) =\left(
\begin{array}{ccc}
2 & -1 & 0 \\
1 & 0 & \zeta \\
0 & -1 & 2%
\end{array}%
\right)
\end{equation}%
The determinant $\det \mathcal{\tilde{J}}_{ij}\left( \zeta \right) =-2\zeta
\left( \zeta +1\right) $ vanishes at $\zeta =-1,0$ while \textrm{the} $\det
\mathcal{\tilde{K}}_{ij}\left( \zeta \right) =2\left( \zeta +1\right) $
vanishes at $\zeta =-1.$ Below, we give the Lie superalgebras at the two
special points $\zeta =\pm 1$:

\begin{description}
\item[A)] \textbf{Orthosymplectic osp(4\TEXTsymbol{\vert}2):} $\zeta =1$ $%
\func{mod}\mathcal{Z}_{2}:$ This is a special point sitting on the real line
of the plan $\mathbb{C}\backslash \{-1,0\}.$ The simple roots are given by%
\begin{equation}
\begin{tabular}{lllll}
$\mathbf{\tilde{\alpha}}_{1}$ & $=$ & $\mathbf{\epsilon }_{1}-\mathbf{%
\epsilon }_{2}$ & $=$ & $\mathbf{\alpha }_{1}^{{\small osp}}$ \\
$\mathbf{\tilde{\alpha}}_{2}$ & $=$ & $\mathbf{\delta }-\mathbf{\epsilon }%
_{1}$ & $=$ & $\mathbf{\alpha }_{2}^{{\small osp}}$ \\
$\mathbf{\tilde{\alpha}}_{3}$ & $=$ & $\mathbf{\epsilon }_{1}+\mathbf{%
\epsilon }_{2}$ & $=$ & $\mathbf{\alpha }_{3}^{{\small osp}}$%
\end{tabular}
\label{osp1}
\end{equation}%
with long root $\mathbf{\psi }^{{\small osp}}=2\mathbf{\delta }.$ The
intersection matrix $\mathcal{\tilde{J}}\left( \zeta \right) |_{\zeta =1}=%
\mathcal{J}^{osp}$ and the associated super Cartan matrix $\mathcal{K}^{osp}$
\textrm{as follows}
\begin{equation}
\mathcal{J}^{osp}=\left(
\begin{array}{ccc}
2 & -1 & 0 \\
-1 & 0 & -1 \\
0 & -1 & 2%
\end{array}%
\right) \qquad ,\qquad \mathcal{K}^{osp}=\left(
\begin{array}{ccc}
2 & -1 & 0 \\
1 & 0 & 1 \\
0 & -1 & 2%
\end{array}%
\right)  \label{kosp}
\end{equation}%
The 4 other composite positive roots are given by%
\begin{equation}
\begin{tabular}{lll}
$\mathbf{\mathbf{\alpha }}_{1}^{{\small osp}}+\mathbf{\alpha }_{2}^{{\small %
osp}}$ & $=$ & $\mathbf{\delta -\mathbf{\epsilon }_{2}}$ \\
$\mathbf{\alpha }_{2}^{{\small osp}}+\mathbf{\alpha }_{3}^{{\small osp}}$ & $%
=$ & $\mathbf{\delta }+\mathbf{\epsilon }_{2}$%
\end{tabular}%
\quad ,\quad
\begin{tabular}{lll}
$\mathbf{\mathbf{\alpha }}_{1}^{{\small osp}}+\mathbf{\alpha }_{2}^{{\small %
osp}}+\mathbf{\alpha }_{3}^{{\small osp}}$ & $=$ & $\mathbf{\delta }+\mathbf{%
\epsilon }_{1}$ \\
$\mathbf{\alpha }_{1}^{{\small osp}}+2\mathbf{\alpha }_{2}^{{\small osp}}+%
\mathbf{\alpha }_{3}^{{\small osp}}$ & $=$ & $2\mathbf{\delta }$%
\end{tabular}
\label{osp2}
\end{equation}

\item[B)] \textbf{Roots} \textbf{at the point} $\zeta =-1$ $\func{mod}%
\mathcal{Z}_{2}:$ This special point corresponds to the Lie superalgebra sl(2%
\TEXTsymbol{\vert}2) with simple roots $\mathbf{\tilde{\alpha}}_{i}|_{\zeta
=-1}=\mathbf{\alpha }_{i}^{{\small sl}_{2|2}}$ as follows%
\begin{eqnarray}
\mathbf{\tilde{\alpha}}_{1}^{{\small sl}_{2|2}} &=&\left( \mathbf{\epsilon }%
_{1}-\mathbf{\epsilon }_{2}\right)  \notag \\
\mathbf{\tilde{\alpha}}_{2}^{{\small sl}_{2|2}} &=&\mathbf{\delta }-\frac{1}{%
2}\left( \mathbf{\epsilon }_{1}-\mathbf{\epsilon }_{2}\right) -\frac{i}{2}%
\left( \mathbf{\epsilon }_{1}+\mathbf{\epsilon }_{2}\right) \\
\mathbf{\tilde{\alpha}}_{3}^{{\small sl}_{2|2}} &=&i\left( \mathbf{\epsilon }%
_{1}+\mathbf{\epsilon }_{2}\right)  \notag
\end{eqnarray}%
where we used $\sqrt{-1}=i.$ These roots are related to the roots of osp(4%
\TEXTsymbol{\vert}2) as follows
\begin{equation}
\begin{tabular}{ccc}
$\mathbf{\alpha }_{1}^{{\small sl}_{2|2}}$ & $=$ & $\mathbf{\alpha }_{1}^{%
{\small osp}}$ \\
$2\mathbf{\tilde{\alpha}}_{2}^{{\small sl}_{2|2}}$ & $=$ & $2\mathbf{\delta }%
-\mathbf{\alpha }_{1}^{{\small osp}}-i\mathbf{\tilde{\alpha}}_{3}^{{\small sl%
}_{2|2}}$ \\
$\mathbf{\alpha }_{3}^{{\small sl}_{2|2}}$ & $=$ & $i\mathbf{\alpha }_{3}^{%
{\small osp}}$%
\end{tabular}%
\end{equation}%
indicating that $\mathbf{\tilde{\alpha}}_{1}^{{\small sl}_{2|2}}+2\mathbf{%
\tilde{\alpha}}_{1}^{{\small sl}_{2|2}}+\mathbf{\tilde{\alpha}}_{3}^{{\small %
sl}_{2|2}}=0$ $\func{mod}\left( 2\mathbf{\delta }\right) $. Here, the long
root $\mathbf{\tilde{\psi}}|_{\zeta =-1}$ vanishes [$\mathbf{\tilde{\psi}}%
|_{\zeta =-1}=0$ $\func{mod}\left( 2\mathbf{\delta }\right) $]. In these
relations, the intersection matrix $\mathcal{J}^{sl(2|2)}$ of the simple
roots and the corresponding super Cartan matrix $\mathcal{K}^{sl(2|2)}$ read
as follows%
\begin{equation}
\mathcal{J}^{sl(2|2)}=\left(
\begin{array}{ccc}
2 & -1 & 0 \\
-1 & 0 & 1 \\
0 & 1 & -2%
\end{array}%
\right) \qquad ,\qquad \mathcal{K}^{sl(2|2)}=\left(
\begin{array}{ccc}
2 & -1 & 0 \\
1 & 0 & -1 \\
0 & -1 & 2%
\end{array}%
\right)  \label{ksl22}
\end{equation}
\end{description}

\subsubsection{Weight diagram of $\tilde{\Phi}_{\mathfrak{d}(2,1;\protect%
\zeta )}^{{\protect\small (2)}}$ from the $\Phi _{{\protect\small osp}(4|2)}$
graph}

\begin{figure}[h]
\begin{center}
\includegraphics[width=10cm]{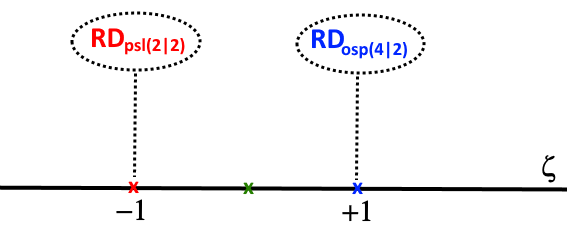}
\end{center}
\par
\vspace{-0.5cm}
\caption{Graphic rpresentation of $\widetilde{\mathfrak{RD}}^{%
{\protect\small (2)}}\left[ \protect\zeta \right] $ as a fibration over the
complex line $\protect\zeta $ of the graph $\mathfrak{RD}_{{\protect\small %
osp}(4|2)}$ sitting at $\protect\zeta =1.$ At the point $\protect\zeta =-1,$
it contracts to the root diagram $\mathfrak{RD}_{{\protect\small sl}(2|2)}.$}
\label{RD0}
\end{figure}

\qquad Because the Lie superalgebra $\mathfrak{d}(2,1;\zeta )$ has four
parametric root systems $\tilde{\Phi}_{\mathfrak{d}(2,1;\zeta )}^{{\small %
(\eta )}}$ in 1:1 with the four possible S$\mathfrak{DD}_{\mathfrak{d}%
(2,1;\zeta )}^{{\small (\eta )}}$ and subsequently in 1:1 with the four
parametric 2-cycles $\mathcal{C}_{\mathfrak{d}(2,1;\zeta )}^{{\small (\eta )}%
},$ i.e:%
\begin{equation}
\begin{tabular}{ccccc}
root systems & : & super Dynkin diagram & : & 2-cycles \\
$\tilde{\Phi}_{\mathfrak{d}(2,1;\zeta )}^{{\small (\eta )}}$ & $%
\leftrightarrow $ & S$\mathfrak{DD}_{\mathfrak{d}(2,1;\zeta )}^{{\small %
(\eta )}}$ & $\leftrightarrow $ & $\mathcal{C}_{\mathfrak{d}(2,1;\zeta )}^{%
{\small (\eta )}}$%
\end{tabular}%
\end{equation}%
one distinguishes 4 types of Root diagrams that we denote like
\begin{equation*}
\mathfrak{RD}_{\mathfrak{d}(2,1;\zeta )}^{{\small (\eta )}}
\end{equation*}%
Using the root system (\ref{DR}), we can draw the diagram $\mathfrak{RD}_{%
\mathfrak{d}(2,1;\zeta )}^{{\small (2)}}$ of the super $\tilde{\Phi}_{%
\mathfrak{d}(2,1;\zeta )}^{{\small (2)}}.$ This is a parametric graph
labeled by $\zeta $
\begin{equation*}
\mathfrak{RD}_{\mathfrak{d}(2,1;\zeta )}^{{\small (2)}}:=\widetilde{%
\mathfrak{RD}}^{{\small (2)}}\left[ \zeta \right] \qquad ,\qquad \eta
=0,1,2,3
\end{equation*}%
As such, it can be thought of as a fibration over the complex line $\zeta $
of the root diagram $\mathfrak{RD}_{osp(4|2)}$ of the orthosymplectic root
system $\Phi _{{\small osp}(4|2)}$ given by eqs(\ref{osp1},\ref{osp2}). This
fibration is depicted by the \textbf{Figure \ref{RD0}} where the $\mathfrak{%
RD}_{sl(2|2)}$ at $\zeta =-1$ is also shown.
The properties of the parametric $\widetilde{\mathfrak{RD}}^{{\small (2)}}%
\left[ \zeta \right] $ descend from those of $\tilde{\Phi}_{\mathfrak{d}%
(2,1;\zeta )}^{{\small (2)}}$ among which we cite the following:

\begin{description}
\item[$\left( \mathbf{1}\right) $] \textbf{values at} $\zeta =\pm 1$: The
simple roots of $\tilde{\Phi}_{\mathfrak{d}(2,1;\zeta )}^{{\small (2)}}$ are
\{$\mathbf{\tilde{\alpha}}_{1}\left( \zeta \right) ,\mathbf{\tilde{\alpha}}%
_{2}\left( \zeta \right) ,\mathbf{\tilde{\alpha}}_{3}\left( \zeta \right) \}$
with odd $\mathbf{\tilde{\alpha}}_{2}\left( \zeta \right) $ and even $%
\mathbf{\tilde{\alpha}}_{1}\left( \zeta \right) ,$ $\mathbf{\tilde{\alpha}}%
_{3}\left( \zeta \right) $; they are functions of $\zeta $ as in (\ref{in}).
Then, the $\tilde{\Phi}_{\mathfrak{d}(2,1;\zeta )}^{{\small (2)}}$ is a
parametric root system with values at $\zeta =\pm 1$ given by%
\begin{equation}
\tilde{\Phi}_{\mathfrak{d}(2,1;+1)}^{{\small (2)}}=\Phi _{{\small osp}%
(4|2)}\qquad ,\qquad \tilde{\Phi}_{\mathfrak{d}(2,1;-1)}^{{\small (2)}}=\Phi
_{{\small psl}(2|2)}
\end{equation}%
From these features, we deduce:%
\begin{equation}
\widetilde{\mathfrak{RD}}^{{\small (2)}}|_{\zeta =1}=\mathfrak{RD}_{{\small %
osp}(4|2)}\qquad ,\qquad \widetilde{\mathfrak{RD}}^{{\small (2)}}|_{\zeta
=-1}=\mathfrak{RD}_{{\small psl}(2|2)}
\end{equation}%
The Cartan-Weyl operators associated with the roots $\mathbf{\tilde{\alpha}}$
include: \newline
$\left( \mathbf{i}\right) $ the three parametric Chevalley triplets ($\tilde{%
H}_{\mathbf{\tilde{\alpha}}_{i}},\tilde{E}_{\pm \mathbf{\tilde{\alpha}}_{i}}$%
) with Cartan charges $\tilde{H}_{\mathbf{\tilde{\alpha}}_{i}}$ given by
\begin{equation*}
\tilde{E}_{+\mathbf{\tilde{\alpha}}_{i}}\tilde{E}_{-\mathbf{\tilde{\alpha}}%
_{i}}-\left( -\right) ^{\left\vert \mathbf{\tilde{\alpha}}_{i}\right\vert
\left\vert \mathbf{\tilde{\alpha}}_{i}\right\vert }\tilde{E}_{-\mathbf{%
\tilde{\alpha}}_{i}}\tilde{E}_{+\mathbf{\tilde{\alpha}}_{i}}
\end{equation*}%
with $\left\vert \mathbf{\tilde{\alpha}}\right\vert $ standing for the
degree of $\mathbf{\tilde{\alpha}}$; and \newline
$\left( \mathbf{ii}\right) $\ the composite operators associated with non
simple roots%
\begin{equation}
\tilde{E}_{\pm \left( \mathbf{\tilde{\alpha}}_{1}+\mathbf{\tilde{\alpha}_{2}}%
\right) }\quad ,\quad \tilde{E}_{\pm \left( \mathbf{\tilde{\alpha}}_{3}+%
\mathbf{\tilde{\alpha}_{2}}\right) }\quad ,\quad \tilde{E}_{\pm \left(
\mathbf{\tilde{\alpha}_{1}}+\mathbf{\tilde{\alpha}}_{3}+\mathbf{\tilde{\alpha%
}}_{2}\right) }\quad ,\quad \tilde{E}_{\pm \left( \mathbf{\tilde{\alpha}_{1}}%
+\mathbf{\tilde{\alpha}}_{3}+2\mathbf{\tilde{\alpha}}_{2}\right) }
\end{equation}%
They are given by the graded Serre relations \textrm{\cite{2B}; and are
functions of }$\zeta $.

\item[$\left( \mathbf{2}\right) $] \textbf{\ the diagram }$\widetilde{%
\mathfrak{RD}}^{{\small (2)}}$: The long root $\mathbf{\tilde{\psi}}=\mathbf{%
\tilde{\psi}}\left( \zeta \right) $ of the system $\tilde{\Phi}_{\mathfrak{d}%
(2,1;\zeta )}^{{\small (2+)}}$ and the long $\mathbf{\psi }^{{\small osp}}$
of the $\Phi _{{\small osp}(4|2)}$ are given by the even roots
\begin{equation*}
\mathbf{\tilde{\psi}}=\mathbf{\mathbf{\tilde{\alpha}_{1}}}+2\mathbf{\mathbf{%
\tilde{\alpha}}_{2}+\mathbf{\tilde{\alpha}}_{3}\qquad ,\qquad \mathbf{\psi }%
^{{\small osp}}}=\mathbf{\mathbf{\alpha }}_{1}^{{\small osp}}+2\mathbf{%
\mathbf{\alpha }}_{2}^{{\small osp}}+\mathbf{\mathbf{\alpha }}_{3}^{{\small %
osp}}
\end{equation*}%
The $\mathbf{\psi }^{{\small osp}}$ and $\mathbf{\tilde{\psi}}$ are also the
highest weight vectors of the "adjoint" representation of ${\small osp}(4|2)$
and $\mathfrak{d}(2,1;\zeta )$ respectively. The root diagram $\mathfrak{RD}%
_{{\small osp}(4|2)}$ and subsequently the diagram $\widetilde{\mathfrak{RD}}%
^{{\small (2)}}$ are depicted by the \textbf{Figure} \textbf{\ref{WD}}. The $%
\widetilde{\mathfrak{RD}}^{{\small (2)}}$ is imagined as the fibration of
the root diagram of ${\small osp}(4|2)$ over the complex line $\mathbb{C}%
\backslash \{-1,0\}$ with variable $\zeta .$
\begin{figure}[h]
\begin{center}
\includegraphics[width=7cm]{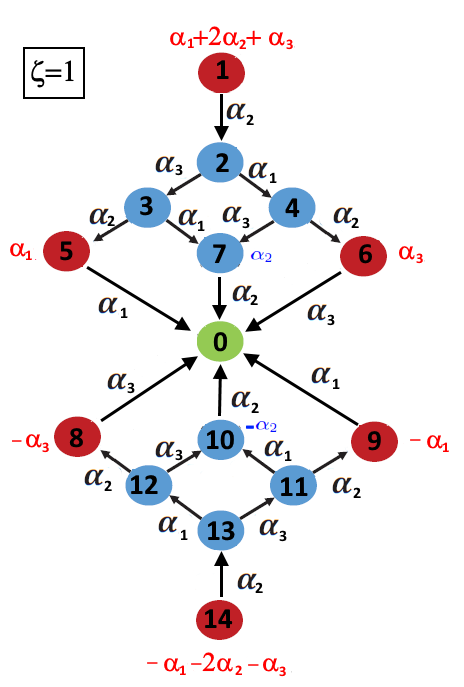}
\end{center}
\par
\vspace{-0.5cm}
\caption{The root diagram of $\widetilde{\mathfrak{RD}}^{{\protect\small (2)}%
}|_{\protect\zeta =1}=\mathfrak{RD}_{{\protect\small osp}(4|2)}$. For
generic $\protect\zeta $, we get the parametric root diagram $\widetilde{%
\mathfrak{RD}}^{{\protect\small (2)}}$. In red the 3+3 even roots of $%
\mathfrak{d}(2,1;\protect\zeta )_{\bar{0}}|_{\protect\zeta =+1}$ and in blue
the 4+4 odd ones. From this diagram, we learn the weight vectors of the
adjoint representation of $\mathfrak{d}(2,1;\protect\zeta )|_{\protect\zeta %
=+1}$ and its highest/lowest weight vectors.}
\label{WD}
\end{figure}
The parametric $\widetilde{\mathfrak{RD}}^{{\small (2)}}\left[ \zeta \right]
$ is built as follows:\newline
We start from the highest weight state \TEXTsymbol{\vert}$\mathbf{\tilde{\psi%
}}$\TEXTsymbol{>} which is: $\left( \mathbf{i}\right) $ an eigenvalue of the
diagonal Cartan charge operators $\tilde{H}_{\mathbf{\tilde{\alpha}}_{i}}|%
\mathbf{\tilde{\psi}}>=\tilde{q}_{i}|\mathbf{\tilde{\psi}}>$; and $\left(
\mathbf{ii}\right) $ annihilated by $\tilde{E}_{-\mathbf{\tilde{\alpha}}%
_{i}} $; that is $\tilde{E}_{-\mathbf{\tilde{\alpha}}_{i}}|\mathbf{\tilde{%
\psi}}>=0 $. \newline
Then, we act successively by appropriate monomials of the step operators $%
\tilde{E}_{\pm \mathbf{\tilde{\alpha}}}$; thus generating the states of the
adjoint representation. For example, we have%
\begin{equation}
\left\vert \mathbf{\mathbf{\tilde{\alpha}_{1}}+\mathbf{\tilde{\alpha}}_{2}+%
\mathbf{\tilde{\alpha}}_{3}}\right\rangle \simeq E_{\mathbf{\mathbf{\tilde{%
\alpha}}_{2}}}\left\vert \mathbf{\tilde{\psi}}\right\rangle
\end{equation}%
and
\begin{equation}
\left\vert \mathbf{\mathbf{\tilde{\alpha}}_{2}}\right\rangle \simeq E_{%
\mathbf{\mathbf{\tilde{\alpha}}_{1}+\mathbf{\tilde{\alpha}}_{3}}}E_{\mathbf{%
\mathbf{\tilde{\alpha}}_{2}}}\left\vert \mathbf{\tilde{\psi}}\right\rangle
\end{equation}%
as well as%
\begin{equation}
\left\vert \mathbf{\mathbf{\tilde{\alpha}_{1}}}\right\rangle \simeq E_{%
\mathbf{\mathbf{\tilde{\alpha}}_{2}+\mathbf{\tilde{\alpha}}_{3}}}E_{+\mathbf{%
\mathbf{\tilde{\alpha}}_{2}}}\left\vert \mathbf{\tilde{\psi}}\right\rangle
\qquad ,\qquad \left\vert \mathbf{\mathbf{\tilde{\alpha}}_{3}}\right\rangle
\simeq E_{\mathbf{\mathbf{\tilde{\alpha}}_{1}+\mathbf{\tilde{\alpha}}_{2}}%
}E_{\mathbf{\mathbf{\tilde{\alpha}}_{2}}}\left\vert \mathbf{\tilde{\psi}}%
\right\rangle
\end{equation}
\end{description}

\section{Geometry of D$(2,1;\protect\zeta )$ as deformation of OSp(4%
\TEXTsymbol{\vert}2)}

\qquad \label{sec4} In this section, we study geometric properties of the 9%
\TEXTsymbol{\vert}8-dim supermanifold describing the exceptional Lie super
group D($2,1;\zeta $) by help of the following features:

\begin{description}
\item[$\left( \mathbf{1}\right) $] the isometry $D(2,1;1)\simeq OSp(4|2)$ at
the point $\zeta =1$ of the complex line $\mathbb{C}\backslash \{-1,0\}.$
The D($2,1;\zeta $) has three parametric even roots (2-cycles) that we
denote like $\mathbf{\mathbf{\tilde{\beta}}}_{1}\left( \zeta \right) ,$ $%
\mathbf{\tilde{\beta}}_{2}\left( \zeta \right) $ and $\mathbf{\tilde{\beta}}%
_{3}\left( \zeta \right) $. Their values at the special point $\zeta =1$
give the roots (2-cycles) $\mathbf{\mathbf{\beta }}_{i}^{{\small osp}}$ of
OSp(4\TEXTsymbol{\vert}2). So, we have $\mathbf{\mathbf{\tilde{\beta}}}%
_{i}|_{\zeta =1}=\mathbf{\mathbf{\beta }}_{i}^{{\small osp}}.$

\item[$\left( \mathbf{2}\right) $] continuous deformation of the Kahler
parameter $\tilde{\varrho}_{3}\left( \zeta \right) $ of the even 2-cycle $%
\mathbf{\mathbf{\tilde{\beta}}}_{3}\left( \zeta \right) $ of the 17D
geometry of D$(2,1;\zeta ).$ We start from the special point
\begin{equation}
\mathbf{\mathbf{\beta }}_{3}^{{\small osp}}=\mathbf{\mathbf{\tilde{\beta}}}%
_{3}\left( \zeta \right) |_{\zeta =1}
\end{equation}%
of $OSp(4|2)$ and we move towards the critical point
\begin{equation}
\mathbf{\mathbf{\beta }}_{3}^{{\small sl(2|2)}}=\mathbf{\mathbf{\tilde{\beta}%
}}_{3}\left( \zeta \right) |_{\zeta =-1}
\end{equation}%
of $PSL(2|2)\times \mathbb{C}^{3}.$ For this continuous deformation, the
Kahler parameter $\tilde{\varrho}_{3}\left( \zeta \right) $ of the 2-cycle $%
\mathcal{\tilde{C}}_{\mathbf{\mathbf{\tilde{\beta}}}_{3}}$ in SL(2,$\mathbb{C%
}$)$_{3}$ is taken like $2/\left\vert \zeta \left( \zeta +1\right)
\right\vert $; its value at $\zeta =1$ is equal to $\varrho _{3}^{{\small osp%
}}=1$ while at $\zeta =-1,$ it is given by the diverging $\varrho _{3}^{%
{\small sl(2|2)}}\rightarrow \infty .$
\end{description}

\ \newline
To undertake this study, we first describe the geometry of the bosonic part
given by%
\begin{equation}
\begin{tabular}{lllllll}
D$(2,1;\zeta )_{\bar{0}}$ & $\simeq $ & $SL(2,\mathbb{C})^{\otimes 3}$ &
\qquad for & $\zeta $ & $\in $ & $\mathbb{C}\backslash \{-1,0\}$ \\
D$(2,1;\zeta )_{\bar{0}}$ & $\simeq $ & $SL(2,\mathbb{C})^{\otimes 2}\otimes
\mathbb{C}^{3}$ & \qquad for & $\zeta $ & $=$ & $\{-1,0\}$%
\end{tabular}%
\end{equation}%
Then, we consider the super extension by implementing the odd part D$%
(2,1;\zeta )_{\bar{1}}$.

\subsection{Bosonic subspace D$(2,1;\protect\zeta )_{\bar{0}}$}

\qquad The complex geometry of the bosonic D($2,1;\zeta $)$_{\bar{0}}$ is
given by a parametric 9 dimensional manifold $SL(2)^{\otimes 3}$ with
parametric roots $\mathbf{\mathbf{\tilde{\beta}}}_{i}$ that are functions of
$\zeta $ \textrm{\cite{SUP}}. It is denoted like $\mathcal{\tilde{X}}_{123}%
\left[ \zeta \right] $ in analogy with the $\mathcal{X}_{120}$ geometry of $%
SL(2|2)_{\bar{0}}\simeq SL(2)^{\otimes 2}\times \mathbb{C}^{\ast }$ sitting
at $\zeta =-1.$

By using (\ref{x12}), the $\mathcal{\tilde{X}}_{123}$ factorises like $%
\mathcal{\tilde{X}}_{1}\times \mathcal{\tilde{X}}_{2}\times \mathcal{\tilde{X%
}}_{3}$ where $\mathcal{\tilde{X}}_{i}\simeq \mathcal{\tilde{C}}_{\mathbf{%
\mathbf{\tilde{\beta}}}_{i}}\times \mathfrak{\tilde{F}}_{i}$ with 2-cycles $%
\mathcal{\tilde{C}}_{\mathbf{\mathbf{\tilde{\beta}}}_{i}}$ (1D complex
curves) and complex surfaces $\mathfrak{\tilde{F}}_{i}.$ The three 2-cycles $%
\mathcal{\tilde{C}}_{\mathbf{\mathbf{\tilde{\beta}}}_{i}}$ are associated
with the even roots of D($2,1;\zeta $); they are given by%
\begin{equation}
\begin{tabular}{lllll}
$\mathbf{\mathbf{\tilde{\beta}}}_{1}$ & $=$ & $\mathbf{\tilde{\alpha}}_{1}$
& $\quad \leftrightarrow \quad $ & $\mathcal{\tilde{C}}_{\mathbf{\mathbf{%
\tilde{\beta}}}_{1}}$ \\
$\mathbf{\mathbf{\tilde{\beta}}}_{2}$ & $=$ & $\mathbf{\tilde{\alpha}}_{3}$
& $\quad \leftrightarrow \quad $ & $\mathcal{\tilde{C}}_{\mathbf{\mathbf{%
\tilde{\beta}}}_{2}}$ \\
$\mathbf{\mathbf{\tilde{\beta}}}_{3}$ & $=$ & $\mathbf{\tilde{\psi}}$ & $%
\quad \leftrightarrow \quad $ & $\mathcal{\tilde{C}}_{\mathbf{\mathbf{\tilde{%
\beta}}}_{3}}$%
\end{tabular}%
\end{equation}%
with $\mathbf{\tilde{\psi}=\tilde{\alpha}}_{1}+2\mathbf{\tilde{\alpha}}_{2}+%
\mathbf{\tilde{\alpha}}_{3}$ and vanishing intersections
\begin{equation}
\mathbf{\mathbf{\tilde{\beta}}}_{1}.\mathbf{\mathbf{\tilde{\beta}}}%
_{2}=0\quad ,\quad \mathbf{\mathbf{\tilde{\beta}}}_{1}.\mathbf{\mathbf{%
\tilde{\beta}}}_{3}=0\quad ,\quad \mathbf{\mathbf{\tilde{\beta}}}_{2}.%
\mathbf{\mathbf{\tilde{\beta}}}_{3}=0
\end{equation}%
By substituting the factorisation $\mathcal{\tilde{C}}_{\mathbf{\mathbf{%
\tilde{\beta}}}_{1}}\times \mathfrak{\tilde{F}}_{i}$ into $\mathcal{\tilde{X}%
}_{1}\times \mathcal{\tilde{X}}_{2}\times \mathcal{\tilde{X}}_{3}$, we can
express the complex $\mathcal{\tilde{X}}_{{\small 123}}$ as the fibration $%
\mathfrak{\tilde{B}}_{{\small 123}}\times \mathfrak{\tilde{F}}_{{\small 123}%
} $ with 6-cycle
\begin{equation}
\mathfrak{\tilde{B}}_{{\small 123}}=\mathcal{\tilde{C}}_{\mathbf{\mathbf{%
\tilde{\beta}}}_{1}}\times \mathcal{\tilde{C}}_{\mathbf{\mathbf{\tilde{\beta}%
}}_{2}}\times \mathcal{\tilde{C}}_{\mathbf{\mathbf{\tilde{\beta}}}_{3}}
\end{equation}%
Because each isospin group of $\otimes _{i=1}^{3}SL(2)_{\mathbf{\mathbf{%
\tilde{\beta}}}_{i}}$ can be realised by complex 2$\times $2 matrices $%
\boldsymbol{X}_{i}$, the geometric description of the super D$(2,1;\zeta )_{%
\bar{0}}$ can be imagined in terms of parametric $6\times 6$ matrix $%
\boldsymbol{D}=\boldsymbol{D}\left( \zeta \right) $ as follows%
\begin{equation}
\boldsymbol{D}=\left(
\begin{array}{ccc}
\boldsymbol{X}_{1} & \mathbf{0} & \mathbf{0} \\
\mathbf{0} & \boldsymbol{X}_{2} & \mathbf{0} \\
\mathbf{0} & \mathbf{0} & \boldsymbol{X}_{3}%
\end{array}%
\right) \qquad with\qquad \boldsymbol{X}_{i}=\left(
\begin{array}{cc}
x_{i} & u_{i} \\
y_{i} & v_{i}%
\end{array}%
\right)  \label{XXX}
\end{equation}%
Moreover, as each $\boldsymbol{X}_{i}\left( \zeta \right) $\ describes an $%
SL(2)_{\mathbf{\mathbf{\tilde{\beta}}}_{i}}$, the complex variables \{$%
\left( x_{i},y_{i},u_{i},v_{i}\right) _{i=1,2,3}$\} parameterising $\left(
\text{End}\mathbb{C}^{2}\right) ^{3}$ are constrained by the conditions $%
\det \boldsymbol{X}_{i}=1$ reading explicitly as%
\begin{equation}
\begin{tabular}{lllll}
$\boldsymbol{X}_{1}$ & $:$ & $x_{1}v_{1}-y_{1}u_{1}$ & $=$ & $1$ \\
$\boldsymbol{X}_{2}$ & $:$ & $x_{2}v_{2}-y_{2}u_{2}$ & $=$ & $1$ \\
$\boldsymbol{X}_{3}$ & $:$ & $x_{3}v_{3}-y_{3}u_{3}$ & $=$ & $1$%
\end{tabular}
\label{3X}
\end{equation}%
Using (\ref{XXX}) and setting $x_{i}=a_{i}+b_{i}$ and $v_{i}=\bar{a}_{i}-%
\bar{b}_{i}$ as well as $y_{i}=c_{i}+d_{i}$ and $u_{i}=\bar{c}_{i}-\bar{d}%
_{i},$ we can expand these relationships (\ref{3X}) as follows
\begin{equation}
\begin{tabular}{lllll}
$\mathfrak{\tilde{B}}_{1}$ & $:$ & $\left\vert a_{1}\right\vert
^{2}+\left\vert d_{1}\right\vert ^{2}-\left\vert b_{1}\right\vert
^{2}-\left\vert c_{1}\right\vert ^{2}$ & $=$ & $1$ \\
$\mathfrak{\tilde{F}}_{1}$ & $:$ & $b_{1}\bar{a}_{1}-a_{1}\bar{b}_{1}+d_{1}%
\bar{c}_{1}-c_{1}\bar{d}_{1}$ & $=$ & $0$%
\end{tabular}%
\end{equation}%
and
\begin{equation}
\begin{tabular}{lllll}
$\mathfrak{\tilde{B}}_{2}$ & $:$ & $\left\vert a_{2}\right\vert
^{2}+\left\vert d_{2}\right\vert ^{2}-\left\vert b_{2}\right\vert
^{2}-\left\vert c_{2}\right\vert ^{2}$ & $=$ & $1$ \\
$\mathfrak{\tilde{F}}_{2}$ & $:$ & $b_{2}\bar{a}_{2}-a_{2}\bar{b}_{2}+d_{2}%
\bar{c}_{2}-c_{2}\bar{d}_{2}$ & $=$ & $0$%
\end{tabular}%
\end{equation}%
as well as%
\begin{equation}
\begin{tabular}{lllll}
$\mathfrak{\tilde{B}}_{3}$ & $:$ & $\left\vert a_{3}\right\vert
^{2}+\left\vert d_{3}\right\vert ^{2}-\left\vert b_{3}\right\vert
^{2}-\left\vert c_{3}\right\vert ^{2}$ & $=$ & $1$ \\
$\mathfrak{\tilde{F}}_{3}$ & $:$ & $b_{3}\bar{a}_{3}-a_{3}\bar{b}_{3}+d_{3}%
\bar{c}_{3}-c_{3}\bar{d}_{3}$ & $=$ & $0$%
\end{tabular}%
\end{equation}%
On the locus $b_{i}=c_{i}=0,$ these relations become%
\begin{equation}
\begin{tabular}{lllll}
$\mathbb{S}_{1}^{2}$ & $:$ & $\left\vert a_{1}\right\vert ^{2}+\left\vert
d_{1}\right\vert ^{2}$ & $=$ & $1$ \\
$\mathbb{S}_{2}^{2}$ & $:$ & $\left\vert a_{2}\right\vert ^{2}+\left\vert
d_{2}\right\vert ^{2}$ & $=$ & $1$ \\
$\mathbb{S}_{3}^{2}$ & $:$ & $\left\vert a_{3}\right\vert ^{2}+\left\vert
d_{3}\right\vert ^{2}$ & $=$ & $1$%
\end{tabular}%
\end{equation}%
which by using the parametrisation $v_{3}=\varrho _{3}\bar{x}_{3}$ and $%
u_{3}=-\varrho _{3}\bar{y}_{3}$ [see also the discussion between eqs(\ref{a1}%
) and (\ref{a})], they take the form
\begin{equation}
\begin{tabular}{lllll}
$\mathbb{S}_{1}^{2}$ & $:$ & $\left\vert a_{1}\right\vert ^{2}+\left\vert
d_{1}\right\vert ^{2}$ & $=$ & $1$ \\
$\mathbb{S}_{2}^{2}$ & $:$ & $\left\vert a_{2}\right\vert ^{2}+\left\vert
d_{2}\right\vert ^{2}$ & $=$ & $1$ \\
$\mathbb{S}_{3}^{2}$ & $:$ & $\left\vert x_{3}\right\vert ^{2}+\left\vert
y_{3}\right\vert ^{2}$ & $=$ & $\frac{1}{\varrho _{3}}$%
\end{tabular}%
\end{equation}%
\textrm{with positive} $\varrho _{3}=\left\vert \zeta \left( \zeta +1\right)
\right\vert /2$ and area
\begin{equation}
vol(\mathbb{S}_{\mathbf{\mathbf{\tilde{\beta}}}_{3}}^{2})=\frac{8\pi }{%
\left\vert \zeta \left( \zeta +1\right) \right\vert }
\end{equation}%
In this geometric description, the contraction $\zeta \rightarrow -1$
reduces the 6+3 dimensional D($2,1;\zeta $)$_{\bar{0}}$ down to the \textrm{6%
} dimensional PSL(2\TEXTsymbol{\vert}2)$_{\bar{0}}$. It corresponds to the
decompactification of the 2-cycle $\mathcal{\tilde{C}}_{\mathbf{\mathbf{%
\tilde{\beta}}}_{3}}|_{\zeta =-1}$ into $\mathbb{R}^{2}\simeq \mathbb{C}$
for 2-sphere or $\mathbb{R}^{1,1}$ for non compact pseudo 2-sphere.

\subsection{Odd part D$(2,1;\protect\zeta )_{\bar{1}}$}

\qquad The contribution of the odd part in the geometric description of the
exceptional supergroup D$(2,1;\zeta )$ is given by the tri-fundamental $%
\left( 1/2,1/2,1/2\right) $ of the even part $D(2,1;\zeta )_{\bar{0}}=\Pi
_{i=1}^{3}SL(2)_{\mathbf{\mathbf{\tilde{\beta}}}_{i}}.$ Because of the Z$%
_{3} $-graded matrix description for D$(2,1;\zeta )$ due to the
tri-fundamental, we cannot apply the method used for SL(2\TEXTsymbol{\vert}%
2). However, we can get around this difficulty by using $\left( \mathbf{i}%
\right) $ the \emph{isomorphism} D$(2,1;1)_{\bar{0}}\simeq OSp(4|2)_{\bar{0}%
} $ with
\begin{equation}
\begin{tabular}{ccc}
$osp(4|2)_{\bar{0}}$ & $\simeq $ & $so_{\mathbb{C}}(4)\oplus sp_{\mathbb{C}%
}(2)$ \\
$so_{\mathbb{C}}(4)$ & $\simeq $ & $sl(2,\mathbb{C})\oplus sl(2,\mathbb{C})$%
\end{tabular}%
\end{equation}%
and $\left( \mathbf{ii}\right) $ the odd sector given by the module $\left(
\mathbf{4},\mathbf{2}\right) $ of the group $SO_{\mathbb{C}}(4)\times SP_{%
\mathbb{C}}(2).$ In this orthosymplectic setting, the parametric geometry of
D$(2,1;1)$ can be investigated by using 6$\times $6 matrices of $OSp(4|2)$.
In this regard, recall that the super $OSp(4|2)$ is a subgroup of the linear
super group GL$(4|2)$ given by super $4|2\times 4|2$ matrices $\boldsymbol{M}
$ and super transpose $\boldsymbol{M}^{st},$
\begin{equation}
\boldsymbol{M}=\left(
\begin{array}{cc}
\boldsymbol{A}_{4\times 4} & \boldsymbol{B}_{4\times 2} \\
\boldsymbol{C}_{2\times 4} & \boldsymbol{D}_{2\times 2}%
\end{array}%
\right) \qquad ,\qquad \boldsymbol{M}^{st}=\left(
\begin{array}{cc}
\boldsymbol{A}_{4\times 4}^{t} & \boldsymbol{C}_{4\times 2}^{t} \\
-\boldsymbol{B}_{2\times 4}^{t} & \boldsymbol{D}_{2\times 2}^{t}%
\end{array}%
\right)  \label{ABCD}
\end{equation}%
constrained as follows%
\begin{equation}
\boldsymbol{M}^{st}\boldsymbol{J}_{4|2}\boldsymbol{M}=\boldsymbol{J}_{4|2}
\end{equation}%
with non vanishing super determinant $Ber\left( \boldsymbol{M}\right) $
factorising like $\det \left( \boldsymbol{A}-\boldsymbol{BD}^{-1}\boldsymbol{%
C}\right) \det \boldsymbol{D}^{-1}.$\ In this relation, the super
orthosymplectic metric $\boldsymbol{J}_{4|2}$ is given by%
\begin{equation}
\boldsymbol{J}_{4|2}=\left(
\begin{array}{cccc}
\boldsymbol{0}_{2\times 2} & \boldsymbol{I}_{2\times 2} & 0 & 0 \\
\boldsymbol{I}_{2\times 2} & \boldsymbol{0}_{2\times 2} & 0 & 0 \\
0 & 0 & 0 & 1 \\
0 & 0 & -1 & 0%
\end{array}%
\right)  \label{J4}
\end{equation}%
with $\boldsymbol{I}_{2\times 2}$ being the identity 2$\times $2 matrix. The
graded $\boldsymbol{M}$ and super transpose $\boldsymbol{M}^{st}$ have even
and odd submatrix blocks; the even block is given by $\boldsymbol{M}_{\bar{0}%
}=\boldsymbol{A}_{4\times 4}\oplus \boldsymbol{D}_{2\times 2};$ its super
determinant
\begin{equation}
\det \boldsymbol{M}_{\bar{0}}=\det \boldsymbol{A}_{4\times 4}\det
\boldsymbol{D}_{2\times 2}^{-1}
\end{equation}%
is invariant under the complex scaling
\begin{equation}
\boldsymbol{A}_{4\times 4}\rightarrow \lambda \boldsymbol{A}_{4\times
4}\qquad ,\qquad \boldsymbol{D}_{2\times 2}\rightarrow \lambda ^{2}%
\boldsymbol{D}_{2\times 2}
\end{equation}%
This is because $\det \boldsymbol{A}_{4\times 4}\rightarrow \lambda ^{4}\det
\boldsymbol{A}_{4\times 4}$ and $\det \boldsymbol{D}_{2\times 2}\rightarrow
\lambda ^{4}\det \boldsymbol{D}_{2\times 2}$. Extending this scaling to the
odd sector $\boldsymbol{M}_{\bar{1}}=\boldsymbol{B}_{4\times 2}\oplus
\boldsymbol{C}_{2\times 4}$, the invariance of the Berezinian $Ber\left(
\boldsymbol{M}\right) $ requires, in addition to the transformation of the
even sector, the scalings%
\begin{equation}
\boldsymbol{B}_{4\times 2}\rightarrow \lambda ^{2}\boldsymbol{B}_{4\times
2}\qquad ,\qquad \boldsymbol{C}_{2\times 4}\rightarrow \lambda \boldsymbol{C}%
_{2\times 4}
\end{equation}%
Furthermore, by putting (\ref{J4}) into the condensed form%
\begin{equation}
\boldsymbol{J}_{4|2}=\left(
\begin{array}{cc}
G_{4\times 4} & 0 \\
0 & \Omega _{2\times 2}%
\end{array}%
\right) ,\quad G_{4\times 4}=\left(
\begin{array}{cc}
\boldsymbol{0}_{2\times 2} & \boldsymbol{I}_{2\times 2} \\
\boldsymbol{I}_{2\times 2} & \boldsymbol{0}_{2\times 2}%
\end{array}%
\right) ,\quad \Omega _{2\times 2}=\left(
\begin{array}{cc}
0 & 1 \\
-1 & 0%
\end{array}%
\right)
\end{equation}%
we can split the condition $\boldsymbol{M}^{st}\boldsymbol{J}_{4|2}%
\boldsymbol{M}=\boldsymbol{J}_{4|2}$ as follows%
\begin{equation}
\begin{tabular}{lll}
$\boldsymbol{A}_{4\times 4}^{t}G_{4\times 4}\boldsymbol{A}_{4\times 4}+%
\boldsymbol{C}_{4\times 2}^{t}\Omega _{2\times 2}\boldsymbol{C}_{2\times 4}$
& $=$ & $G_{4\times 4}$ \\
$\boldsymbol{D}_{2\times 2}^{t}\Omega _{2\times 2}\boldsymbol{D}_{2\times 2}-%
\boldsymbol{B}_{2\times 4}^{t}G_{4\times 4}\boldsymbol{B}_{4\times 2}$ & $=$
& $\Omega _{2\times 2}$ \\
$\boldsymbol{A}_{4\times 4}^{t}G_{4\times 4}\boldsymbol{B}_{4\times 2}+%
\boldsymbol{C}_{4\times 2}^{t}\Omega _{2\times 2}\boldsymbol{D}_{2\times 2}$
& $=$ & $0$ \\
$\boldsymbol{D}_{2\times 2}^{t}\Omega _{2\times 2}\boldsymbol{C}_{2\times 4}-%
\boldsymbol{B}_{2\times 4}^{t}G_{4\times 4}\boldsymbol{A}_{4\times 4}$ & $=$
& $0$%
\end{tabular}
\label{ctt}
\end{equation}%
Notice that from (\ref{ctt}), we learn that $\left( \mathbf{i}\right) $ the
third and the fourth relations are related under transpose involution; and $%
\left( \mathbf{ii}\right) $ the odd blocks $\boldsymbol{B}_{4\times 2}\ $and
$\boldsymbol{C}_{4\times 2}$ are linked like%
\begin{equation}
\boldsymbol{C}_{4\times 2}^{t}=\boldsymbol{A}_{4\times 4}^{t}G_{4\times 4}%
\boldsymbol{B}_{4\times 2}\boldsymbol{D}_{2\times 2}^{-1}\Omega _{2\times 2}
\label{ct}
\end{equation}%
leaving free the block matrices $\boldsymbol{A}_{4\times 4}$ and $%
\boldsymbol{B}_{4\times 2}$\ as well as $\boldsymbol{D}_{2\times 2}$. The
remaining relations give $10+1+8=19$ constraint relations amongst the $%
6\times 6=36$ variables in eq(\ref{ABCD}); thus leaving $17$ free degrees of
freedom parameterising the complex variety of super $OSp(4|2).$

In summary, the even sector has 9 degrees of freedom; 6 degrees coming from 4%
$\times $4 (antisymmetric) matrix $\boldsymbol{A}_{4\times 4}$ and 3 ones
from the 2$\times $2 (symmetric) matrix $\boldsymbol{D}_{2\times 2}.$ The
odd sector has 8 degrees of freedom coming from the $4\times 2$ matrix $%
\boldsymbol{B}$ or equivalently from the $2\times 4$ matrix $\boldsymbol{C}%
_{4\times 2}\sim \left( \boldsymbol{B}_{2\times 4}\right) ^{t}$ as in (\ref%
{ct}).

The link of the orthosymplectic description $OSp(4|2)$ to the 15 dimensional
SL(2\TEXTsymbol{\vert}2) is obtained by imposing two conditions: First, by
fixing the super determinant like
\begin{equation*}
Ber\left( \boldsymbol{M}\right) =1
\end{equation*}
this reduces the dimension from 17 down to 16. Second, by taking the coset
by the scaling transformation%
\begin{equation}
\left(
\begin{array}{cc}
\boldsymbol{A} & \boldsymbol{B} \\
\boldsymbol{C} & \boldsymbol{D}%
\end{array}%
\right) \qquad \rightarrow \qquad \left(
\begin{array}{cc}
\lambda \boldsymbol{A} & \lambda ^{2}\boldsymbol{B} \\
\lambda \boldsymbol{C} & \lambda ^{2}\boldsymbol{D}%
\end{array}%
\right)  \label{lam}
\end{equation}%
this identification reduces further the dimension 16 down to 15. Eqs(\ref%
{ctt}) are invariant under (\ref{lam}).

\section{Integrable D(2,1; $\protect\zeta $) superspin chain}

\qquad \label{sec5} In this section, we give an application of the above
analysis on the parametric $\tilde{\Phi}_{\mathfrak{d}(2,1;\zeta )}^{{\small %
(2)}}$ and the S$\mathfrak{DD}^{{\small (2)}}\left[ \zeta \right] $ in the
study of integrable superspin chains by using 4D Chern-Simons (CS) theory
with D(2,1; $\zeta $) symmetry. This will be done by calculating one of the
basic operators in integrable spin systems namely the parametric Lax
operator
\begin{equation*}
\mathcal{L}_{\mathfrak{d}(2,1;\zeta )}^{\left( \mathbf{\mu }\right) }:=%
\boldsymbol{L}\left( \zeta \right)
\end{equation*}%
which, in the 4D CS gauge theory, is given by the Costello- Gaiotto- Yagi
formula $e^{X}z^{\mathbf{\mu }}e^{Y}$ \textrm{\cite{Yama}}. Useful details
on the Chern-Simons construction \textrm{\cite{1AA,2AA,BE}} and on the
triplet ($z^{\mathbf{\mu }},X,Y$) are reported in the \textrm{\autoref{appD}%
. Here, we focus on the calculation of }$\mathcal{L}_{\mathfrak{d}(2,1;\zeta
)}^{\left( \mathbf{\mu }\right) }$ by using interpolation ideas motivated by
algebraic geometry properties between the local data on $\mathcal{L}%
_{sl(2|2)}^{\left( \mathbf{\mu }\right) }$ and $\mathcal{L}%
_{osp(4|2)}^{\left( \mathbf{\mu }\right) }$ living at $\zeta =\pm 1.$

To undertake this study, we recall that integrable distinguished superspin
chain $\mathfrak{d}$(2,1; $\zeta $) is characterised by a root system $%
\tilde{\Phi}_{\mathfrak{d}(2,1;\zeta )}^{{\small (2)}}$ generated by three
simple roots $\left( \mathbf{\mathbf{\tilde{\alpha}}}_{1},\mathbf{\mathbf{%
\tilde{\alpha}}}_{2},\mathbf{\mathbf{\tilde{\alpha}}}_{3}\right) $ with odd $%
\mathbf{\mathbf{\tilde{\alpha}}}_{2}$ and even $\mathbf{\mathbf{\tilde{\alpha%
}}}_{1},\mathbf{\mathbf{\tilde{\alpha}}}_{3}$. This distinguished root basis
is realised in terms of the canonical weights ($\mathbf{\epsilon }_{1},%
\mathbf{\epsilon }_{2},\mathbf{\delta }$) as in (\ref{se}) namely,
\begin{eqnarray}
\mathbf{\tilde{\alpha}}_{1} &=&\left( \mathbf{\epsilon }_{1}-\mathbf{%
\epsilon }_{2}\right)  \notag \\
\mathbf{\tilde{\alpha}}_{2} &=&\frac{1}{2}\sqrt{2\left( 1+\zeta \right) }%
\mathbf{\delta }-\frac{1}{2}\left( \mathbf{\epsilon }_{1}-\mathbf{\epsilon }%
_{2}\right) -\frac{1}{2}\sqrt{\zeta }\left( \mathbf{\epsilon }_{1}+\mathbf{%
\epsilon }_{2}\right) \\
\mathbf{\tilde{\alpha}}_{3} &=&\sqrt{\zeta }\left( \mathbf{\epsilon }_{1}+%
\mathbf{\epsilon }_{2}\right)  \notag
\end{eqnarray}%
where $\zeta $ belongs to $\mathbb{C}\backslash \{-1,0\}.$ For these
parametric simple roots, the intersection matrix $\mathbf{\mathbf{\tilde{%
\alpha}}}_{i}.\mathbf{\mathbf{\tilde{\alpha}}}_{j}$ is a parametric 3$\times
$3 matrix $\mathcal{\tilde{J}}_{ij}\left( \zeta \right) =\mathbf{\mathbf{%
\tilde{\alpha}}}_{i}.\mathbf{\mathbf{\tilde{\alpha}}}_{j}$ given by%
\begin{equation}
\mathcal{\tilde{J}}_{ij}\left( \zeta \right) =\left(
\begin{array}{ccc}
2 & -1 & 0 \\
-1 & 0 & -\zeta \\
0 & -\zeta & 2\zeta%
\end{array}%
\right) \qquad ,\qquad \det \mathcal{\tilde{J}}_{ij}\left( \zeta \right)
=-2\zeta \left( 1+\zeta \right)  \label{ala}
\end{equation}%
with vanishing value ($\det \mathcal{\tilde{J}}_{ij}\left( \zeta \right) =0$%
) at $\zeta =-1,0;$ and $\det \mathcal{\tilde{J}}_{ij}\left( \zeta \right)
|_{\zeta =1}=-4$ for osp(4\TEXTsymbol{\vert}2) and super Cartan matrices as
in (\ref{kosp}) and (\ref{ksl22}). Eq(\ref{ala}) contains as subsystems two
particular super symmetries located at $\zeta =\pm 1$ and then two
integrable superspin sub-chains namely:

\begin{itemize}
\item the integrable sl(2\TEXTsymbol{\vert}2) superspin chain with Lax
operator $\mathcal{L}_{sl(2|2)}^{\left( \mathbf{\mu }\right) }$ at $\zeta
=-1 $; and

\item the integrable osp(4\TEXTsymbol{\vert}2) spin chain with L-operator $%
\mathcal{L}_{osp(4|2)}^{\left( \mathbf{\mu }\right) }$ at $\zeta =1$.
\end{itemize}

To determine the $\mathcal{L}_{\mathfrak{d}(2,1;\zeta )}^{\left( \mathbf{\mu
}\right) }$, we use the following features:

\begin{description}
\item[$\left( \mathbf{i}\right) $] the $\mathcal{L}_{\mathfrak{d}(2,1;\zeta
)}^{\left( \mathbf{\mu }\right) }$ must inherit properties of $\mathfrak{d}%
(2,1;\zeta )$; it should be a parametric operator with parameter $\zeta ;$
and must have critical values at $\zeta =-1,0$ where $\det \left( \mathbf{%
\mathbf{\tilde{\alpha}}}_{i}.\mathbf{\mathbf{\tilde{\alpha}}}_{j}\right) =0$
and where $\mathfrak{d}(2,1;\zeta )$ contracts to semi-simple Lie sub-
superalgebras.

\item[$\left( \mathbf{ii}\right) $] the expression of the parametric $%
\mathcal{L}_{\mathfrak{d}(2,1;\zeta )}^{\left( \mathbf{\mu }\right) }$ must
reproduce the values of $\mathcal{L}_{sl(2|2)}^{\left( \mathbf{\mu }\right)
} $ and $\mathcal{L}_{osp(4|2)}^{\left( \mathbf{\mu }\right) }$ at $\zeta
=\pm 1;$ that is:%
\begin{equation}
\left. \mathcal{L}_{\mathfrak{d}(2,1;\zeta )}^{\left( \mathbf{\mu }\right)
}\right\vert _{\zeta =-1}=\mathcal{L}_{sl(2|2)}^{\left( \mathbf{\mu }\right)
}\qquad ,\qquad \left. \mathcal{L}_{\mathfrak{d}(2,1;\zeta )}^{\left(
\mathbf{\mu }\right) }\right\vert _{\zeta =+1}=\mathcal{L}%
_{osp(4|2)}^{\left( \mathbf{\mu }\right) }  \label{spt}
\end{equation}
\end{description}

Using these local features, we first calculate $\mathcal{L}%
_{sl(2|2)}^{\left( \mathbf{\mu }\right) }$; thus completing results in
\textrm{\cite{BF,3AA}} concerning $\mathcal{L}_{sl(m|n)}^{\left( \mathbf{\mu
}\right) }$ for $m\neq n$. Then, we compute $\mathcal{L}_{osp(4|2)}^{\left(
\mathbf{\mu }\right) }$ which also completes results for $\mathcal{L}%
_{osp(2m|2n)}^{\left( \mathbf{\mu }\right) }$ obtained in \textrm{\cite{3AA}}%
\ for $m\geq 3.$ After that, we give an approach to connect the $\mathcal{L}%
_{sl(2|2)}^{\left( \mathbf{\mu }\right) }$ and $\mathcal{L}%
_{osp(4|2)}^{\left( \mathbf{\mu }\right) }$ and then deduce a parametric
realisation of $\mathcal{L}_{\mathfrak{d}(2,1;\zeta )}^{\left( \mathbf{\mu }%
\right) }.$

\subsection{Integrable sl(2\TEXTsymbol{\vert}2) superspin chain}

\qquad We begin by recalling that the semi-simple Lie superalgebra $sl(2|2)$
[resp. the simple $psl(2|2)$] has rank $r=3$ [resp. $2$] and dimensions $15$
[resp. $14$]. It has two sectors: $\left( \mathbf{i}\right) $ an even sector
$sl(2|2)_{\bar{0}}$ [resp. $psl(2|2)_{\bar{0}}$] describing bosons and
splitting as\textrm{\ }%
\begin{equation}
\begin{tabular}{lll}
$sl(2|2)_{\bar{0}}$ & $=$ & $sl(2)\oplus sl(2)\oplus gl(1)_{z}$ \\
$psl(2|2)_{\bar{0}}$ & $=$ & $sl(2)\oplus sl(2)$%
\end{tabular}%
\end{equation}%
where $gl(1)_{z}$ is generated by the diagonal central element (identity
operator). $\left( \mathbf{ii}\right) $ an odd sector\textrm{\ }$sl(2|2)_{%
\bar{1}}$ [resp. $psl(2|2)_{\bar{1}}$] given by the module\textrm{\ }$2|\bar{%
2}\oplus \bar{2}|2$ describing fermions\textrm{. }

The Levi- decomposition of the $sl(2|2)$\ superalgebra reads in general as
follows%
\begin{equation}
\begin{tabular}{ccc}
$sl(2|2)$ & $\rightarrow $ & $\mathbf{N}_{+}^{(k,l)}\oplus \boldsymbol{l}%
_{\mu }^{(k,l)}\oplus \mathbf{N}_{-}^{(k,l)}$ \\
$\boldsymbol{l}_{\mu }^{(k,l)}$ & $=$ & $sl(k|l)\oplus sl(2-k|2-l)\oplus
gl(1)_{\mathbf{\mu }}$%
\end{tabular}
\label{gr}
\end{equation}%
where $\boldsymbol{l}_{\mathbf{\mu }}^{(k,l)}$ is the Levi sub-superalgebra
with\textrm{\ }$0\leq k\leq 2$ and $0\leq l\leq 2$ and where the nilpotent
subalgebra $\mathbf{N}_{\pm }^{(k,l)}$ have the dimensions%
\begin{equation}
\dim \mathbf{N}_{+}^{(k,l)}=\dim \mathbf{N}_{-}^{(k,l)}=\left( k+l\right)
\left( 4-k-l\right)  \label{6}
\end{equation}%
In graphical setting, the 3-grading (\ref{gr}) correspond to the cutting of
any of the $3$ nodes (bosonic or fermionic) of the four possible super
Dynkin diagram reported in \textrm{\autoref{appA}}. This is a special
property of $sl(m|n)$ where all nodes act like minuscule coweights; and as
such the $\mathbf{\mu }$ can be labeled like $\mathbf{\mu }_{k,l}$ \textrm{%
\cite{BE,BF,3AA}}.

\subsubsection{Distinguished sl(2\TEXTsymbol{\vert}2) spin chain}

\begin{figure}[tbph]
\begin{center}
\includegraphics[width=12cm]{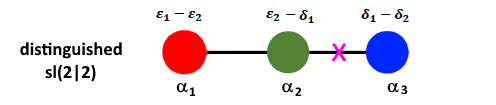}
\end{center}
\par
\vspace{-0.5cm}
\caption{Distinguished super Dynkin diagram of the Lie superalgebra $%
sl(2|2). $ The odd root is given by the Green node $\mathbf{\protect\alpha }%
_{2}$ with vanishing $\mathbf{\protect\alpha }_{2}.\mathbf{\protect\alpha }%
_{2}=0$. }
\label{ASL}
\end{figure}

\qquad For our concern, we will be interested into the distinguished super
Dynkin diagram of sl(2\TEXTsymbol{\vert}2) given by the \textbf{Figure}
\textbf{\ref{ASL}},
and integers $\left( k,l\right) =\left( 0,1\right) $. Here, the Levi-
decomposition of $sl(2|2)$ corresponds to the cutting of the (blue) node $%
\mathbf{\alpha }_{3};$ it is formulated like
\begin{equation}
\begin{tabular}{ccc}
$sl(2|2)$ & $:$ & $gl(1)_{\mathbf{\mu }_{3}}\oplus sl(2|1)\oplus \mathbf{N}%
_{+}\oplus \mathbf{N}_{-}$ \\
$15$ dim & $:$ & $1\quad +\quad 8\quad +\quad 3\quad +\quad \bar{3}$ \\
$12$\textbf{\ }roots & $:$ & $0\quad +\quad 6\quad +\quad 3\quad +\quad \bar{%
3}$%
\end{tabular}
\label{rg}
\end{equation}%
with $sl(2|1)_{\bar{0}}=s\left[ gl(2)\oplus gl(1)\right] $ and Levi-
superalgebra $l_{\mathbf{\mu }_{3}}^{sl_{2|2}}=gl(1)_{\mathbf{\mu }%
_{3}}\oplus sl(2|1).$ The interaction matrix $\mathbf{\alpha }_{i}.\mathbf{%
\alpha }_{j}$ of the simple roots making the diagram of the \textbf{Figure}
\textbf{\ref{Asl2} }reads as
\begin{equation}
\mathbf{\alpha }_{i}.\mathbf{\alpha }_{j}=\left(
\begin{array}{ccc}
2 & -1 & 0 \\
-1 & 0 & 1 \\
0 & 1 & -2%
\end{array}%
\right) \qquad ,\qquad \det \left( \mathbf{\alpha }_{i}.\mathbf{\alpha }%
_{j}\right) =0
\end{equation}%
Under this decomposition, we have the following features:

\begin{description}
\item[$\left( \mathbf{i}\right) $] the positive root system $\Phi
_{sl(2|2)}^{+}$ gets splitted like%
\begin{equation}
\begin{tabular}{|c||c|c|}
\hline
$\Phi _{sl(2|2)}^{+}$ & $\Phi _{sl(2|1)}^{+}$ & $\mathbf{N}_{+}$ \\ \hline
$\left.
\begin{array}{c}
\mathbf{\alpha }_{1},\quad \mathbf{\alpha }_{2},\quad \mathbf{\alpha }_{3},
\\
\mathbf{\mathbf{\alpha }_{2}+\mathbf{\alpha }_{1},\quad \alpha }_{2}+\mathbf{%
\alpha }_{3} \\
\mathbf{\mathbf{\alpha }}_{1}+\mathbf{\alpha }_{2}+\mathbf{\alpha }_{3}%
\end{array}%
\right. $ & $\left.
\begin{array}{c}
\mathbf{\alpha }_{1},\quad \mathbf{\alpha }_{2}, \\
\mathbf{\mathbf{\alpha }_{2}+\mathbf{\alpha }_{1}} \\
\mathbf{-}%
\end{array}%
\right. $ & $\left.
\begin{array}{c}
\mathbf{\alpha }_{3} \\
\mathbf{\alpha }_{2}+\mathbf{\alpha }_{3} \\
\mathbf{\mathbf{\alpha }}_{1}+\mathbf{\alpha }_{2}+\mathbf{\alpha }_{3}%
\end{array}%
\right. $ \\ \hline
\end{tabular}%
\end{equation}%
and similarly for negative roots. For those 3+3 roots characterising the
nilpotent subalgebras $\mathbf{N}_{\pm }$ in (\ref{rg}), their realisations
in terms of the canonical weight vectors $\left( \mathbf{\varepsilon }_{1},%
\mathbf{\varepsilon }_{2};\mathbf{\delta }_{1},\mathbf{\delta }_{2}\right) $
read as%
\begin{equation}
\mathbf{\alpha }_{3}=\mathbf{\delta }_{1}-\mathbf{\delta }_{2},\qquad
\mathbf{\alpha }_{2}+\mathbf{\alpha }_{3}=\mathbf{\varepsilon }_{2}-\mathbf{%
\delta }_{2},\qquad \mathbf{\mathbf{\alpha }}_{1}+\mathbf{\alpha }_{2}+%
\mathbf{\alpha }_{3}=\mathbf{\varepsilon }_{1}-\mathbf{\delta }_{2}
\end{equation}%
From this realisation, we deduce the 4$\times $4 matrix representation of
the step operators namely the even E$_{\pm \mathbf{\alpha }_{3}}$ and the
odd F$_{\pm \left( \mathbf{\alpha }_{2}+\mathbf{\alpha }_{3}\right) },$ F$%
_{\pm \left( \mathbf{\mathbf{\alpha }}_{1}+\mathbf{\alpha }_{2}+\mathbf{%
\alpha }_{3}\right) }$. For the example of the step operators associated
with the positive roots $+\mathbf{\alpha }$, we have the matrix
representation%
\begin{equation}
\begin{tabular}{lllll}
F$_{\mathbf{\mathbf{\alpha }}_{1}+\mathbf{\alpha }_{2}+\mathbf{\alpha }_{3}}$
& $=$ & $\left\vert \varepsilon _{1}\right\rangle \left\langle \delta
_{2}\right\vert $ & $\equiv $ & $\mathfrak{F}_{1}$ \\
F$_{\mathbf{\alpha }_{2}+\mathbf{\alpha }_{3}}$ & $=$ & $\left\vert
\varepsilon _{2}\right\rangle \left\langle \delta _{2}\right\vert $ & $%
\equiv $ & $\mathfrak{F}_{2}$ \\
E$_{+\mathbf{\alpha }_{3}}$ & $=$ & $\left\vert \delta _{1}\right\rangle
\left\langle \delta _{2}\right\vert $ & $\equiv $ & $\mathcal{E}$%
\end{tabular}%
\end{equation}%
from which we see that the odd F$_{\mathbf{\mathbf{\alpha }}_{1}+\mathbf{%
\alpha }_{2}+\mathbf{\alpha }_{3}}$ and F$_{\mathbf{\alpha }_{2}+\mathbf{%
\alpha }_{3}}$ form an isodoublet
\begin{equation*}
\mathfrak{F}_{i}=\left( \mathfrak{F}_{1},\mathfrak{F}_{2}\right)
\end{equation*}
while the even E$_{+\mathbf{\alpha }_{3}}$ is an isosinglet $\mathcal{E}.$
Similar features hold for the operators F$_{-\mathbf{\mathbf{\alpha }}_{1}-%
\mathbf{\alpha }_{2}-\mathbf{\alpha }_{3}},$ F$_{-\mathbf{\alpha }_{2}-%
\mathbf{\alpha }_{3}}$and E$_{-\mathbf{\alpha }_{3}}$ associated with
negative roots. From this description, we also learn that the nilpotent
representation $\mathbf{N}_{+}=\mathbf{3}$ reduces as
\begin{equation*}
\mathbf{3}=\mathbf{2}\oplus \mathbf{1}
\end{equation*}
describing a fermionic sl(2) isodoublet $\mathfrak{F}_{i}\sim \mathbf{2}$
and a bosonic isosinglet $\mathcal{E}\sim \mathbf{1}.$ The same thing is
valid for $\mathbf{N}_{-}=\mathbf{\bar{3}}$ decomposing as $\mathbf{\bar{2}}%
\oplus \mathbf{\bar{1}}$.

\item[$\left( \mathbf{ii}\right) $] the fundamental representation $\mathbf{%
2|2}$ of the super $sl(2|2)$ decomposes into irreps of $sl(2|1)$ as follows%
\begin{equation}
\mathbf{2{\LARGE |}2\qquad \rightarrow \qquad 2{\LARGE |}}\left( \mathbf{1}%
_{1/2}\oplus \mathbf{1}_{-1/2}\right)  \label{5}
\end{equation}%
\textrm{The subscripts }$\pm 1/2$\textrm{\ designate the charges under the }$%
gl(1)_{\mathbf{\mu }_{3}}$\textrm{\ corresponding to the cutted node }$%
\mathbf{\mu }_{3}$ dual to $\mathbf{\alpha }_{3}$\textrm{. These
representations }are labeled by the four basis states $\left\vert e_{\text{%
\textsc{a}}}\right\rangle $ for $sl(2|2),$ the triplet $\left\vert
e_{a}\right\rangle $ and the singlet $\left\vert e_{\lambda }\right\rangle $
for $sl(2|1).$ This convention notation is illustrated in the following table%
\begin{equation}
\begin{tabular}{c|c||c}
algebra & $sl(2|2)$ & $sl(2|1)\oplus gl(1)_{\mathbf{\mu }_{3}}$ \\
\hline\hline
repres & $\mathbf{2}|\mathbf{2}$ & $\mathbf{2{\LARGE |}1}_{1/2}\oplus
\mathbf{0{\LARGE |}1}_{-1/2}$ \\ \hline
basis states & $\left\vert e_{\text{\textsc{a}}}\right\rangle $ & $%
\left\vert e_{a}\right\rangle \oplus \left\vert e_{\lambda }\right\rangle $
\\ \hline\hline
\end{tabular}
\label{51}
\end{equation}%
with the gradings \
\begin{equation}
\left\vert e_{\text{\textsc{a}}}\right\rangle =\left\vert \varepsilon
_{i}\right\rangle \oplus \left\vert \delta _{j}\right\rangle \mathbf{\qquad
,\qquad }\left\vert e_{a}\right\rangle =\left\vert \varepsilon
_{i}\right\rangle \oplus \left\vert \delta _{1}\right\rangle \mathbf{\qquad
,\qquad }\left\vert e_{\lambda }\right\rangle =\left\vert \delta
_{2}\right\rangle
\end{equation}%
where $i,j=1,2.$ Using these quantum states ordered as $\left( \left\vert
\varepsilon _{1}\right\rangle ,\left\vert \varepsilon _{2}\right\rangle
,\left\vert \delta _{1}\right\rangle ,\left\vert \delta _{2}\right\rangle
\right) $, we can write down the action of the coweight $\mathbf{\mu }_{3}$;
it is given by help of the projectors $\Pi _{1}=\sum_{a=1}^{3}\left\vert
e_{a}\right\rangle \left\langle e_{a}\right\vert $ and $\Pi _{2}=\left\vert
\delta _{2}\right\rangle \left\langle \delta _{2}\right\vert $ as follows
\begin{eqnarray}
\mathbf{\mu }_{3} &=&\frac{1}{2}\left( \left\vert \varepsilon
_{1}\right\rangle \left\langle \varepsilon _{1}\right\vert +\left\vert
\varepsilon _{2}\right\rangle \left\langle \varepsilon _{2}\right\vert
+\left\vert \delta _{1}\right\rangle \left\langle \delta _{1}\right\vert
\right) -\frac{1}{2}\left\vert \delta _{2}\right\rangle \left\langle \delta
_{2}\right\vert  \notag \\
&=&\frac{1}{2}\Pi _{1}-\frac{1}{2}\Pi _{2}  \label{m}
\end{eqnarray}%
The $\mathbf{\mu }_{3}$ has a vanishing super trace $str(\mathbf{\mu }%
_{3})=0.$
\end{description}

\subsubsection{Calculating L-operator $\mathcal{L}_{sl_{2|2}}^{\mathbf{%
\protect\mu }_{3}}$}

\qquad Having the above tools at hand, we can now calculate the Lax operator
of the integrable distinguished sl(2\TEXTsymbol{\vert}2) superspin chain by
using the Costello- Gaiotto- Yagi (CGY) formula \textrm{\cite{Yama,BE,BF}}%
\begin{equation}
\mathcal{L}_{sl_{2|2}}^{\mathbf{\mu }_{3}}=e^{X}z^{\mathbf{\mu }_{3}}e^{Y}
\label{mxy}
\end{equation}%
The $X$ and $Y$ are nilpotent operators belonging to $\mathbf{N}_{+}$ and $%
\mathbf{N}_{-}$ appearing in (\ref{rg}); their explicit expressions are
needed for the computation of the Lax operator. The $z^{\mathbf{\mu }_{3}}$
is diagonal given by (\ref{m}) while the $X$ and $Y$ are determined by
resolving the Levi-constraints%
\begin{equation}
\left[ \mathbf{\mu }_{3},X\right] =X\qquad ,\qquad \left[ \mathbf{\mu }_{3},Y%
\right] =-Y\qquad ,\qquad \left[ X,Y\right] =\mathbb{C}\mathbf{\mu }_{3}
\label{LR}
\end{equation}%
These conditions are \textrm{solved} in terms of three pairs of harmonic
oscillators: an even $\left( b,c\right) $ ---a singlet---; and two odd $%
\left( \mathrm{\beta }^{i},\mathrm{\gamma }_{i}\right) $ ---a doublet--- as
follows
\begin{equation}
\begin{tabular}{lll}
$X$ & $=$ & $\mathrm{b}\left\vert \delta _{1}\right\rangle \left\langle
\delta _{2}\right\vert +\dsum\limits_{i=1}^{2}\mathrm{\beta }^{i}\left\vert
\varepsilon _{i}\right\rangle \left\langle \delta _{2}\right\vert $ \\
$Y$ & $=$ & $\mathrm{c}\left\vert \delta _{2}\right\rangle \left\langle
\delta _{1}\right\vert +\dsum\limits_{i=1}^{2}\mathrm{\gamma }_{j}\left\vert
\delta _{2}\right\rangle \left\langle \varepsilon ^{j}\right\vert $%
\end{tabular}
\label{xy}
\end{equation}%
which can be shortly expressed like
\begin{equation}
X=\left\vert \Psi _{X}\right\rangle \left\langle \delta _{2}\right\vert
\qquad ,\qquad Y=\left\vert \delta _{2}\right\rangle \left\langle \Psi
_{Y}\right\vert
\end{equation}%
with $\left\vert \Psi _{X}\right\rangle =\mathrm{\beta }^{i}\left\vert
\varepsilon _{i}\right\rangle +\mathrm{b}\left\vert \delta _{1}\right\rangle
$ and $\left\langle \Psi _{Y}\right\vert =\mathrm{\gamma }_{j}\left\langle
\varepsilon ^{j}\right\vert +\mathrm{c}\left\langle \delta _{1}\right\vert $%
. In this regard, recall that in the 4D Chern-Simons description of
integrable superspin chains, the pair $\left( b,c\right) $ is a bosonic
harmonic oscillator while the two pairs $\left( \mathrm{\beta }^{i},\mathrm{%
\gamma }_{i}\right) $\ are two fermioinc\ oscillators. Quantum mechanically,
these pairs obey the graded commutation relations
\begin{equation*}
\left[ b,c\right] =1\qquad ,\qquad \{\mathrm{\beta }^{i},\mathrm{\gamma }%
_{j}\}=\delta _{j}^{i}
\end{equation*}%
instead of the classical orthosymplectic Poisson brakets. For the explicit
calculations given below, we treat them classically, quantum corrections can
be implemented straightforwardly as done in \textrm{\cite{BE}}. Using the
above relationships, we can compute the expressions of monomials of X and Y
for the calculation of (\ref{mxy}). We have%
\begin{equation}
\begin{tabular}{lllllll}
$X^{2}$ & $=$ & $0$ & $\qquad ,\qquad $ & $X\Pi _{1}=\Pi _{1}Y$ & $=$ & $0$
\\
$Y^{2}$ & $=$ & $0$ & $\qquad ,\qquad $ & $\Pi _{2}X=Y\Pi _{2}$ & $=$ & $0$%
\end{tabular}%
\end{equation}%
and $XY=\left\vert \Psi _{X}\right\rangle \left\langle \Psi _{Y}\right\vert $
expanding like%
\begin{equation}
XY=\mathrm{\beta }^{i}\mathrm{\gamma }_{j}\left\vert \varepsilon
_{i}\right\rangle \left\langle \varepsilon ^{j}\right\vert +\mathrm{bc}%
\left\vert \delta _{1}\right\rangle \left\langle \delta _{1}\right\vert +%
\mathrm{b\gamma }_{j}\left\vert \delta _{1}\right\rangle \left\langle
\varepsilon ^{j}\right\vert +\mathrm{c\beta }^{i}\left\vert \varepsilon
_{i}\right\rangle \left\langle \delta _{1}\right\vert
\end{equation}%
Using these relations, the Lax operator $e^{X}z^{\mathbf{\mu }_{3}}e^{Y}$
expands like $\left( 1+X\right) z^{\mathbf{\mu }_{3}}\left( 1+Y\right) .$ By
substituting (\ref{m}-\ref{xy}), we get
\begin{equation}
\mathcal{L}_{sl_{2|2}}^{\mathbf{\mu }_{3}}=z^{\mathbf{\mu }_{3}}+Xz^{\mathbf{%
\mu }_{3}}+z^{\mathbf{\mu }_{3}}Y+Xz^{\mathbf{\mu }_{3}}Y
\end{equation}%
In terms of the matrix representation $\mathcal{L}_{sl_{2|2}}^{\mathbf{\mu }%
_{3}}=\sum \left\vert e_{\text{\textsc{a}}}\right\rangle L_{\text{\textsc{b}}%
}^{\text{\textsc{a}}}\left\langle e^{\text{\textsc{b}}}\right\vert $ in the
basis $\left( \left\vert \varepsilon _{1}\right\rangle ,\left\vert
\varepsilon _{2}\right\rangle ,\left\vert \delta _{1}\right\rangle
,\left\vert \delta _{2}\right\rangle \right) ,$ the L-operator reads in
block submatrices as follows
\begin{equation}
L_{\text{ \textsc{b}}}^{\text{\textsc{a}}}=\left(
\begin{array}{ccc}
z^{1/2}\delta _{j}^{i}+z^{-1/2}\mathrm{\beta }^{i}\mathrm{\gamma }_{j} &
z^{-1/2}\mathrm{c\beta }^{i} & z^{-1/2}\mathrm{\beta }^{i} \\
z^{-1/2}\mathrm{b\gamma }_{j} & z^{1/2}+z^{-1/2}\mathrm{bc} & z^{-1/2}%
\mathrm{b} \\
z^{-1/2}\mathrm{\gamma }_{j} & z^{-1/2}\mathrm{c} & z^{-1/2}%
\end{array}%
\right)  \label{20}
\end{equation}%
This L-matrix agrees with the solution obtained in the superspin chain
literature in \textrm{\cite{FRC} }using algebraic methods to solve the RLL
equation.

\subsection{Integrable osp(4\TEXTsymbol{\vert}2) superspin chain}

\begin{figure}[tbph]
\begin{center}
\includegraphics[width=12cm]{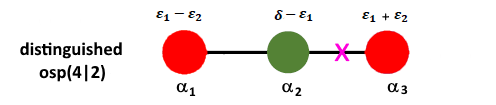}
\end{center}
\par
\vspace{-0.5cm}
\caption{Distinguished super Dynkin diagram of the Lie superalgebra $%
osp(4|2).$ The odd root is given by the Green node $\mathbf{\protect\alpha }%
_{2}$ with vanishing $\mathbf{\protect\alpha }_{2}.\mathbf{\protect\alpha }%
_{2}=0$.}
\label{osp4}
\end{figure}

\qquad In this subsection, we calculate the Lax operator $\mathcal{L}_{osp}^{%
\mathbf{\mu }_{3}}$ of distinguished integrable osp(4\TEXTsymbol{\vert}2)
superchain by using the 4D Chern-Simons formalism; for an alternative
algebraic method, see \textrm{\cite{Fra}}. Here, the L-operator is given by
the CGY formula $e^{X}z^{\mathbf{\mu }_{3}}e^{Y}$ with ($z^{\mathbf{\mu }%
_{3}},X,Y$) valued in osp(4\TEXTsymbol{\vert}2). Recall that the
distinguished super Dynkin diagram of the orthosymplectic osp(4\TEXTsymbol{%
\vert}2) superalgebra is given by the \textbf{Figure} \textbf{\ref{osp4}}.
This super diagram can be thought of in terms of the intersection matrix of
the simple roots $\left( \mathbf{\alpha }_{1},\mathbf{\alpha }_{2},\mathbf{%
\alpha }_{3}\right) $ labeling its three nodes with odd $\mathbf{\alpha }%
_{2} $. This intersection matrix reads as follows%
\begin{equation}
\mathbf{\alpha }_{i}.\mathbf{\alpha }_{j}=\left(
\begin{array}{ccc}
2 & -1 & 0 \\
-1 & 0 & -1 \\
0 & -1 & 2%
\end{array}%
\right) \qquad ,\qquad \det \left( \mathbf{\alpha }_{i}.\mathbf{\alpha }%
_{j}\right) =-4
\end{equation}%
The super diagram \textbf{\ref{osp4}} has an outer automorphism $\mathcal{Z}%
_{2}$ permuting the $\mathbf{\alpha }_{1}$ and the $\mathbf{\alpha }_{3}$
simple roots (red nodes in the diagram) with coweights $\mathbf{\mu }_{1}$\
and $\mathbf{\mu }_{3}$ which are needed for the calculation of the
associated Lax operators $\mathcal{L}_{osp}^{\mathbf{\mu }}=e^{X}z^{\mathbf{%
\mu }}e^{Y}.$

\subsubsection{Distinguished orthosymplectic spin chain osp(4\TEXTsymbol{%
\vert}2)}

\qquad The Levi-decomposition (3-grading) for the distinguished
orthosymplectic osp(4\TEXTsymbol{\vert}2) chain is obtained by the \emph{%
cutting} the node $\mathbf{\alpha }_{3}$ in the super Dynkin diagram of the
\textbf{Figure} \textbf{\ref{osp4}}. This decomposition is given by%
\begin{equation}
\begin{tabular}{ccc}
$osp(4|2)$ & $:$ & $gl(1)_{\mathbf{\mu }_{3}}\oplus sl(2|1)\oplus \mathbf{N}%
_{+}^{osp}\oplus \mathbf{N}_{-}^{osp}$ \\
$17$ dim & $:$ & $1\quad \ \ +\quad 8\quad +\quad 4\quad +\quad \bar{4}$ \\
$14$ roots & $:$ & $0\quad $\ $\ \ +\quad $\ $6\quad $\ $+\quad $\ $4\quad
+\quad \bar{4}$%
\end{tabular}
\label{spsl}
\end{equation}%
\begin{equation*}
\end{equation*}%
with $sl(2|1)\simeq osp(2|2)$ and $gl(1)_{\mathbf{\mu }_{3}}\simeq so(2,%
\mathbb{C})$ and where$\ \boldsymbol{l}_{\mathbf{\mu }_{3}}^{osp}=gl(1)_{%
\mathbf{\mu }_{3}}\oplus sl(2|1)$ is the Levi-subalgebra. Comparing the
3-grading in (\ref{spsl}) with the 3-grading (\ref{rg}), we learn that we
have the isomorphism $\boldsymbol{l}_{\mathbf{\mu }_{3}}^{sl_{2|2}}\simeq
\boldsymbol{l}_{\mathbf{\mu }_{3}}^{osp}$ while the nilpotent subalgebras $%
\mathbf{N}_{\pm }^{osp}$ and $\mathbf{N}_{\pm }^{sl_{2|2}}$ are related like
\begin{equation}
\mathbf{N}_{\pm }^{osp}\simeq \mathbf{N}_{\pm }^{sl_{2|2}}\oplus \mathbf{1}%
_{\pm }\mathbf{\qquad \Leftrightarrow \qquad 4}_{\pm }^{osp}\simeq \mathbf{3}%
_{\pm }^{sl_{2|2}}\oplus \mathbf{1}_{\pm }
\end{equation}%
showing that $\mathbf{N}_{\pm }^{osp}$ has one dimension bigger with respect
to $\mathbf{N}_{\pm }^{sl_{2|2}}.$

Below, we use the 6-dimensional vector representation of $osp(4|2)$ labeled
by six canonical weight vectors \{\text{\textbf{e}}$_{1},...,$\textbf{e}$%
_{6} $\} thought of in terms of the $\pm \epsilon _{i}$ and $\pm \delta $
weight vectors ($\delta ^{2}=\epsilon _{i}^{2}=-1$) as follows
\begin{equation}
\mathbf{e}_{1}=\epsilon _{1},\quad \mathbf{e}_{2}=\epsilon _{2},\quad
\mathbf{e}_{3}=\delta ,\quad \mathbf{e}_{4}=\bar{\delta},\quad \mathbf{e}%
_{5}=\bar{\epsilon}_{2},\quad \mathbf{e}_{6}=\bar{\epsilon}_{1}  \label{qs}
\end{equation}%
This set of quantum states can be combined into two subset like $\left\vert
e_{a}\right\rangle \oplus \left\vert \bar{e}^{\bar{a}}\right\rangle $ with $%
\left\vert e_{a}\right\rangle =(\left\vert \epsilon _{1}\right\rangle
,\left\vert \epsilon _{2}\right\rangle ,\left\vert \delta \right\rangle )$
and $\left\vert \bar{e}^{\bar{a}}\right\rangle =(\left\vert \bar{\delta}%
\right\rangle ,\left\vert \bar{\epsilon}^{2}\right\rangle ,\left\vert \bar{%
\epsilon}^{1}\right\rangle ).$ For the bras, we have $\left\langle
e^{a}\right\vert =(\left\langle \epsilon ^{1}\right\vert ,\left\langle
\epsilon ^{2}\right\vert ,\left\langle \delta \right\vert )$ and $\left\vert
\bar{e}_{\bar{a}}\right\rangle =(\left\langle \bar{\delta}\right\vert
,\left\langle \bar{\epsilon}_{2}\right\vert ,\left\langle \bar{\epsilon}%
_{1}\right\vert )$. Under the decomposition (\ref{spsl}), we have the
following interesting features:

\begin{description}
\item[$\left( \mathbf{a}\right) $] the positive root system $\Phi
_{osp(4|2)}^{+}$ gets splitted into $\Phi _{sl(2|1)}^{+}\oplus \mathbf{N}%
_{+}^{osp}$ like%
\begin{equation}
\begin{tabular}{|c||c|c|}
\hline
$\Phi _{osp(4|2)}^{+}$ & $\Phi _{sl(2|1)}^{+}$ & $\mathbf{N}_{+}^{osp}$ \\
\hline
$\left.
\begin{array}{c}
\mathbf{\alpha }_{1},\quad \mathbf{\alpha }_{2},\quad \mathbf{\alpha }_{3},
\\
\mathbf{\mathbf{\alpha }_{2}+\mathbf{\alpha }_{1},\quad \mathbf{\alpha }}%
_{1}+\mathbf{\alpha }_{2}+\mathbf{\alpha }_{3} \\
\mathbf{\alpha }_{2}+\mathbf{\alpha }_{3},\quad \mathbf{\mathbf{\alpha }}%
_{1}+2\mathbf{\alpha }_{2}+\mathbf{\alpha }_{3}%
\end{array}%
\right. $ & $\left.
\begin{array}{c}
\mathbf{\alpha }_{1},\quad \mathbf{\alpha }_{2}, \\
\mathbf{\mathbf{\alpha }_{2}+\mathbf{\alpha }_{1}} \\
\mathbf{-}%
\end{array}%
\right. $ & $\left.
\begin{array}{c}
\mathbf{\alpha }_{3},\quad \mathbf{\mathbf{\alpha }}_{1}+2\mathbf{\alpha }%
_{2}+\mathbf{\alpha }_{3} \\
\mathbf{\alpha }_{2}+\mathbf{\alpha }_{3},\mathbf{\quad } \\
\mathbf{\mathbf{\alpha }}_{1}+\mathbf{\alpha }_{2}+\mathbf{\alpha }_{3}%
\end{array}%
\right. $ \\ \hline
\end{tabular}%
\end{equation}%
\begin{equation*}
\end{equation*}%
with the simple roots realised as $\mathbf{\alpha }_{1}=\mathbf{\epsilon }%
_{1}-\mathbf{\epsilon }_{2},\quad \mathbf{\alpha }_{2}=\mathbf{\delta
-\epsilon }_{1}$ and $\mathbf{\alpha }_{3}=\mathbf{\epsilon }_{1}+\mathbf{%
\epsilon }_{2}$. For the roots characterising the nilpotent sector $\mathbf{N%
}_{+}^{osp}$, we have the roots
\begin{equation}
\begin{tabular}{lll}
$\mathbf{\alpha }_{3}$ & $=$ & $\mathbf{\epsilon }_{1}+\mathbf{\epsilon }%
_{2} $ \\
$\mathbf{\alpha }_{3}+\mathbf{\alpha }_{2}$ & $=$ & $\mathbf{\delta }+%
\mathbf{\epsilon }_{2}$ \\
$\mathbf{\alpha }_{3}+\mathbf{\alpha }_{2}+\mathbf{\mathbf{\alpha }}_{1}$ & $%
=$ & $\mathbf{\delta +\epsilon }_{1}$ \\
$\mathbf{\alpha }_{3}+2\mathbf{\alpha }_{2}+\mathbf{\mathbf{\alpha }}_{1}$ &
$=$ & $2\mathbf{\delta }$%
\end{tabular}
\label{213}
\end{equation}%
from which we learn that the $\mathbf{N}_{+}=\mathbf{4}$ reduces like
\begin{equation*}
\mathbf{4}=\mathbf{2}\oplus \mathbf{1}\oplus \mathbf{1}
\end{equation*}
where the doublet $\mathbf{2}$ describes the two odd roots $\mathbf{\delta }+%
\mathbf{\epsilon }_{i}$ and the two singlets $\mathbf{1}\oplus \mathbf{1}$
corresponding to the even $\mathbf{\epsilon }_{1}+\mathbf{\epsilon }_{2}$
and the even long root $2\mathbf{\delta }.$

\item[$\left( \mathbf{b}\right) $] the fundamental representation $\mathbf{%
4|2}$ of the orthosymplectic Lie algebra $osp(4|2)$ decomposes as the sum of
two irreps of $sl(2|1)$ as follows%
\begin{equation}
\left( \mathbf{4|2}\right) =\left( \mathbf{2|1}\right) _{+\frac{1}{2}}\oplus
\left( \mathbf{\bar{2}|\bar{1}}\right) _{-\frac{1}{2}}
\end{equation}%
\textrm{The subscripts }$\pm 1/2$\textrm{\ designate the charges under the
Cartan subalgebra }$gl(1)_{\mathbf{\mu }_{3}}$\textrm{. These
representations }are described by the basis states $\left\vert
e_{a}\right\rangle \oplus \left\vert \bar{e}^{\bar{a}}\right\rangle $ like%
\begin{equation}
\begin{tabular}{c|c||c}
algebra & osp$(4|2)$ & $sl(2|1)\oplus gl(1)_{\mathbf{\mu }_{3}}$ \\
\hline\hline
repres & $\mathbf{4}|\mathbf{2}$ & $\mathbf{2{\LARGE |}1}_{1/2}\quad \oplus
\quad \mathbf{\bar{2}{\LARGE |}\bar{1}}_{-1/2}$ \\ \hline
basis states & $\left\vert e_{\text{\textsc{a}}}\right\rangle $ & $%
\left\vert e_{a}\right\rangle \quad \oplus \quad \left\vert \bar{e}^{\bar{a}%
}\right\rangle $ \ \  \\ \hline\hline
\end{tabular}
\label{15}
\end{equation}
\end{description}

\subsubsection{ Computing the orthosymplectic $\mathcal{L}_{osp(4|2)}^{%
\mathbf{\protect\mu }_{3}}$}

\qquad Given the above quantum states ordered as in (\ref{qs}), we can:

$\left( \mathbf{i}\right) $ write down the action of the coweight $\mathbf{%
\mu }_{3}^{{\small osp}}$ (dual to $\mathbf{\alpha }_{3}^{{\small osp}}$);
it is given by using projectors $\Pi $ and $\bar{\Pi}$ as follows%
\begin{equation}
\mathbf{\mu }_{3}^{{\small osp}}=\frac{1}{2}\Pi -\frac{1}{2}\bar{\Pi}
\label{mu3}
\end{equation}%
with%
\begin{equation}
\begin{tabular}{lll}
$\Pi $ & $=$ & $\left\vert \epsilon _{1}\right\rangle \left\langle \epsilon
^{1}\right\vert +\left\vert \epsilon _{2}\right\rangle \left\langle \epsilon
^{2}\right\vert +\left\vert \delta _{1}\right\rangle \left\langle \delta
^{1}\right\vert $ \\
$\bar{\Pi}$ & $=$ & $\left\vert \bar{\delta}^{1}\right\rangle \left\langle
\bar{\delta}_{1}\right\vert +\left\vert \bar{\epsilon}^{2}\right\rangle
\left\langle \bar{\epsilon}_{2}\right\vert +\left\vert \bar{\epsilon}%
^{1}\right\rangle \left\langle \bar{\epsilon}_{1}\right\vert $%
\end{tabular}%
\end{equation}%
It has a vanishing super trace $str(\mathbf{\mu }_{3}^{{\small osp}})=0.$

$\left( \mathbf{ii}\right) $ calculate the super L-operator $\mathcal{L}%
_{osp}^{\mathbf{\mu }_{3}}$ (or equivalently $\mathcal{L}_{osp}^{\mathbf{\mu
}_{1}}$); it is given by a $6\times 6$ matrix having the sub-blocks form
\begin{equation}
\left( L_{{\small osp}}\right) _{\text{ \textsc{b}}}^{\text{\textsc{a}}%
}=\left(
\begin{array}{cc}
L_{b}^{a} & L_{b}^{\bar{a}} \\
L_{\bar{b}}^{a} & L_{\bar{b}}^{\bar{a}}%
\end{array}%
\right)
\end{equation}%
with basis vectors $\left\vert e_{a}\right\rangle $ for the representation $%
\left( \mathbf{2|1}\right) _{+1/2}$ ($\left\vert e_{a}\right\rangle
=(\left\vert \epsilon _{1}\right\rangle ,\left\vert \epsilon
_{2}\right\rangle ,\left\vert \delta \right\rangle $); and $\left\vert \bar{e%
}^{\bar{a}}\right\rangle $ for the representation $\left( \mathbf{\bar{2}|%
\bar{1}}\right) _{-1/2}$ with $\left\vert \bar{e}^{\bar{a}}\right\rangle
=(\left\vert \bar{\delta}\right\rangle ,\left\vert \bar{\epsilon}%
^{2}\right\rangle ,\left\vert \bar{\epsilon}^{1}\right\rangle ).$

Using eqs(\ref{213}-\ref{15}), we can determine the super Lax operator $%
\mathcal{L}_{osp}^{\mathbf{\mu }_{3}}$ by help of the representation $e^{X_{%
{\small osp}}}z^{\mathbf{\mu }_{3}}e^{Y_{{\small osp}}}$ with $\mathbf{\mu }%
_{3}^{{\small osp}}$ as in (\ref{mu3}). For the nilpotent $X_{{\small osp}}$
and $Y_{{\small osp}}$, they expand like
\begin{equation}
\begin{tabular}{lll}
$X_{{\small osp}}$ & $=$ & $\mathrm{\beta }^{i}\mathcal{X}_{i}+\mathrm{b}%
X_{0}+\mathrm{b}^{\prime }X_{0}^{\prime }$ \\
$Y_{{\small osp}}$ & $=$ & $\mathrm{\gamma }_{i}\mathcal{Y}^{i}+\mathrm{c}%
Y_{0}+\mathrm{c}^{\prime }Y_{0}^{\prime }$%
\end{tabular}
\label{sp}
\end{equation}%
and obey the Levi- constraint relations (\ref{LR}) namely $\left[ \mathbf{%
\mu }_{3},X\right] =X$ and $\left[ \mathbf{\mu }_{3},Y\right] =-Y$. In these
relations the pairs (\textrm{b,c}) and (\textrm{b',c'}) are bosonic
oscillators and the pairs ($\mathrm{\beta }^{i},\mathrm{\gamma }_{i}$) are
fermionic ones. The above mentioned Levi- conditions can be solved as follows%
\begin{equation}
\begin{tabular}{lll}
$\mathcal{X}_{i}$ & $=$ & $\left\vert \epsilon _{i}\right\rangle
\left\langle \bar{\delta}\right\vert -\left\vert \delta \right\rangle
\left\langle \bar{\epsilon}_{i}\right\vert $ \\
$X_{0}$ & $=$ & $\left\vert \delta \right\rangle \left\langle \bar{\delta}%
\right\vert $ \\
$X_{0}^{\prime }$ & $=$ & $\left\vert \epsilon _{1}\right\rangle
\left\langle \bar{\epsilon}_{2}\right\vert -\left\vert \epsilon
_{2}\right\rangle \left\langle \bar{\epsilon}_{1}\right\vert $%
\end{tabular}%
\qquad ,\qquad
\begin{tabular}{lll}
$\mathcal{Y}^{i}$ & $=$ & $\left\vert \bar{\delta}\right\rangle \left\langle
\epsilon ^{i}\right\vert -\left\vert \bar{\epsilon}^{i}\right\rangle
\left\langle \delta \right\vert $ \\
$Y_{0}$ & $=$ & $\left\vert \bar{\delta}\right\rangle \left\langle \delta
\right\vert $ \\
$Y_{0}^{\prime }$ & $=$ & $\left\vert \bar{\epsilon}^{2}\right\rangle
\left\langle \epsilon ^{1}\right\vert -\left\vert \bar{\epsilon}%
^{1}\right\rangle \left\langle \epsilon ^{2}\right\vert $%
\end{tabular}
\label{ps}
\end{equation}%
where $\left\vert \epsilon _{1}\right\rangle \left\langle \bar{\epsilon}%
_{2}\right\vert -\left\vert \epsilon _{2}\right\rangle \left\langle \bar{%
\epsilon}_{1}\right\vert $ can be expressed shortly as $\varepsilon
^{kl}|\epsilon _{k}><\bar{\epsilon}_{l}|$ in terms of the usual
antisymmetric tensor $\varepsilon ^{kl}$. We also have $|\bar{\epsilon}%
^{l}><\epsilon ^{k}|\varepsilon _{kl}$ for the singlet $\left\vert \bar{%
\epsilon}^{2}\right\rangle \left\langle \epsilon ^{1}\right\vert -\left\vert
\bar{\epsilon}^{1}\right\rangle \left\langle \epsilon ^{2}\right\vert .$
Substituting these expressions into (\ref{sp}), we can present $X_{{\small %
osp}}$ and $Y_{{\small osp}}$ like%
\begin{equation}
\begin{tabular}{lll}
$X_{{\small osp}}$ & $=$ & $\mathrm{\beta }^{k}\left( \left\vert \epsilon
_{k}\right\rangle \left\langle \bar{\delta}\right\vert -\left\vert \delta
\right\rangle \left\langle \bar{\epsilon}_{k}\right\vert \right) +\mathrm{b}%
\left\vert \delta \right\rangle \left\langle \bar{\delta}\right\vert +%
\mathrm{b}^{\prime }\varepsilon ^{kl}\left\vert \epsilon _{k}\right\rangle
\left\langle \bar{\epsilon}_{l}\right\vert $ \\
$Y_{{\small osp}}$ & $=$ & $\mathrm{\gamma }_{n}\left( \left\vert \bar{\delta%
}\right\rangle \left\langle \epsilon ^{n}\right\vert -\left\vert \bar{%
\epsilon}^{n}\right\rangle \left\langle \delta \right\vert \right) +\mathrm{c%
}\left\vert \bar{\delta}\right\rangle \left\langle \delta \right\vert +%
\mathrm{c}^{\prime }\left\vert \bar{\epsilon}^{m}\right\rangle \left\langle
\epsilon ^{n}\right\vert \varepsilon _{nm}$%
\end{tabular}
\label{xosp}
\end{equation}%
From these relations, we learn the properties $X_{{\small osp}}^{2}=0=Y_{%
{\small osp}}^{2}$ indicating that $e^{X_{{\small osp}}}=1+X_{{\small osp}}$
and $e^{Y_{{\small osp}}}=1+Y_{{\small osp}}$. We also learn the useful
features $X_{{\small osp}}\Pi =0=\Pi Y_{{\small osp}}$ as well as $\Pi X_{%
{\small osp}}=X_{{\small osp}}$ and $\Pi Y_{{\small osp}}=Y_{{\small osp}}$.
For convenience, we express the above $X_{{\small osp}}$ and $Y_{{\small osp}%
}$ as follows%
\begin{equation}
\begin{tabular}{lll}
$X_{{\small osp}}$ & $=$ & $\left\vert \Phi _{X}\right\rangle \left\langle
\bar{\delta}\right\vert -\left\vert \Psi _{X}^{k}\right\rangle \left\langle
\bar{\epsilon}_{k}\right\vert $ \\
$Y_{{\small osp}}$ & $=$ & $\left\vert \bar{\delta}\right\rangle
\left\langle \Phi _{Y}\right\vert -\left\vert \bar{\epsilon}%
^{l}\right\rangle \left\langle \Psi _{Yl}\right\vert $%
\end{tabular}%
\end{equation}%
with
\begin{equation}
\begin{tabular}{lll}
$\left\vert \Psi _{X}^{k}\right\rangle $ & $=$ & $\mathrm{b}^{\prime
}\varepsilon ^{kl}\left\vert \epsilon _{l}\right\rangle +\mathrm{\beta }%
^{k}\left\vert \delta \right\rangle $ \\
$\left\langle \Psi _{Yn}\right\vert $ & $=$ & $\mathrm{c}^{\prime
}\left\langle \epsilon ^{m}\right\vert \varepsilon _{mn}+\mathrm{\gamma }%
_{n}\left\langle \delta \right\vert $%
\end{tabular}%
\qquad ,\qquad
\begin{tabular}{lll}
$\left\vert \Phi _{X}\right\rangle $ & $=$ & $\mathrm{\beta }^{k}\left\vert
\epsilon _{k}\right\rangle +\mathrm{b}\left\vert \delta \right\rangle $ \\
$\left\langle \Phi _{Y}\right\vert $ & $=$ & $\mathrm{\gamma }%
_{n}\left\langle \epsilon ^{n}\right\vert +\mathrm{c}\left\langle \delta
\right\vert $%
\end{tabular}%
\end{equation}%
leading to%
\begin{equation}
\begin{tabular}{lll}
$X_{{\small osp}}Y_{{\small osp}}$ & $=$ & $\left\vert \Phi
_{X}\right\rangle \left\langle \Phi _{Y}\right\vert +\left\vert \Psi
_{X}^{k}\right\rangle \left\langle \Psi _{Yk}\right\vert $ \\
$\left\vert \Phi _{X}\right\rangle \left\langle \Phi _{Y}\right\vert $ & $=$
& $\mathrm{\beta }^{i}\mathrm{\gamma }_{j}\left\vert \epsilon
_{i}\right\rangle \left\langle \epsilon ^{j}\right\vert +\mathrm{c\beta }%
^{i}\left\vert \epsilon _{i}\right\rangle \left\langle \delta \right\vert +%
\mathrm{b\gamma }_{j}\left\vert \delta \right\rangle \left\langle \epsilon
^{j}\right\vert +\mathrm{bc}\left\vert \delta \right\rangle \left\langle
\delta \right\vert $ \\
$\left\vert \Psi _{X}^{k}\right\rangle \left\langle \Psi _{Yk}\right\vert $
& $=$ & $\mathrm{b}^{\prime }\mathrm{c}^{\prime }\left\vert \epsilon
_{l}\right\rangle \left\langle \epsilon ^{l}\right\vert +\mathrm{\beta }^{k}%
\mathrm{c}^{\prime }\varepsilon _{mk}\left\vert \delta \right\rangle
\left\langle \epsilon ^{m}\right\vert +\mathrm{b}^{\prime }\mathrm{\gamma }%
_{k}\varepsilon ^{kl}\left\vert \epsilon _{l}\right\rangle \left\langle
\delta \right\vert +\mathrm{\beta }^{k}\mathrm{\gamma }_{k}\left\vert \delta
\right\rangle \left\langle \delta \right\vert $%
\end{tabular}%
\end{equation}%
\begin{equation*}
\end{equation*}%
Using these relations, the Lax operator $e^{X}z^{\mathbf{\mu }_{3}}e^{Y}$
expands like $\left( 1+X\right) z^{\mathbf{\mu }_{3}}\left( 1+Y\right) .$ By
substituting (\ref{m}-\ref{xy}), we get $\mathcal{L}_{osp}^{\mathbf{\mu }%
_{3}}=z^{\mathbf{\mu }_{3}}+z^{-1/2}\left( X+Y+XY\right) .$ The matrix
elements $L_{\text{\textsc{b}}}^{\text{\textsc{a}}}$ in the basis $\left(
\left\vert e_{a}\right\rangle ,\left\vert \bar{e}^{\bar{a}}\right\rangle
\right) $ reads in block matrices as follows%
\begin{equation}
\mathcal{L}_{osp}^{\mathbf{\mu }_{3}}=\left(
\begin{array}{cccc}
z^{1/2}\delta _{j}^{i}+z^{-1/2}(\mathrm{b}^{\prime }\mathrm{c}^{\prime
}\delta _{j}^{i}+\mathrm{\beta }^{i}\mathrm{\gamma }_{j}) & z^{-1/2}\left(
\mathrm{c\beta }^{i}+\mathrm{b}^{\prime }\mathrm{\gamma }_{k}\varepsilon
^{ki}\right) & z^{-1/2}\mathrm{\beta }^{i} & z^{-1/2}\mathrm{b}^{\prime
}\varepsilon ^{ij} \\
z^{-1/2}(\mathrm{b\gamma }_{j}+\mathrm{c}^{\prime }\varepsilon _{jl}\mathrm{%
\beta }^{l}) & z^{1/2}+z^{-1/2}\left( \mathrm{bc+\beta }^{l}\mathrm{\gamma }%
_{l}\right) & z^{-1/2}\mathrm{b} & -z^{-1/2}\mathrm{\beta }^{j} \\
z^{-1/2}\mathrm{\gamma }_{j} & z^{-1/2}\mathrm{c} & z^{-1/2} & 0 \\
z^{-1/2}\mathrm{c}^{\prime }\varepsilon _{ij} & -z^{-1/2}\mathrm{\gamma }_{i}
& 0 & z^{-1/2}\delta _{i}^{j}%
\end{array}%
\right)  \label{osp}
\end{equation}%
\begin{equation*}
\end{equation*}%
This L-matrix completes the series of the expressions of the L-operators for
osp(2m\TEXTsymbol{\vert}2n) given in \textrm{\cite{3AA}} for $m\geq 3$ and $%
n\geq 1.$ Notice that by setting $\mathrm{b}^{\prime }=\mathrm{c}^{\prime
}=0 $ in (\ref{osp}), it reduces down to $\left. \mathcal{L}_{osp}^{\mathbf{%
\mu }_{3}}\right\vert _{\mathrm{b}^{\prime }=\mathrm{c}^{\prime }=0}$
reading as follows%
\begin{equation}
\left(
\begin{array}{cccc}
z^{1/2}\delta _{j}^{i}+z^{-1/2}\mathrm{\beta }^{i}\mathrm{\gamma }_{j} &
z^{-1/2}\mathrm{c\beta }^{i} & z^{-1/2}\mathrm{\beta }^{i} & 0 \\
z^{-1/2}\mathrm{b\gamma }_{j} & z^{1/2}+z^{-1/2}\left( \mathrm{bc+\beta }^{l}%
\mathrm{\gamma }_{l}\right) & z^{-1/2}\mathrm{b} & -z^{-1/2}\mathrm{\beta }%
^{j} \\
z^{-1/2}\mathrm{\gamma }_{j} & z^{-1/2}\mathrm{c} & z^{-1/2} & 0 \\
0 & -z^{-1/2}\mathrm{\gamma }_{i} & 0 & z^{-1/2}\delta _{i}^{j}%
\end{array}%
\right)
\end{equation}%
containing in turns $\mathcal{L}_{sl(2|2)}^{\mathbf{\mu }_{3}}$ as a 4$%
\times $4 submatrix as follows%
\begin{equation}
\left(
\begin{array}{cccc}
&  &  & 0 \\
& \left[ \mathcal{L}_{sl(2|2)}^{\mathbf{\mu }_{3}}\right] _{4\times 4} &  &
-z^{-1/2}\mathrm{\beta }^{j} \\
&  &  & 0 \\
0 & -z^{-1/2}\mathrm{\gamma }_{i} & 0 & z^{-1/2}\delta _{i}^{j}%
\end{array}%
\right)
\end{equation}

\subsection{Lax operator of D(2,1;$\protect\zeta $) superspin chain}

\qquad Here, we use the obtained L-operators $\mathcal{L}_{sl(2|2)}^{\mathbf{%
\mu }_{3}}$ and $\mathcal{L}_{osp}^{\mathbf{\mu }_{3}}$ given by eqs(\ref{20}%
) and (\ref{osp}) to determine $\mathcal{L}_{\mathfrak{d}(2,1;\zeta )}^{%
\mathbf{\mu }_{3}}$. We think about (\ref{20}) as the value $\mathcal{L}_{%
\mathfrak{d}(2,1;\zeta )}^{\mathbf{\mu }_{3}}|_{\zeta =-1}$ and about (\ref%
{osp}) like the value of $\mathcal{L}_{\mathfrak{d}(2,1;\zeta )}^{\mathbf{%
\mu }_{3}}|_{\zeta =+1}$ as indicated by (\ref{spt}).

\subsubsection{$\mathcal{L}_{\mathfrak{d}(2,1;\protect\zeta )}^{\mathbf{%
\protect\mu }_{3}}$ as deformation of $\mathcal{L}_{osp}^{\mathbf{\protect%
\mu }_{3}}$}

\qquad Despite our knowledge of the values $\mathcal{L}_{\mathfrak{d}%
(2,1;\zeta )}^{\mathbf{\mu }_{3}}|_{\zeta =\pm 1}$ given by (\ref{20}) and (%
\ref{osp}), this is not enough to construct $\mathcal{L}_{\mathfrak{d}%
(2,1;\zeta )}^{\mathbf{\mu }_{3}}$ for generic values of $\zeta .$ To get
around this difficulty, we use the fact that $\mathfrak{d}(2,1;\zeta )$ can
be seen as a deformation of osp(4\TEXTsymbol{\vert}2)=$\mathfrak{d}(2,1;1)$.
In this vision, we propose to construct $\mathcal{L}_{\mathfrak{d}(2,1;\zeta
)}^{\mathbf{\mu }_{3}}$ starting from the orthosymplectic solutions (\ref%
{xosp}) that led to $\mathcal{L}_{osp}^{\mathbf{\mu }_{3}}$ and consider
their deformations towards $\mathcal{L}_{\mathfrak{d}(2,1;\zeta )}^{\mathbf{%
\mu }_{3}}$ with the conditions (\ref{spt}); i.e:%
\begin{equation}
\mathcal{L}_{\mathfrak{d}(2,1;\zeta )}^{\mathbf{\mu }_{3}}=\mathcal{L}%
_{osp}^{\mathbf{\mu }_{3}}+\delta \mathcal{L}\left( \zeta \right)
\end{equation}%
So, to determine the expression of the L-operator of the integrable
distinguishable superspin chain $\mathfrak{d}(2,1;\zeta )$, we start from
the solution (\ref{xosp}) and use the fact that $\mathfrak{d}(2,1;\zeta )$
and osp(4\TEXTsymbol{\vert}2) have the same dimensions ($\dim =17$) and the
same rank ($r=3$) to think about $\mathbf{\mu }_{3}^{\mathfrak{d}{\small %
(2,1;\zeta )}}$ like%
\begin{equation}
\mathbf{\mu }_{3}^{\mathfrak{d}{\small (2,1;\zeta )}}=\frac{1}{2}\Pi -\frac{1%
}{2}\bar{\Pi}  \label{d1}
\end{equation}%
and the nilpotent operators $X_{\mathfrak{d}{\small (2,1;\zeta )}}$ and $Y_{%
\mathfrak{d}{\small (2,1;\zeta )}}$ as follows%
\begin{eqnarray}
X_{\mathfrak{d}{\small (2,1;\zeta )}} &=&\mathrm{\beta }^{k}\left( \mathrm{%
\kappa }_{1}\left\vert \epsilon _{k}\right\rangle \left\langle \bar{\delta}%
\right\vert +\mathrm{\kappa }_{2}\left\vert \delta \right\rangle
\left\langle \bar{\epsilon}_{k}\right\vert \right) +\mathrm{b}\left\vert
\delta \right\rangle \left\langle \bar{\delta}\right\vert +\mathrm{b}%
^{\prime }\mathrm{\kappa }_{0}\varepsilon ^{kl}\left\vert \epsilon
_{k}\right\rangle \left\langle \bar{\epsilon}_{l}\right\vert  \notag \\
Y_{\mathfrak{d}{\small (2,1;\zeta )}} &=&\mathrm{\gamma }_{n}\left( \mathrm{%
\kappa }_{1}\left\vert \bar{\delta}\right\rangle \left\langle \epsilon
^{n}\right\vert +\mathrm{\kappa }_{1}\left\vert \bar{\epsilon}%
^{n}\right\rangle \left\langle \delta \right\vert \right) +\mathrm{c}%
\left\vert \bar{\delta}\right\rangle \left\langle \delta \right\vert +%
\mathrm{c}^{\prime }\mathrm{\bar{\kappa}}_{0}\left\vert \bar{\epsilon}%
^{m}\right\rangle \left\langle \epsilon ^{n}\right\vert \varepsilon _{nm}
\label{d2}
\end{eqnarray}%
In the above relations, the 3+3 moduli
\begin{equation}
\mathrm{\kappa }_{i}\left( \zeta \right) \quad ,\quad \mathrm{\bar{\kappa}}%
_{i}\left( \zeta \right)  \label{kk}
\end{equation}%
capture the deformation away from the osp(4\TEXTsymbol{\vert}2) solution (%
\ref{sp}-\ref{ps}). These are\textrm{\ functions of }$\zeta $\textrm{\ that
we determine by using the two following conditions: }

\begin{description}
\item[$\left( \mathbf{i}\right) $] Eqs\textrm{\ (\ref{d1}-\ref{d2})} must
satisfy the Levi-constraint relations $\left[ \mathbf{\mu }_{3},X\right] =X$
and $\left[ \mathbf{\mu }_{3},Y\right] =-Y.$ Actually, eqs(\ref{d2}) are
particular deformations of (\ref{xosp}) because a general deformation would
involve more free parameters than the six $\left( \mathrm{\kappa }_{i},%
\mathrm{\bar{\kappa}}_{i}\right) $ in eq(\ref{kk}). Eq(\ref{d2}) were chosen
because they are the general ones effectively solving $\left[ \mathbf{\mu }%
_{3},X\right] =X$ and $\left[ \mathbf{\mu }_{3},Y\right] =-Y.$ A reduced
family of solutions in given by setting $\mathrm{\kappa }_{i}\left( \zeta
\right) =\mathrm{\bar{\kappa}}_{i}\left( \zeta \right) $; thus restricting
the number of free deformation functions to 3. Further simpler solutions are
given by putting extra constraints on the three $\mathrm{\kappa }_{i}\left(
\zeta \right) $ as we will see below.

\item[$\left( \mathbf{ii}\right) $] the $\mathcal{L}_{\mathfrak{d}(2,1;\zeta
)}^{\mathbf{\mu }_{3}}$ based on (\ref{d2}) is a function of $\mathrm{\kappa
}_{i}\left( \zeta \right) ,\ \mathrm{\bar{\kappa}}_{i}\left( \zeta \right) $%
; its value should reproduce the solutions obtained previously in deriving
L-operators $\mathcal{L}_{sl(2|2)}^{\mathbf{\mu }_{3}}$ and $\mathcal{L}%
_{osp}^{\mathbf{\mu }_{3}}$ given by eqs(\ref{20}) and (\ref{osp}). So, we
must have%
\begin{equation}
\begin{tabular}{|c|c|c|c|c|c|c|c|}
\hline
$\mathfrak{d}{\small (2,1;\zeta )}$ & : & $\quad \mathrm{\kappa }_{1}\quad $
& $\quad \mathrm{\kappa }_{2}\quad $ & $\quad \mathrm{\kappa }_{0}\quad $ & $%
\quad \mathrm{\bar{\kappa}}_{1}\quad $ & $\quad \mathrm{\bar{\kappa}}%
_{2}\quad $ & $\quad \mathrm{\bar{\kappa}}_{0}\quad $ \\ \hline
$sl(2|2)$ & : & $1$ & $0$ & $0$ & $1$ & $0$ & $0$ \\ \hline
osp(4\TEXTsymbol{\vert}2) & : & $1$ & $-1$ & $1$ & $1$ & $-1$ & $1$ \\ \hline
\end{tabular}%
\end{equation}%
indicating that $\mathrm{\kappa }_{2}$ and $\mathrm{\kappa }_{0}$ may be
related like $\mathrm{\kappa }_{2}=-\mathrm{\kappa }_{0}$; thus restricting
the number of free functions.
\end{description}

\subsubsection{Calculation of $\mathcal{L}_{\mathfrak{d}(2,1;\protect\zeta %
)}^{\mathbf{\protect\mu }_{3}}$}

To determine the L-operator $\mathcal{L}_{\mathfrak{d}(2,1;\zeta )}^{\mathbf{%
\mu }_{3}}$ while using \textrm{(\ref{d1}-\ref{d2})}, we rewrite eq(\textrm{%
\ref{d2}}) as follows%
\begin{equation}
\begin{tabular}{lll}
$X_{\mathfrak{d}(2,1;\zeta )}$ & $=$ & $\left\vert \Phi _{X}\right\rangle
\left\langle \bar{\delta}\right\vert +\left\vert \Psi _{X}^{k}\right\rangle
\left\langle \bar{\epsilon}_{k}\right\vert $ \\
$Y_{\mathfrak{d}(2,1;\zeta )}$ & $=$ & $\left\vert \bar{\delta}\right\rangle
\left\langle \Phi _{Y}\right\vert +\left\vert \bar{\epsilon}%
^{l}\right\rangle \left\langle \Psi _{Yl}\right\vert $%
\end{tabular}
\label{xd}
\end{equation}%
with%
\begin{equation}
\begin{tabular}{lll}
$\left\vert \Phi _{X}\right\rangle $ & $=$ & $\mathrm{\kappa }_{1}\mathrm{%
\beta }^{k}\left\vert \epsilon _{k}\right\rangle +\mathrm{b}\left\vert
\delta \right\rangle $ \\
$\left\langle \Phi _{Y}\right\vert $ & $=$ & $\mathrm{\bar{\kappa}}_{1}%
\mathrm{\gamma }_{n}\left\langle \epsilon ^{n}\right\vert +\mathrm{c}%
\left\langle \delta \right\vert $%
\end{tabular}%
\end{equation}%
and
\begin{equation}
\begin{tabular}{lll}
$\left\vert \Psi _{X}^{k}\right\rangle $ & $=$ & $\mathrm{\kappa }_{0}%
\mathrm{b}^{\prime }\varepsilon ^{kl}\left\vert \epsilon _{l}\right\rangle +%
\mathrm{\kappa }_{2}\mathrm{\beta }^{k}\left\vert \delta \right\rangle $ \\
$\left\langle \Psi _{Yn}\right\vert $ & $=$ & $\mathrm{\bar{\kappa}}_{0}%
\mathrm{c}^{\prime }\left\langle \epsilon ^{m}\right\vert \varepsilon _{mn}+%
\mathrm{\bar{\kappa}}_{2}\mathrm{\gamma }_{n}\left\langle \delta \right\vert
$%
\end{tabular}%
\end{equation}%
where $\left( \mathrm{\beta }^{k},\mathrm{\gamma }_{k}\right) $ are
fermionic oscillators, and $\left( \mathrm{b,c}\right) $ and $\left( \mathrm{%
b}^{\prime }\mathrm{,c}^{\prime }\right) $ are bosonic oscillator; they are
the same oscillators as for osp(4\TEXTsymbol{\vert}2). From these relations,
we calculate the product $XY=\left\vert \Phi _{X}\right\rangle \left\langle
\Phi _{Y}\right\vert +\left\vert \Psi _{X}^{n}\right\rangle \left\langle
\Psi _{Yn}\right\vert $ expanding as follows%
\begin{eqnarray}
\left\vert \Phi _{X}\right\rangle \left\langle \Phi _{Y}\right\vert
&=&\left( \mathrm{\kappa }_{1}\mathrm{\bar{\kappa}}_{1}\right) \mathrm{\beta
}^{k}\mathrm{\gamma }_{n}\left\vert \epsilon _{k}\right\rangle \left\langle
\epsilon ^{n}\right\vert +\mathrm{bc}\left\vert \delta \right\rangle
\left\langle \delta \right\vert + \\
&&\mathrm{\kappa }_{1}\mathrm{c\beta }^{k}\left\vert \epsilon
_{k}\right\rangle \left\langle \delta \right\vert +\mathrm{\bar{\kappa}}_{1}%
\mathrm{b\gamma }_{n}\left\vert \delta \right\rangle \left\langle \epsilon
^{n}\right\vert  \notag
\end{eqnarray}%
and%
\begin{eqnarray}
\left\vert \Psi _{X}^{n}\right\rangle \left\langle \Psi _{Yn}\right\vert
&=&\left( \mathrm{\kappa }_{0}\mathrm{\bar{\kappa}}_{0}\right) \mathrm{b}%
^{\prime }\mathrm{c}^{\prime }\left\vert \epsilon _{m}\right\rangle
\left\langle \epsilon ^{m}\right\vert +\left( \mathrm{\kappa }_{2}\mathrm{%
\bar{\kappa}}_{2}\right) \mathrm{\beta }^{n}\mathrm{\gamma }_{n}\left\vert
\delta \right\rangle \left\langle \delta \right\vert +  \notag \\
&&\left( \mathrm{\kappa }_{0}\mathrm{\bar{\kappa}}_{2}\right) \mathrm{b}%
^{\prime }\mathrm{\gamma }_{n}\varepsilon ^{nl}\left\vert \epsilon
_{l}\right\rangle \left\langle \delta \right\vert +\left( \mathrm{\kappa }%
_{2}\mathrm{\bar{\kappa}}_{0}\right) \mathrm{c}^{\prime }\varepsilon _{mn}%
\mathrm{\beta }^{n}\left\vert \delta \right\rangle \left\langle \epsilon
^{m}\right\vert
\end{eqnarray}%
Moreover, using the properties and $X^{2}=Y^{2}=0,$ the L-operator $\mathcal{%
L}_{\mathfrak{d}(2,1;\zeta )}^{\mathbf{\mu }_{3}}$ reduces to $\left(
1+X\right) z^{\mathbf{\mu }_{3}}\left( 1+Y\right) $ with $z^{\mathbf{\mu }%
_{3}}=z^{1/2}\Pi +z^{-1/2}\bar{\Pi}\ $and $X,$ $Y$ as in (\ref{xd}). By help
of the properties%
\begin{equation}
\begin{tabular}{lll}
$\Pi X$ & $=$ & $X$ \\
$\bar{\Pi}Y$ & $=$ & $Y$%
\end{tabular}%
\quad ,\quad
\begin{tabular}{lll}
$X\bar{\Pi}$ & $=$ & $X$ \\
$Y\Pi $ & $=$ & $Y$%
\end{tabular}%
\quad ,\quad
\begin{tabular}{lll}
$X\Pi =\Pi Y$ & $=$ & $0$ \\
$Y\bar{\Pi}=\bar{\Pi}X$ & $=$ & $0$%
\end{tabular}%
\end{equation}%
it can be further reduced down to the form $z^{\mathbf{\mu }%
_{3}}+z^{-1/2}\left( X+Y+XY\right) .$ By substituting, the $\mathcal{L}_{%
\mathfrak{d}(2,1;\zeta )}^{\mathbf{\mu }_{3}}$ reads as follows
\begin{eqnarray}
&&\left(
\begin{array}{cccc}
(z^{\frac{1}{2}}+\mathrm{\kappa }_{0}\mathrm{\bar{\kappa}}_{0}z^{-\frac{1}{2}%
}\mathrm{b}^{\prime }\mathrm{c}^{\prime })\delta _{j}^{i}+\mathrm{\kappa }%
_{1}\mathrm{\bar{\kappa}}_{1}z^{-\frac{1}{2}}\mathrm{\beta }^{i}\mathrm{%
\gamma }_{j} & z^{-\frac{1}{2}}(\mathrm{\kappa }_{1}\mathrm{c\beta }^{i}+%
\mathrm{\kappa }_{0}\mathrm{\bar{\kappa}}_{2}\mathrm{b}^{\prime }\mathrm{%
\gamma }_{n}\varepsilon ^{ni}) & \mathrm{\kappa }_{1}z^{-\frac{1}{2}}\mathrm{%
\beta }^{i} & \mathrm{\kappa }_{0}z^{-\frac{1}{2}}\mathrm{b}^{\prime
}\varepsilon ^{ji} \\
z^{-\frac{1}{2}}(\mathrm{\bar{\kappa}}_{1}\mathrm{b\gamma }_{j}+\mathrm{\bar{%
\kappa}}_{0}\mathrm{\kappa }_{2}\mathrm{c}^{\prime }\varepsilon _{jn}\mathrm{%
\beta }^{n}) & z^{\frac{1}{2}}+z^{-\frac{1}{2}}(\mathrm{bc}+\mathrm{\kappa }%
_{2}\mathrm{\bar{\kappa}}_{2}\mathrm{\beta }^{n}\mathrm{\gamma }_{n}) & z^{-%
\frac{1}{2}}\mathrm{b} & \mathrm{\kappa }_{2}z^{-\frac{1}{2}}\mathrm{\beta }%
^{j} \\
z^{-\frac{1}{2}}\mathrm{\bar{\kappa}}_{1}\mathrm{\gamma }_{j} & z^{-\frac{1}{%
2}}\mathrm{c} & z^{-\frac{1}{2}} & 0 \\
\mathrm{\bar{\kappa}}_{0}z^{-\frac{1}{2}}\mathrm{c}^{\prime }\varepsilon
_{ji} & \mathrm{\bar{\kappa}}_{2}z^{-\frac{1}{2}}\mathrm{\gamma }_{i} & 0 &
z^{-\frac{1}{2}}\delta _{i}^{j}%
\end{array}%
\right)  \notag \\
&&\text{ \ \ }  \label{ed}
\end{eqnarray}%
By taking $\mathrm{\bar{\kappa}}_{i}\left( \zeta \right) =\mathrm{\kappa }%
_{i}\left( \zeta \right) $, the L-operator has three free functions
constrained like%
\begin{equation}
\begin{tabular}{|ccc||ccc|}
\hline
\multicolumn{3}{|c||}{$sl(2|2)$} & \multicolumn{3}{||c|}{$osp(4|2)$} \\
\hline
$\qquad \left. \mathrm{\kappa }_{1}\left( \zeta \right) \right\vert _{\zeta
=-1}$ & $=$ & $1\qquad $ & $\qquad \left. \mathrm{\kappa }_{1}\left( \zeta
\right) \right\vert _{\zeta =+1}$ & $=$ & $+1\qquad $ \\ \hline
$\qquad \left. \mathrm{\kappa }_{2}\left( \zeta \right) \right\vert _{\zeta
=-1}$ & $=$ & $0\qquad $ & $\qquad \left. \mathrm{\kappa }_{2}\left( \zeta
\right) \right\vert _{\zeta =+1}$ & $=$ & $-1\qquad $ \\ \hline
$\qquad \left. \mathrm{\kappa }_{0}\left( \zeta \right) \right\vert _{\zeta
=-1}$ & $=$ & $0\qquad $ & $\qquad \left. \mathrm{\kappa }_{0}\left( \zeta
\right) \right\vert _{\zeta =+1}$ & $=$ & $+1\qquad $ \\ \hline
\end{tabular}
\label{kv}
\end{equation}%
\begin{equation*}
\end{equation*}%
A particular solution is given by taking $\mathrm{\kappa }_{1}\left( \zeta
\right) =1$ and $\mathrm{\kappa }_{2}\left( \zeta \right) =-\mathrm{\kappa }%
_{0}\left( \zeta \right) $. Using the property $\det \left( \mathbf{\mathbf{%
\tilde{\alpha}}}_{i}.\mathbf{\mathbf{\tilde{\alpha}}}_{j}\right) =-2\zeta
\left( 1+\zeta \right) ,$ it is then natural to think about the deformation
function $\mathrm{\kappa }_{0}\left( \zeta \right) $ as given by $\frac{1}{2}%
\zeta \left( 1+\zeta \right) .$

\section{Conclusion and comments}

\qquad \label{conc} To conclude this study, we summarise our main finds
along with some additional comments. Motivated by getting more insight into
the properties of D($2,1;\zeta $) and its applications in $\left( \mathbf{i}%
\right) $ integrable exotic superspin chains; and $\left( \mathbf{ii}\right)
$ in building type II string 3D models, we revisited useful algebraic and
geometric aspects on this 9\TEXTsymbol{\vert}8 dimensional exceptional
parametric symmetry. In the present work, we focused on the algebraic
geometry of complexified D($2,1;\zeta $) and its application in the study of
the integrable superspin chain with Lie superalgebra $\mathfrak{d}$($%
2,1;\zeta $) within the framework of 4D Chern-Simons theory. The real forms
of this parametric super symmetry will be considered in a forthcoming
occasion focussing on applications aimed at the study of parametric AdS$_{3}$
\textrm{strings along the lines of \cite{1AB,1AC,VB1,VB2}.}

To undertake the investigation on the algebraic geometry of the parametric D(%
$2,1;\zeta $) and its applications in integrable super systems \`{a} la
Chern-Simons, we took advantage of its local structures at some special
points of the complex variable $\zeta ;$ in particular:

$\left( \mathbf{1}\right) $ at $\zeta =1$ where D($2,1;\zeta $) identifies
with the 9\TEXTsymbol{\vert}8-dim orthosymplectic OSp(4\TEXTsymbol{\vert}2).

$\left( \mathbf{2}\right) $ at the singular point $\zeta =-1$ where D($%
2,1;\zeta $) contracts to PSL(2\TEXTsymbol{\vert}2)$\times \mathbb{C}^{3}.$
\newline
In order to describe geometrical 2-cycles of D($2,1;\zeta $), we considered
it useful to start by analysing SL(2\TEXTsymbol{\vert}2) since $\left(
\mathbf{i}\right) $ it is a subspace of it; and $\left( \mathbf{ii}\right) $
it contains two SL(2) submanifolds permitting to make use toric geometry to
address SL(2\TEXTsymbol{\vert}2) and then D($2,1;\zeta $). In \textrm{%
\autoref{sec2}}, we showed that PSL(2\TEXTsymbol{\vert}2) can be thought of
in term of the toric diagram given by the \textbf{Figure} \textbf{\ref{CP2}}%
. In this regard, it would be interesting to extend this construction to
OSp(4\TEXTsymbol{\vert}2) and D($2,1;\zeta $) which have not been addressed
in present study.

Using the correspondence (\ref{cor}) and the ansatz (\ref{az}), we studied
the four graded homological 2-cycles sets $\mathcal{C}_{\mathfrak{d}%
(2,1;\zeta )}^{(\mathfrak{\eta })}$ in the supermanifold D($2,1;\zeta $)
with label $\mathfrak{\eta }=0,1,2,3$; see eqs(\ref{s2}-\ref{ss}). For that,
we described the four parametric super Dynkin graphs S$\mathfrak{DD}_{%
\mathfrak{d}(2,1;\zeta )}^{(\mathfrak{\eta })}$ of complex $\mathfrak{d}%
(2,1;\zeta )$ and their discrete outer automorphisms. These diagrams are of
two kinds given by $\left( \mathbf{i}\right) $ the close graph S$\mathfrak{DD%
}_{\mathfrak{d}(2,1;\zeta )}^{(0)}$ with three odd simple roots; and $\left(
\mathbf{ii}\right) $ the 3 open diagrams S$\mathfrak{DD}_{\mathfrak{d}%
(2,1;\zeta )}^{(i)}$ related by outer automorphisms as shown by the \textbf{%
Figure \ref{3D}}. By extending to $\mathfrak{d}$($2,1;\zeta $) the usual
correspondence $\mathbf{\alpha }\leftrightarrow \mathcal{C}_{\mathbf{\alpha }%
}$ in ADE geometries linking roots $\mathbf{\alpha }$ of ADE Lie algebras
with 2-cycle $\mathcal{C}_{\mathbf{\alpha }}$ in asymptotic locally
Euclidean (ALE) spaces, we end up with the four parametric sets $\mathcal{C}%
_{\mathfrak{d}(2,1;\zeta )}^{(\mathfrak{\eta })}$ and their relationships
with $\mathcal{C}_{osp(4|2)}$ and $\mathcal{C}_{psl(2|2)}$.

By help of these specific features and thinking about D($2,1;\zeta $) as a
continuous deviation of OSp(4\TEXTsymbol{\vert}2) with deformation parameter
$\zeta ,$ we derived interesting properties on exceptional symmetry. These
specific features allowed us to shed more light on the relationships with
OSp(4\TEXTsymbol{\vert}2) and with the 6\TEXTsymbol{\vert}8-dim PSL(2%
\TEXTsymbol{\vert}2). They also made it possible to calculate the Lax
operator for the superspin chain by using 4D Chern-Simons (CS) theory with
D(2,1; $\zeta $) symmetry. In this regard, we computed the Lax operator $%
\mathcal{L}_{\mathfrak{d}(2,1;\zeta )}^{\left( \mathbf{\mu }\right) }$
having 3 free parameters $\mathrm{\kappa }_{1}\left( \zeta \right) ,$ $%
\mathrm{\kappa }_{2}\left( \zeta \right) $\ and $\mathrm{\kappa }_{0}\left(
\zeta \right) $ which are functions of $\zeta $. It is given by eq(\ref{ed})
which recovers as particular values of the distinguished $\mathcal{L}%
_{sl(2|2)}^{\left( \mathbf{\mu }\right) }$ and $\mathcal{L}%
_{osp(4|2)}^{\left( \mathbf{\mu }\right) }$ by taking $\mathrm{\kappa }_{i}$
as in (\ref{kv}). By setting $\mathrm{\kappa }_{1}=1$ and $\mathrm{\kappa }%
_{2}=-\mathrm{\kappa }_{0},$ one is left with a simple expression of
L-operator having one free function parameter $\mathrm{\kappa }_{0}$ and
reading as follows
\begin{equation}
z^{-1/2}\left(
\begin{array}{cccc}
(z+\mathrm{\kappa }_{0}^{2}\mathrm{b}^{\prime }\mathrm{c}^{\prime })\delta
_{j}^{i}+\mathrm{\beta }^{i}\mathrm{\gamma }_{j} & \mathrm{c\beta }^{i}-%
\mathrm{\kappa }_{0}^{2}\mathrm{b}^{\prime }\mathrm{\gamma }_{n}\varepsilon
^{ni} & \mathrm{\beta }^{i} & \mathrm{\kappa }_{0}\mathrm{b}^{\prime
}\varepsilon ^{ji} \\
(\mathrm{b\gamma }_{j}-\mathrm{\kappa }_{0}^{2}\mathrm{c}^{\prime
}\varepsilon _{jn}\mathrm{\beta }^{n} & z+\mathrm{bc}+\mathrm{\kappa }%
_{0}^{2}\mathrm{\beta }^{n}\mathrm{\gamma }_{n} & \mathrm{b} & -\mathrm{%
\kappa }_{0}\mathrm{\beta }^{j} \\
\mathrm{\gamma }_{j} & \mathrm{c} & 1 & 0 \\
\mathrm{\kappa }_{0}\mathrm{c}^{\prime }\varepsilon _{ji} & -\mathrm{\kappa }%
_{0}\mathrm{\gamma }_{i} & 0 & \delta _{i}^{j}%
\end{array}%
\right)  \label{c1}
\end{equation}%
where the pairs $\left( \mathrm{\beta }^{i},\mathrm{\gamma }_{i}\right) $
are fermionic oscillators and where $\left( \mathrm{b,c}\right) $ and $%
\left( \mathrm{b}^{\prime }\mathrm{,c}^{\prime }\right) $ are bosonic
oscillators; they satisfy the usual graded commutation relations. An
interesting expression of $\mathrm{\kappa }_{0}\left( \zeta \right) $ is
given by $\frac{1}{2}\zeta \left( 1+\zeta \right) $ interpreted in terms of
the determinant $\det \left( \mathbf{\mathbf{\tilde{\alpha}}}_{i}.\mathbf{%
\mathbf{\tilde{\alpha}}}_{j}\right) $ which is equal $-2\zeta \left( 1+\zeta
\right) ;$ is also the area of the 2-cycle $\mathcal{\tilde{C}}_{\mathbf{%
\tilde{\psi}}}$. In terms of the \emph{Kaplansky} parameters \textsc{s}$_{i}$
with the realisation $\left( 2\text{\textsc{s}}_{1},2\text{\textsc{s}}_{2},2%
\text{\textsc{s}}_{3}\right) =\left( 1,-1-\zeta ,\zeta \right) ,$\ the value
of $\mathrm{\kappa }_{0}$ is given by $-4$\textsc{s}$_{1}$\textsc{s}$_{2}$%
\textsc{s}$_{3}$.

From the above Lax matrix, we learn that by setting $\mathrm{\kappa }_{0}=1$
(corresponding to $\zeta =1$), we obtain the L-operator $\mathcal{L}%
_{osp(4|2)}^{\left( \mathbf{\mu }\right) }$ given by (\ref{osp}). Recall
that at $\zeta =1$, we have the orthosymplectic point $\mathfrak{d}%
(2,1;+1)\simeq osp(4|2)$. Similarly, for the case $\mathrm{\kappa }_{0}=0$ ($%
\zeta =-1$), eq(\ref{c1}) becomes
\begin{equation}
\left. \mathcal{L}_{\mathfrak{d}(2,1;\zeta )}^{\left( \mathbf{\mu }\right)
}\right\vert _{\zeta =-1}=\left(
\begin{array}{cc}
\mathcal{L}_{sl_{2|2}}^{\mathbf{\mu }} & 0 \\
0 & z^{-\frac{1}{2}}\delta _{i}^{j}%
\end{array}%
\right)
\end{equation}%
with $\mathcal{L}_{sl_{2|2}}^{\mathbf{\mu }}$ given by eq(\ref{20}).

In the end, we would like to add that along with the three super Dynkin
diagrams S$\mathfrak{DD}_{\mathfrak{d}(2,1;\zeta )}^{(1)},$ S$\mathfrak{DD}_{%
\mathfrak{d}(2,1;\zeta )}^{(2)}$ and S$\mathfrak{DD}_{\mathfrak{d}(2,1;\zeta
)}^{(3)}$ related by $\mathcal{S}_{3}$ outer automorphisms, one also has S$%
\mathfrak{DD}_{\mathfrak{d}(2,1;\zeta )}^{(0)}$ behaving as a singlet under $%
\mathcal{S}_{3}$ as shown by the \textbf{Figure \ref{4DD} }and the \textbf{%
Figure \ref{3D}}. Here we have studied the integrable superspin chain
associated with the open super Dynkin diagrams; it would be interesting to
look for the remaning expression of Lax operator associated with the closed S%
$\mathfrak{DD}_{\mathfrak{d}(2,1;\zeta )}^{(0)}$.

\section*{\textbf{{Acknowledgements}:}}

I thank Y. Boujakhrout, R. Ahl Laamara and L.B Drissi for previous
collaboration on integrable quantum chains and 4D Chern-Simons theory.

\section*{ \textbf{{Conflict of interest}:}}

The author has no conflicts to disclose. \appendix

\section*{Appendices}

\qquad We give four appendices A, B, C and D where we report some technical
details completing the analysis given in this paper. \textrm{In \autoref%
{appA}}, we give useful characteristic data on the algebraic geometry of SL(2%
\TEXTsymbol{\vert}2). \textrm{In \autoref{appB}}, we describe the
orthosymplectic osp(4\TEXTsymbol{\vert}2) Lie algebra and its realisation
using graded oscillators. \textrm{In \autoref{appC}}, we recall interesting
aspects on the Scheneurt description of $\mathfrak{d}$(2,1;$\zeta $); it
uses a realisation of the simple roots $\mathbf{\tilde{\alpha}}_{i}$
different from (\ref{se}). \textrm{In \autoref{appD}}, we provide tools for
the study of integrable superspin chains by the help of CS gauge potentials
valued in $\mathfrak{d}$(2,1;$\zeta $).

\section{Characteristic data of SL(2\TEXTsymbol{\vert}2)}

\qquad \label{appA} In this appendix, we give properties of the complex
semi-simple sl(2\TEXTsymbol{\vert}2) and its simple subalgebra psl(2%
\TEXTsymbol{\vert}2) used in the main text in link with the parametric super
algebra $\mathfrak{d}(2,1;\zeta )$ at $\zeta =-1.$ The 14 dimensional psl(2%
\TEXTsymbol{\vert}2) is defined like $psl(2|2)=sl(2|2)/\left\langle
H_{0}\right\rangle $ where $H_{0}=\mathbb{C}\mathtt{I}_{2|2}$ is the central
element of sl(2\TEXTsymbol{\vert}2).

\subsection*{Complex superalgebra and real forms}

\qquad The even part $sl(2|2)_{\bar{0}}$ of the Lie super algebra $sl(2|2)$
is 7 dimensional ($\sup \dim $ 7\TEXTsymbol{\vert}8); it given by the $%
sl(2)_{1}\oplus $ $sl(2)_{2}\oplus $ $\mathbb{C}$ with matrix generators $%
J_{b_{i}}^{a_{i}}=\sum_{x=1}^{3}\boldsymbol{J}_{x}\left( \mathbf{\sigma }%
^{x}\right) _{b_{i}}^{a_{i}}$ as%
\begin{equation}
\begin{tabular}{|c|c|c|c|}
\hline
symmetry & $sl(2,\mathbb{C})_{1}$ & $sl(2,\mathbb{C})_{3}$ & $\mathbb{C}:%
\mathbb{=C}_{3}$ \\ \hline
generators & $J_{b_{1}}^{a_{1}}:=J_{b}^{a}$ & $R_{b_{2}}^{a_{2}}:=R_{\mathrm{%
\beta }}^{\mathrm{\alpha }}$ & $H_{0}$ \\ \hline
action & rotation & rotation & central charge \\ \hline
\end{tabular}
\label{2sl2}
\end{equation}%
obeying the commutation relations%
\begin{equation}
\begin{tabular}{lll}
$\left[ J_{b}^{a},J_{d}^{c}\right] $ & $=$ & $\delta
_{b}^{c}J_{d}^{a}-\delta _{d}^{a}J_{b}^{c}$ \\
$\left[ R_{\mathrm{\beta }}^{\mathrm{\alpha }},R_{\mathrm{\delta }}^{\mathrm{%
\gamma }}\right] $ & $=$ & $\delta _{\mathrm{\beta }}^{\mathrm{\gamma }}R_{%
\mathrm{\delta }}^{\mathrm{\alpha }}-\delta _{\mathrm{\delta }}^{\mathrm{%
\alpha }}R_{\mathrm{\beta }}^{\mathrm{\gamma }}$ \\
$\left[ H_{0},J_{b}^{a}\right] $ & $=$ & $\left[ H_{0},R_{\mathrm{\beta }}^{%
\mathrm{\alpha }}\right] =0$%
\end{tabular}%
\end{equation}%
The odd part $sl(2|2)_{\bar{1}}$ has 4+4 fermionic generators $Q_{a}^{%
\mathrm{\alpha }}$ and $G_{\mathrm{\alpha }}^{a}$ transforming in the ($%
2,2^{\ast }$) and ($2^{\ast },2$) representations of $SL(2)\otimes SL(2)$;
they obey the anticommutation relations%
\begin{equation}
\begin{tabular}{lll}
$\left\{ Q_{a}^{\mathrm{\alpha }},G_{\mathrm{\beta }}^{b}\right\} $ & $=$ & $%
\delta _{\mathrm{\beta }}^{\mathrm{\alpha }}J_{a}^{b}+\delta _{a}^{b}R_{%
\mathrm{\beta }}^{\mathrm{\alpha }}+\delta _{a}^{b}\delta _{\mathrm{\beta }%
}^{\mathrm{\alpha }}H_{0}$ \\
$\left\{ Q_{a}^{\mathrm{\alpha }},Q_{b}^{\mathrm{\beta }}\right\} $ & $=$ & $%
\varepsilon ^{\mathrm{\alpha \beta }}\varepsilon _{ab}H_{+}$ \\
$\left\{ G_{\mathrm{\alpha }}^{a},G_{\mathrm{\beta }}^{b}\right\} $ & $=$ & $%
\varepsilon ^{ab}\varepsilon _{\mathrm{\alpha \beta }}H_{-}$%
\end{tabular}%
\end{equation}%
where $H_{+}$ and $H_{-}$ are two additional central charges generating two
extra directions. Together with the original central charge, they form a
triplet $P_{i}=(H_{0},H_{\pm })$ having an interpretation in the 17
dimensional $\mathfrak{d}(2,1;\zeta )$ at $\zeta =-1$. Because the $SL(2,%
\mathbb{C})$ group has two real forms given by the compact $SU(2)$ and the
non compact $sl(2,\mathbb{R})$; the above relations should be adapted
accordingly. For example, eq(\ref{2sl2}) can be interpreted either as
\begin{equation}
\begin{tabular}{|c|c|c|c|}
\hline
symmetry & $su(2)$ & $su(2)$ & $\mathbb{R}$ \\ \hline
generators & $J_{b}^{a}$ & $R_{\mathrm{\beta }}^{\mathrm{\alpha }}$ & $H_{0}$
\\ \hline
\end{tabular}
\label{21}
\end{equation}%
with label $a,b=1,2,3$ (first 3-sphere $\mathbb{S}_{1}^{3}$) and $\mathrm{%
\alpha },\mathrm{\beta }=4,5,6$ (second 3-sphere $\mathbb{S}_{2}^{3}$); or
like%
\begin{equation}
\begin{tabular}{|c|c|c|c|}
\hline
symmetry & $sl(2,\mathbb{R})$ & $su(2)$ & $\mathbb{R}$ \\ \hline
generators & $L_{\nu }^{\mu }$ & $J_{b}^{a}$ & $H_{0}$ \\ \hline
\end{tabular}
\label{22}
\end{equation}%
with spacetime label $\mu ,\nu =0,1,2$ (AdS$_{3}$) and internal $a,b=3,4,5$
(3-sphere $\mathbb{S}^{3}$). Along with (\ref{21}-\ref{22}), one might also
consider the situation with symmetry $sl(2,\mathbb{R})\oplus sl(2,\mathbb{R}%
)\oplus \mathbb{R}$ (two pseudo-spheres $\mathbb{S}_{i}^{1,2}\simeq
SU(1,1)_{i}$).

\subsection*{Distinguished Cartan Weyl operators}

\qquad The distinguished root system $\Phi _{sl(2|2)}^{\text{{\small (}dis)}%
} $ of the semi-simple Lie super algebra $sl(2|2)$ has 12 roots: 6 positive
roots and 6 negative ones. They are generated by three simple roots ($%
\mathbf{\alpha }_{1},\mathbf{\alpha }_{2},\mathbf{\alpha }_{3}$) with parity
as follows%
\begin{equation}
\begin{tabular}{|c|c|}
\hline
even parity (bosonic) & odd parity (fermionic) \\ \hline
$\mathbf{\alpha }_{1}\quad ,\quad \mathbf{\alpha }_{3}$ & $\mathbf{\alpha }%
_{2}\quad ,\quad \mathbf{\mathbf{\alpha }_{1}+\mathbf{\alpha }_{2}\quad
,\quad \alpha }_{2}+\mathbf{\alpha }_{3}\quad ,\quad \mathbf{\mathbf{\alpha }%
}_{1}+\mathbf{\alpha }_{2}+\mathbf{\alpha }_{3}$ \\ \hline
\end{tabular}
\label{4f}
\end{equation}%
In the Cartan Weyl basis, the generators of the even part of the Lie super
algebra $sl(2|2)_{\bar{0}}$ is labeled by two bosonic simple roots $\mathbf{%
\alpha }_{1}$ and $\mathbf{\alpha }_{3}$ as follows%
\begin{equation}
\begin{tabular}{|c|c|c|c|}
\hline
symmetry & $sl(2,\mathbb{C})_{\text{\textsc{a}}}$ & $sl(2,\mathbb{C})_{\text{%
\textsc{b}}}$ & $\mathbb{C}$ \\ \hline
generators & $E_{\pm \mathbf{\alpha }_{1}},H_{\mathbf{\alpha }_{1}}$ & $%
E_{\pm \mathbf{\alpha }_{3}},H_{\mathbf{\alpha }_{3}}$ & $H_{0}$ \\ \hline
\end{tabular}%
\end{equation}%
obeying the commutation relations%
\begin{equation}
\begin{tabular}{lll}
$\left[ E_{+\mathbf{\alpha }_{i}},E_{-\mathbf{\alpha }_{i}}\right] $ & $=$ &
$H_{\mathbf{\alpha }_{i}}$ \\
$\left[ H_{\mathbf{\alpha }_{i}},E_{\pm \mathbf{\alpha }_{j}}\right] $ & $=$
& $\pm \mathcal{K}_{ij}E_{\pm \mathbf{\alpha }_{j}}$%
\end{tabular}%
\end{equation}%
with super Cartan matrix $\mathcal{K}_{ij}.$ The fermionic operators $G_{a}^{%
\mathrm{\alpha }}$ and $Q_{\mathrm{\alpha }}^{a}$ generating $sl(2|2)_{\bar{1%
}}$ are labeled by the 4 fermionic roots (\ref{4f}) which for convenience we
set as follows%
\begin{equation}
\begin{tabular}{lll}
$\mathbf{\gamma }_{1}$ & $=$ & $\mathbf{\alpha }_{2}$ \\
$\mathbf{\gamma }_{2}$ & $=$ & $\mathbf{\mathbf{\alpha }_{1}+\mathbf{\alpha }%
_{2}}$ \\
$\mathbf{\gamma }_{3}$ & $=$ & $\mathbf{\alpha }_{2}+\mathbf{\alpha }_{3}$
\\
$\mathbf{\gamma }_{4}$ & $=$ & $\mathbf{\mathbf{\alpha }}_{1}+\mathbf{\alpha
}_{2}+\mathbf{\alpha }_{3}$%
\end{tabular}%
\qquad ,\qquad
\begin{tabular}{lll}
$Q_{\pm \mathbf{\gamma }_{1}}$ & $=$ & $Q_{\pm \mathbf{\alpha }_{2}}$ \\
$Q_{\pm \mathbf{\gamma }_{2}}$ & $=$ & $Q_{\pm \left( \mathbf{\mathbf{\alpha
}_{1}+\mathbf{\alpha }_{2}}\right) }$ \\
$Q_{\pm \mathbf{\gamma }_{3}}$ & $=$ & $Q_{\pm \left( \mathbf{\alpha }_{2}+%
\mathbf{\alpha }_{3}\right) }$ \\
$Q_{\pm \mathbf{\gamma }_{4}}$ & $=$ & $Q_{\pm \left( \mathbf{\mathbf{\alpha
}}_{1}+\mathbf{\alpha }_{2}+\mathbf{\alpha }_{3}\right) }$%
\end{tabular}%
\end{equation}%
The Chevalley generators are given by%
\begin{equation}
\begin{tabular}{|c|c|c|c|}
\hline
symmetry & $sl(2,\mathbb{C})_{\text{\textsc{a}}}$ & $sl(1|1)_{\text{\textsc{%
ab}}}$ & $sl(2,\mathbb{C})_{\text{\textsc{b}}}$ \\ \hline
generators & $E_{\pm \mathbf{\alpha }_{1}},H_{\mathbf{\alpha }_{1}}$ & $%
E_{\pm \mathbf{\alpha }_{2}},H_{\mathbf{\alpha }_{2}}$ & $E_{\pm \mathbf{%
\alpha }_{3}},H_{\mathbf{\alpha }_{3}}$ \\ \hline
\end{tabular}
\label{dec}
\end{equation}%
obeying the graded commutation relations $\left[ H_{\mathbf{\alpha }%
_{i}},E_{\pm \mathbf{\alpha }_{j}}\right\} =\pm \mathcal{K}_{ij}E_{\pm
\mathbf{\alpha }_{j}}$ with super Cartan matrix $\mathcal{K}=\left( \mathcal{%
K}_{ij}\right) $ factorising like $\mathcal{K}=\mathcal{J}^{diag}\mathcal{J}%
^{sym}$ with
\begin{equation}
\mathcal{J}^{diag}=\left(
\begin{array}{ccc}
{\small 1} & {\small 0} & {\small 0} \\
{\small 0} & {\small 1} & {\small 0} \\
{\small 0} & {\small 0} & {\small -1}%
\end{array}%
\right) ,\quad \mathcal{J}^{sym}=\left(
\begin{array}{ccc}
{\small 2} & {\small -1} & {\small 0} \\
{\small -1} & {\small 0} & {\small 1} \\
{\small 0} & {\small 1} & {\small -2}%
\end{array}%
\right) ,\quad \mathcal{K}=\left(
\begin{array}{ccc}
{\small 2} & {\small -1} & {\small 0} \\
{\small -1} & {\small 0} & {\small 1} \\
{\small 0} & {\small -1} & {\small 2}%
\end{array}%
\right)
\end{equation}%
where the symmetric part $\mathcal{J}^{sym}$ is given by the intersection
matrix $\mathbf{\alpha }_{i}.\mathbf{\alpha }_{j}.$ The other generators are
given by the graded Serre relations $(Ad_{E_{\pm \alpha
_{j}}})^{n_{jk}}\left( E_{\pm \alpha _{k}}\right) $ \cite{2B}.

\subsubsection*{Super oscillator realisation of $sl(2|2)$}

\qquad A realisation of the Chevalley generators is given by 2+2 pairs of
graded quantum oscillators $(b_{i},b_{i}^{\dagger })$ and $%
(f_{i},f_{i}^{\dagger })$\ obeying commutation relations $%
[b_{i},b_{j}^{\dagger }]=\delta _{ij}$ and anticommutations $%
\{f_{i},f_{j}^{\dagger }\}=\delta _{ij}$. Depending on whether we are
considering the complexified sl(2\TEXTsymbol{\vert}2) or its real forms su(2%
\TEXTsymbol{\vert}2) or su(1,1\TEXTsymbol{\vert}2), we distinguish different
realisations. For the example su(2\TEXTsymbol{\vert}2), we have the
following realisation of its subalgebras $su(2)_{\text{\textsc{a}}},$ $%
su(1|1)_{\text{\textsc{ab}}}$ and $su(2)_{\text{\textsc{b}}}$ respectively
labelled by the simple roots $\mathbf{\mathbf{\alpha }}_{1},$ $\mathbf{%
\alpha }_{2}$ and $\mathbf{\alpha }_{3}$:

$\bullet $ \textbf{Subalgebra} $\mathbf{su(2)}_{\text{\textsc{a}}}$: It is
contained in su(2\TEXTsymbol{\vert}2) and realised in terms of bosonic
oscillators as follows
\begin{equation}
E_{+\mathbf{\alpha }_{1}}=b_{1}^{\dagger }b_{2}\quad ,\quad E_{-\mathbf{%
\alpha }_{1}}=b_{2}^{\dagger }b_{1}\quad ,\quad H_{\mathbf{\alpha }%
_{1}}=b_{1}^{\dagger }b_{1}-b_{2}^{\dagger }b_{2}  \label{E1}
\end{equation}%
Notice that for the case $sl(2,\mathbb{C})$ contained into sl(2\TEXTsymbol{%
\vert}2), we have instead of $(b_{i},b_{i}^{\dagger })$ the complexified
oscillators $(b_{i},c_{i})$ with $[b_{i},c_{j}]=\delta _{ij}$. For this
case, the above realisation extends like%
\begin{equation}
E_{+\mathbf{\alpha }_{1}}=c_{1}b_{2}\quad ,\quad E_{-\mathbf{\alpha }%
_{1}}=c_{2}b_{1}\quad ,\quad H_{\mathbf{\alpha }_{1}}=c_{1}b_{1}-c_{2}b_{2}
\end{equation}

$\bullet $ \textbf{Case of }$\mathbf{su(1|1)}_{\text{\textsc{ab}}}:$ We have%
\begin{equation}
E_{+\mathbf{\alpha }_{2}}=b_{2}^{\dagger }f_{1}\quad ,\quad E_{-\mathbf{%
\alpha }_{2}}=f_{1}^{+}b_{2}\quad ,\quad H_{\mathbf{\alpha }%
_{2}}=b_{2}^{\dagger }b_{2}+f_{1}^{\dagger }f_{1}
\end{equation}%
satisfying the cross commutators $\left[ H_{\mathbf{\alpha }_{1}},E_{+%
\mathbf{\alpha }_{2}}\right] =-E_{+\mathbf{\alpha }_{2}}$ and $\left[ H_{%
\mathbf{\alpha }_{2}},E_{+\mathbf{\alpha }_{1}}\right] =-E_{+\mathbf{\alpha }%
_{1}}$ giving the link with the $su(2)_{\text{\textsc{a}}}$ inside $su(2|2).$
Similar relationships can be written down for $sl(1|1)_{\text{\textsc{ab}}}$
by using the bosonic $(b_{i},c_{i})$ and the fermionic $(f_{i},g_{i}).$

$\bullet $ \textbf{Subalgebra} $\mathbf{su(2)}_{3}:$ It is realised in terms
of fermionic oscillators like%
\begin{equation}
E_{+\mathbf{\alpha }_{3}}=f_{1}^{+}f_{2}\quad ,\quad E_{-\mathbf{\alpha }%
_{3}}=f_{2}^{+}f_{1}\quad ,\quad H_{\mathbf{\alpha }%
_{3}}=f_{1}^{+}f_{1}-f_{2}^{+}f_{2}  \label{E2}
\end{equation}%
with cross commutations $\left[ H_{\mathbf{\alpha }_{2}},E_{+\mathbf{\alpha }%
_{3}}\right] =+E_{+\mathbf{\alpha }_{3}}$ and $\left[ H_{\mathbf{\alpha }%
_{3}},E_{+\mathbf{\alpha }_{2}}\right] =-E_{+\mathbf{\alpha }_{2}}$. \newline
Notice that for su(2\TEXTsymbol{\vert}2) the total hermitian hamiltonian
\begin{equation}
H_{0}=b_{1}^{\dagger }b_{1}+b_{2}^{\dagger
}b_{2}+f_{1}^{+}f_{1}+f_{2}^{+}f_{2}  \label{dia}
\end{equation}%
has the properties: $\left( \mathbf{i}\right) $ it commutes with the three
triplets of Chevalley generators $E_{\pm \mathbf{\alpha }_{i}},$ $H_{\mathbf{%
\alpha }_{i}}$ (central element). $\left( \mathbf{ii}\right) $\ it is given
by the linear combination of the Cartan generators
\begin{equation}
H_{0}=H_{\mathbf{\alpha }_{1}}+2H_{\mathbf{\alpha }_{2}}-H_{\mathbf{\alpha }%
_{3}}  \label{diag}
\end{equation}

\subsubsection*{Distinguished root diagram of $psl(2|2)$}

\begin{figure}[h]
\begin{center}
\includegraphics[width=10cm]{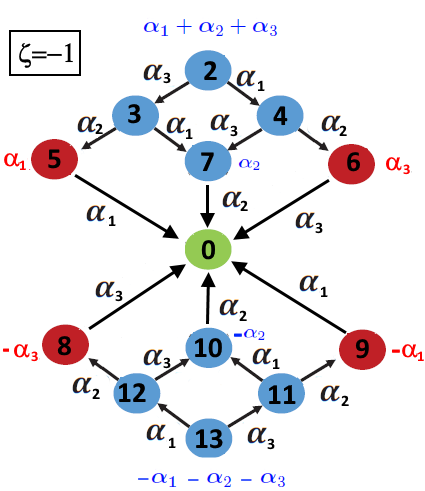}
\end{center}
\par
\vspace{-0.5cm}
\caption{Root diagram of psl(2\TEXTsymbol{\vert}2). In red the 2+2 even
roots of psl(2\TEXTsymbol{\vert}2)$_{\bar{0}}$ and in blue the 4+4 odd ones.
From this diagram, we learn (i) its highest (lowest) weight vectors, and
(ii) the list of the weight vectors of the adjoint representation of psl(2%
\TEXTsymbol{\vert}2).}
\label{WDD}
\end{figure}

\qquad Here, we construct the root diagram of psl(2\TEXTsymbol{\vert}2). It\
has 12 roots (6 positive and 6 negative). Its highest weight state $%
\left\vert hws\right\rangle $ is given by the long root $\mathbf{\psi }=%
\mathbf{\mathbf{\alpha }_{1}+\mathbf{\alpha }}_{2}+\mathbf{\alpha }_{3}.$
The weight diagram relating these roots is depicted by the \textbf{Figure}
\textbf{\ref{WDD}}.
It describes the adjoint representation of psl(2\TEXTsymbol{\vert}2) with
weight vectors as
\begin{equation}
\begin{tabular}{lll}
w$_{1}$ & $=$ & $\left( 1,1,1\right) $ \\
w$_{2}$ & $=$ & $\left( 1,1,0\right) $ \\
w$_{3}$ & $=$ & $\left( 0,1,1\right) $ \\
w$_{4}$ & $=$ & $\left( 0,1,0\right) $ \\
w$_{5}$ & $=$ & $\left( 1,0,0\right) $ \\
w$_{6}$ & $=$ & $\left( 0,0,1\right) $ \\
w$_{7}$ & $=$ & $\left( 0,0,0\right) $%
\end{tabular}%
\qquad ,\qquad
\begin{tabular}{lll}
w$_{8}$ & $=$ & $\left( 0,0,0\right) $ \\
w$_{9}$ & $=$ & $\left( 0,0,-1\right) $ \\
w$_{10}$ & $=$ & $\left( -1,0,0\right) $ \\
w$_{11}$ & $=$ & $\left( 0,-1,0\right) $ \\
w$_{12}$ & $=$ & $\left( 0,-1,-1\right) $ \\
w$_{13}$ & $=$ & $\left( -1,-1,0\right) $ \\
w$_{14}$ & $=$ & $\left( -1,-1,-1\right) $%
\end{tabular}%
\end{equation}%
Starting from the highest weight vector $\left\vert \mathbf{\psi }%
\right\rangle $ satisfying $E_{-\mathbf{\alpha }}\left\vert \mathbf{\psi }%
\right\rangle =0$, one can generate the descendent states by applying the
(creation/annihilation) step operators $E_{\pm \mathbf{\alpha }}$. For
example, the state $\left\vert \mathbf{\mathbf{\alpha }_{1}+\mathbf{\alpha }}%
_{2}\right\rangle $ is reached by the action%
\begin{equation}
E_{\mathbf{\alpha }_{3}}\left\vert \mathbf{\psi }\right\rangle
\end{equation}%
and the state $\left\vert \mathbf{\mathbf{\alpha }}_{2}+\mathbf{\alpha }%
_{3}\right\rangle $ is reached as $E_{\mathbf{\mathbf{\alpha }_{1}}%
}\left\vert \mathbf{\psi }\right\rangle $ while there no direct way to go
from $\left\vert \mathbf{\psi }\right\rangle $\ towards $\left\vert \mathbf{%
\mathbf{\alpha }}_{2}\right\rangle $ since we have $\left\vert \mathbf{%
\mathbf{\alpha }}_{2}\right\rangle =E_{\mathbf{\mathbf{\alpha }_{1}}}E_{%
\mathbf{\alpha }_{3}}\left\vert \mathbf{\psi }\right\rangle $. The full set
of transitions can be read from the diagram of the \textbf{Figure} \textbf{%
\ref{WDD}}.

\subsection*{Four super Dynkin diagrams}

\begin{figure}[tbph]
\begin{center}
\includegraphics[width=12cm]{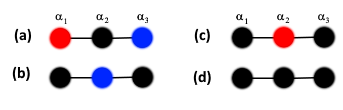}
\end{center}
\par
\vspace{-0.5cm}
\caption{Four graded Dynkin diagrams (super DD) of the Lie superalgebra $%
sl(2|2).$ The first (top left) has one odd simple root (black node $\mathbf{%
\mathbf{\protect\alpha }_{2}}$). The fourth graph (bottom right) has three
(black) odd roots. The four graphs are associated with the orderings $bbff,$
$fbbf,bffb,$ $bfbf.$ Here, we have drawn the intersections $I_{ij}^{%
{\protect\small (\protect\mu )}}=\mathbf{\protect\alpha }_{i}^{%
{\protect\small (\protect\mu )}}.\mathbf{\protect\alpha }_{j}^{%
{\protect\small (\protect\mu )}}.$}
\label{Asl2}
\end{figure}

\qquad The super Lie algebra sl(2\TEXTsymbol{\vert}2)\ has four super Dynkin
diagrams that we denote like SDD$_{1},$ SDD$_{2},$ SDD$_{3}$ and SDD$_{0}$
and which are given by the \textbf{Figure} \textbf{\ref{Asl2}}.
Each SDD$_{\mu }^{{\small sl(2|2)}}$ has three simple roots \{$\mathbf{%
\alpha }_{1}^{{\small (\mu )}},\mathbf{\alpha }_{2}^{{\small (\mu )}},%
\mathbf{\alpha }_{3}^{{\small (\mu )}}$\} whose parity depends on the
grading of the canonical weight vectors \{$\mathbf{e}_{1},\mathbf{e}_{2},%
\mathbf{e}_{3},\mathbf{e}_{4}$\}. For convenience, we denote these canonical
weight vectors like \{$\mathbf{b}_{1},\mathbf{b}_{2},\mathbf{f}_{1},\mathbf{f%
}_{2}$\}; they generate the 2+2 graded space $\mathbb{C}^{2|2}$ with
pairings as $\mathbf{b}_{i}.\mathbf{b}_{j}=\delta _{ij}$ and $\mathbf{f}_{i}.%
\mathbf{f}_{j}=-\delta _{ij}$ as well as $\mathbf{b}_{i}.\mathbf{f}_{j}=0.$
Along with these canonical weight vectors, we can introduce the following
"mean weight vector"
\begin{equation}
\epsilon =\frac{1}{4}\left( \mathbf{b}_{1}+\mathbf{b}_{2}-\mathbf{f}_{1}-%
\mathbf{f}_{2}\right)  \label{pe}
\end{equation}%
that turns out to play an important role in sl(2\TEXTsymbol{\vert}2); it
labels the central element H$_{0}$.

\subsubsection*{Four basis sets of simple roots}

\qquad Because of the intimate relationship between roots $\mathbf{\alpha }$
of Lie algebras and 2-cycles $\mathcal{C}_{\mathbf{\alpha }}$ of complex
manifolds like the familiar ADE geometries of ALE surfaces \textrm{\cite%
{LEUNG,SEBB, AbouN}}, we think it interesting to describe explicitly the
four sets of simple roots \{$\mathbf{\alpha }_{1}^{{\small (\mu )}},\mathbf{%
\alpha }_{2}^{{\small (\mu )}},\mathbf{\alpha }_{3}^{{\small (\mu )}}$\} and
the associated 2-cycle basis sets of the SL(2\TEXTsymbol{\vert}2) geometry \{%
$\mathcal{C}_{\mathbf{\alpha }_{1}^{{\small (\mu )}}},\mathcal{C}_{\mathbf{%
\alpha }_{2}^{{\small (\mu )}}},\mathcal{C}_{\mathbf{\alpha }_{3}^{{\small %
(\mu )}}}$\}. Due to the different orderings of the canonical weight vectors
\{$\mathbf{b}_{1},\mathbf{b}_{2},\mathbf{f}_{1},\mathbf{f}_{2}$\}, the
superalgebra sl(2\TEXTsymbol{\vert}2) has four basis sets of simple roots
systems $\Pi _{\mathrm{\mu }}$ labelled by $\mathrm{\mu }=0,1,2,3;$%
\begin{equation}
\Pi _{\mathrm{0}}\quad ,\quad \Pi _{\mathrm{1}}\mathbf{\quad ,\quad }\Pi _{%
\mathrm{2}}\quad ,\quad \Pi _{\mathrm{3}}  \label{rb}
\end{equation}%
These four sets can be determined from the four possible realisations of the
long roots. We denote these four realisations like $\mathbf{\psi }_{1}$, $%
\mathbf{\psi }_{2},$ $\mathbf{\psi }_{3}$ and $\mathbf{\psi }_{0}$.
According to the orderings (bbff), (bffb), (fbbf) and (bfbf), we have the
following:

\begin{itemize}
\item \textbf{Ordering (bbff)}: long root $\mathbf{\psi }_{1}=\mathbf{b}_{1}-%
\mathbf{f}_{2}$
\begin{equation}
\begin{tabular}{lll}
$\mathbf{b}_{1}-\mathbf{f}_{2}$ & $=$ & $\left( \mathbf{b}_{1}-\mathbf{b}%
_{2}\right) +\left( \mathbf{b}_{2}-\mathbf{f}_{1}\right) +\left( \mathbf{f}%
_{1}-\mathbf{f}_{2}\right) $ \\
$\ \ \mathbf{\psi }_{1}$ & $=$ & $\ \ \mathbf{\alpha }_{1}\qquad +\qquad
\mathbf{\alpha }_{2}\quad +\quad \mathbf{\alpha }_{3}$%
\end{tabular}%
\end{equation}%
from which we read $\mathbf{\alpha }_{1}=\mathbf{b}_{1}-\mathbf{b}_{2}$
(even) and $\mathbf{\alpha }_{2}=\mathbf{b}_{2}-\mathbf{f}_{1}$ (odd) as
well as $\mathbf{\alpha }_{3}=\mathbf{f}_{1}-\mathbf{f}_{2}$ (even).\

\item \textbf{Ordering (bffb)}: long root $\mathbf{\psi }_{2}=\mathbf{b}_{1}-%
\mathbf{b}_{2}$%
\begin{equation}
\begin{tabular}{lll}
$\mathbf{b}_{1}-\mathbf{b}_{2}$ & $=$ & $\left( \mathbf{b}_{1}-\mathbf{f}%
_{1}\right) +\left( \mathbf{f}_{1}-\mathbf{f}_{2}\right) +\left( \mathbf{f}%
_{2}+\mathbf{b}_{2}\right) $ \\
$\ \ \mathbf{\psi }_{2}$ & $=$ & $\ \ \mathbf{\alpha }_{1}^{\prime }\qquad
+\quad \mathbf{\alpha }_{2}^{\prime }\quad +\quad \mathbf{\alpha }%
_{3}^{\prime }$%
\end{tabular}
\label{O2}
\end{equation}%
with odd .

\item \textbf{Ordering (fbbf)}: long root $\mathbf{\psi }_{3}=\mathbf{f}_{1}-%
\mathbf{f}_{2}$
\begin{equation}
\begin{tabular}{lll}
$\mathbf{f}_{1}-\mathbf{f}_{2}$ & $=$ & $\left( \mathbf{f}_{1}-\mathbf{b}%
_{1}\right) +\left( \mathbf{b}_{1}-\mathbf{b}_{2}\right) +\left( \mathbf{b}%
_{2}-\mathbf{f}_{2}\right) $ \\
$\ \ \mathbf{\psi }_{1}$ & $=$ & $\ \ \mathbf{\alpha }_{1}^{\prime \prime
}\qquad +\quad \mathbf{\alpha }_{2}^{\prime \prime }\quad +\quad \mathbf{%
\alpha }_{3}^{\prime \prime }$%
\end{tabular}%
\end{equation}

\item \textbf{Ordering (bfbf)}: long root $\mathbf{\psi }_{0}=\mathbf{b}_{1}-%
\mathbf{f}_{2}$
\begin{equation}
\begin{tabular}{lll}
$\mathbf{b}_{1}-\mathbf{f}_{2}$ & $=$ & $\left( \mathbf{b}_{1}-\mathbf{f}%
_{1}\right) +\left( \mathbf{f}_{1}-\mathbf{b}_{2}\right) +\left( \mathbf{b}%
_{2}-\mathbf{f}_{2}\right) $ \\
$\ \ \mathbf{\psi }_{0}$ & $=$ & $\ \ \mathbf{\alpha }_{1}^{\prime \prime
\prime }\qquad +\quad \mathbf{\alpha }_{2}^{\prime \prime \prime }\quad
+\quad \mathbf{\alpha }_{3}^{\prime \prime \prime }$%
\end{tabular}
\label{O4}
\end{equation}
\end{itemize}

So, the four basis sets of the root system $\Phi _{sl(2|2)}$ are as follows
\begin{equation}
\begin{tabular}{|c||c|c|c||c|c|}
\hline
{\small basis sets} & $\mathbf{\alpha }_{1}$ & $\mathbf{\alpha }_{2}$ & $%
\mathbf{\alpha }_{3}$ & {\small odd roots} & {\small even roots} \\
\hline\hline
$\Pi _{1}\left( {\small b,b,f,f}\right) $ & $\mathbf{b}_{1}-\mathbf{b}_{2}$
& $\mathbf{b}_{2}-\mathbf{f}_{1}$ & $\mathbf{f}_{1}-\mathbf{f}_{2}$ & $%
\mathbf{\alpha }_{2}$ & $\mathbf{\alpha }_{1},$ $\mathbf{\alpha }_{3}$ \\
\hline
$\Pi _{2}\left( {\small b,f,f,b}\right) $ & $\mathbf{b}_{1}-\mathbf{f}_{1}$
& $\mathbf{f}_{1}-\mathbf{f}_{2}$ & $\mathbf{f}_{2}-\mathbf{b}_{2}$ & $%
\mathbf{\alpha }_{1}\mathbf{,\alpha }_{3}$ & $\mathbf{\alpha }_{2}$ \\ \hline
$\Pi _{3}\left( {\small f,b,b,f}\right) $ & $\mathbf{f}_{1}-\mathbf{b}_{1}$
& $\mathbf{b}_{1}-\mathbf{b}_{2}$ & $\mathbf{b}_{2}-\mathbf{f}_{2}$ & $%
\mathbf{\alpha }_{1}\mathbf{,\alpha }_{3}$ & $\mathbf{\alpha }_{2}$ \\ \hline
$\Pi _{0}\left( {\small b,f,b,f}\right) $ & $\mathbf{b}_{1}-\mathbf{f}_{1}$
& $\mathbf{f}_{1}-\mathbf{b}_{2}$ & $\mathbf{b}_{2}-\mathbf{f}_{2}$ & $%
\mathbf{\alpha }_{1}\mathbf{,\alpha }_{2}\mathbf{,\alpha }_{3}$ & - \\
\hline\hline
\end{tabular}
\label{pi}
\end{equation}%
where the $\Pi _{2}$ and $\Pi _{3}$ are distinguished by the different
realisation of $\mathbf{\alpha }_{2}$ (red and blue colors in the \textbf{%
Figure} \textbf{\ref{Asl2}}). By using the correspondence algebraic roots/
2-cycles ($\mathbf{\alpha }$ $\mathbf{\leftrightarrow }$ $\mathcal{C}_{%
\mathbf{\alpha }}$), we have the associated 2-cycles basis sets%
\begin{equation}
\begin{tabular}{|c||c|c|}
\hline
{\small 2-cycles basis sets} & {\small odd roots} & {\small even roots} \\
\hline\hline
$\Pi _{1}$ & $\mathcal{C}_{\mathbf{\alpha }_{2}}$ & $\mathcal{C}_{\mathbf{%
\alpha }_{1}},$ $\mathcal{C}_{\mathbf{\alpha }_{3}}$ \\ \hline
$\Pi _{2}$ & $\mathcal{C}_{\mathbf{\alpha }_{1}},$ $\mathcal{C}_{\mathbf{%
\alpha }_{3}}$ & $\mathcal{C}_{\mathbf{\alpha }_{2}}$ \\ \hline
$\Pi _{3}$ & $\mathcal{C}_{\mathbf{\alpha }_{1}},$ $\mathcal{C}_{\mathbf{%
\alpha }_{3}}$ & $\mathcal{C}_{\mathbf{\alpha }_{2}}$ \\ \hline
$\Pi _{0}$ & $\mathcal{C}_{\mathbf{\alpha }_{1}}\mathbf{,\mathcal{C}_{%
\mathbf{\alpha }_{2}},\mathcal{C}_{\mathbf{\alpha }_{3}}}$ & - \\
\hline\hline
\end{tabular}%
\end{equation}

\subsubsection*{Super Dynkin graphs}

\qquad Because $\Phi _{{\small sl(2|2)}}$ is given by $\Phi _{{\small sl(2|2)%
}}^{+}\cup \Phi _{{\small sl(2|2)}}^{-},$ the four root systems $\Phi _{%
{\small sl(2|2)}}^{\Pi _{\mu }}$ of sl(2\TEXTsymbol{\vert}2) have each $6$
positive and $6$ negative graded roots. For $\Phi _{{\small sl(2|2)}}^{+},$
we have%
\begin{equation}
\begin{tabular}{|c||c|c|}
\hline
{\small roots } & {\small fermionic subset }${\small sl(2|2)}_{\bar{1}}$ &
{\small bosonic subset }${\small sl(2|2)}_{0}$ \\ \hline\hline
$\Phi _{{\small sl(2|2)}}^{\Pi _{1}}$ & $\left.
\begin{array}{c}
\mathbf{\alpha }_{2},\quad \mathbf{\mathbf{\alpha }}_{1}+\mathbf{\alpha }%
_{2}+\mathbf{\alpha }_{3}, \\
\mathbf{\mathbf{\alpha }_{2}+\mathbf{\alpha }_{1},\quad \alpha }_{2}+\mathbf{%
\alpha }_{3}%
\end{array}%
\right. $ & $\mathbf{\alpha }_{1},\quad \mathbf{\alpha }_{3}$ \\ \hline
$\Phi _{{\small sl(2|2)}}^{\Pi _{2}}$ & $\quad \mathbf{\mathbf{\alpha }%
_{1},\quad \alpha }_{3},\quad \mathbf{\mathbf{\alpha }_{1}+\mathbf{\alpha }%
_{2},\quad \alpha }_{2}+\mathbf{\alpha }_{3}\quad $ & $\quad \mathbf{\mathbf{%
\alpha }_{2},\quad \mathbf{\alpha }}_{1}+\mathbf{\alpha }_{2}+\mathbf{\alpha
}_{3}\quad $ \\ \hline
$\Phi _{{\small sl(2|2)}}^{\Pi _{3}}$ & $\mathbf{\mathbf{\alpha }_{1},\quad
\alpha }_{3},\quad \mathbf{\mathbf{\alpha }_{1}+\mathbf{\alpha }_{2},\quad
\alpha }_{2}+\mathbf{\alpha }_{3}$ & $\mathbf{\mathbf{\alpha }_{2}},\quad
\mathbf{\mathbf{\alpha }}_{1}+\mathbf{\alpha }_{2}+\mathbf{\alpha }_{3}$ \\
\hline
$\Phi _{{\small sl(2|2)}}^{\Pi _{0}}$ & $\mathbf{\mathbf{\alpha }_{1},\quad
\mathbf{\alpha }_{2},\quad \alpha }_{3},\quad \mathbf{\mathbf{\alpha }}_{1}+%
\mathbf{\alpha }_{2}+\mathbf{\alpha }_{3}$ & $\mathbf{\mathbf{\alpha }_{1}+%
\mathbf{\alpha }_{2},\quad \alpha }_{2}+\mathbf{\alpha }_{3}$ \\ \hline\hline
\end{tabular}%
\end{equation}%
\begin{equation*}
\end{equation*}%
By using the relations (\ref{pi}), one can write down the realisation of the
positive roots $\mathbf{\alpha }$ of the system $\Phi _{{\small sl(2|2)}%
}^{\Pi _{\mathrm{\mu }}}$ in terms of the canonical weight vectors. For the
example of the graded root system $\Phi _{{\small sl(2|2)}}^{\Pi _{1}}$ with
weight vectors ordered like ($\mathbf{b}_{1},\mathbf{b}_{2},\mathbf{f}_{1},%
\mathbf{f}_{2}$)$,$ we have%
\begin{equation}
\begin{tabular}{lllllll}
$\mathbf{\alpha }_{1}$ & $=$ & $\mathbf{b}_{1}-\mathbf{b}_{2}$ & $,\quad $ &
$\mathbf{\mathbf{\alpha }_{1}+\mathbf{\alpha }_{2}}$ & $\mathbf{=}$ & $%
\mathbf{b}_{1}-\mathbf{f}_{1}$ \\
$\mathbf{\alpha }_{2}$ & $=$ & $\mathbf{b}_{2}-\mathbf{f}_{1}$ & $,\quad $ &
$\mathbf{\alpha }_{2}+\mathbf{\alpha }_{3}$ & $\mathbf{=}$ & $\mathbf{b}_{2}-%
\mathbf{f}_{2}$ \\
$\mathbf{\alpha }_{3}$ & $=$ & $\mathbf{f}_{1}-\mathbf{f}_{2}$ & $,\quad $ &
$\mathbf{\mathbf{\alpha }}_{1}+\mathbf{\alpha }_{2}+\mathbf{\alpha }_{3}$ & $%
=$ & $\mathbf{b}_{1}-\mathbf{f}_{2}$%
\end{tabular}
\label{rsy}
\end{equation}%
Similar realisations can be written down for the other orderings; see eqs(%
\ref{O2}-\ref{O4}). The four super Dynkin diagrams DD$_{\mathrm{\mu }}^{%
{\small sl(2|2)}}$ associated with the four orderings of the canonical
weights are depicted by the \textbf{Figure} \textbf{\ref{Asl2}}.

\section{Orthosymplectic osp(4\TEXTsymbol{\vert}2)}

\qquad \label{appB} In this appendix, we give some tools of osp(4\TEXTsymbol{%
\vert}2) and study its oscillator realisation as well as its link with the
exceptional super D$(2,1;\zeta )$ at $\zeta =1.$ Recall that the
distinguished super Cartan matrix $\mathcal{K}_{D(2,1;\zeta )}$ and the
intersection matrix $\mathbf{\tilde{\alpha}}_{i}.\mathbf{\tilde{\alpha}}_{j}$
with odd $\mathbf{\tilde{\alpha}}_{2}$ given by\
\begin{equation}
\mathcal{K}_{D(2,1;\zeta )}=\left(
\begin{array}{ccc}
2 & -1 & 0 \\
1 & 0 & \zeta \\
0 & -1 & 2%
\end{array}%
\right) \quad ,\quad \mathbf{\tilde{\alpha}}_{i}.\mathbf{\tilde{\alpha}}%
_{j}=\left(
\begin{array}{ccc}
2 & -1 & 0 \\
-1 & 0 & -\zeta \\
0 & -\zeta & 2\zeta%
\end{array}%
\right)  \label{kf}
\end{equation}%
take the following osp(4\TEXTsymbol{\vert}2) values%
\begin{equation}
\mathcal{K}_{D(2,1;\zeta )}=\left(
\begin{array}{ccc}
2 & -1 & 0 \\
1 & 0 & 1 \\
0 & -1 & 2%
\end{array}%
\right) \quad ,\quad \mathbf{\tilde{\alpha}}_{i}.\mathbf{\tilde{\alpha}}%
_{j}=\left(
\begin{array}{ccc}
2 & -1 & 0 \\
-1 & 0 & -1 \\
0 & -1 & 2%
\end{array}%
\right)
\end{equation}

\subsection*{Root system of osp(4\TEXTsymbol{\vert}2)\ }

\qquad Here, we use the isomorphism between the triple-sum of isospins $%
sl(2)_{\mathbf{\mathbf{\beta }}_{1}}\oplus sl(2)_{\mathbf{\beta }_{2}}\oplus
sl(2)_{\mathbf{\beta }_{3}}$ and the bi-sum $so(4)_{\mathbf{\mathbf{\beta }}%
_{1},\mathbf{\beta }_{2}}\oplus sp(2)_{\mathbf{\beta }_{3}}$ to realise $%
\mathfrak{d}(2,1;\zeta )$ at $\zeta =1.$\ The root system $\Phi _{{\small %
osp(4|2)}}$ has 14 roots; 7 positive and 7 negative. The set of the 7
positive splits into an even subset $\left\{ \mathbf{\mathbf{\beta }}_{1},%
\mathbf{\beta }_{2},\mathbf{\beta }_{3}\right\} $ of three bosonic roots;
and an odd subset $\left\{ \mathbf{\mathbf{\gamma }}_{1},\mathbf{\gamma }%
_{2},\mathbf{\gamma }_{3},\mathbf{\gamma }_{4}\right\} $ of fermionic
roots.\

\begin{figure}[h]
\begin{center}
\includegraphics[width=12cm]{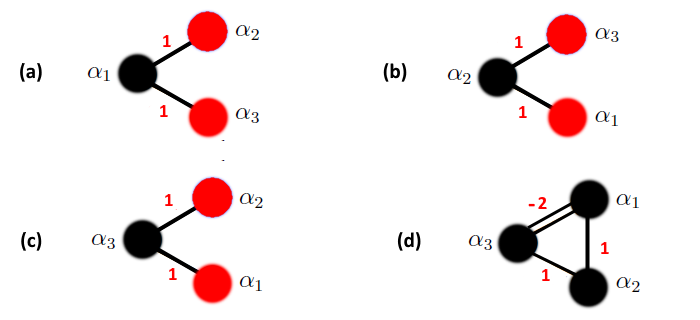}
\end{center}
\par
\vspace{-0.5cm}
\caption{Super Dynkin diagrams of the orthosymplectic Lie superalgebra osp(4%
\TEXTsymbol{\vert}2).}
\label{4DOSP}
\end{figure}

$\bullet $ \emph{Super Dynkin diagrams of osp(4\TEXTsymbol{\vert}2)}\newline
The structure of the super Dynkin diagrams of osp(4\TEXTsymbol{\vert}2)
follow from the parametric super Dynkin diagrams of $\mathfrak{d}(2,1;\zeta
) $ at $\zeta =1$. These super diagrams are given by the four graphs of the
\textbf{Figure \ref{3D}} with Kaplansky parameters (\textsc{s}$_{1},$\textsc{%
s}$_{2},$\textsc{s}$_{3}$) realised for open graphs as follows%
\begin{equation}
\begin{tabular}{|c|c|c|c|}
\hline
odd root & 2\textsc{s}$_{1}$ & 2\textsc{s}$_{2}$ & 2\textsc{s}$_{3}$ \\
\hline
$\mathbf{\alpha }_{1}$ & $-2$ & $1$ & $1$ \\ \hline
$\mathbf{\alpha }_{2}$ & $1$ & $-2$ & $1$ \\ \hline
$\mathbf{\alpha }_{3}$ & $1$ & $1$ & $-2$ \\ \hline
\end{tabular}%
\end{equation}%
\ These super diagrams are given by \textbf{the Figure \ref{4DOSP}}.
For the three open diagrams, the simple roots can be realised in terms of
the graded canonical weights $\mathbf{\mathbf{\epsilon }}_{i}$ and $\mathbf{%
\delta }$ as follow%
\begin{equation}
\begin{tabular}{|c|c|c|c|}
\hline
simple root & $\mathbf{\mathbf{\alpha }}_{1}$ & $\mathbf{\mathbf{\alpha }}%
_{2}$ & $\mathbf{\mathbf{\alpha }}_{3}$ \\ \hline
\textbf{Fig \ref{3D}}-(a) & $\quad \mathbf{\delta -\mathbf{\epsilon }_{1}}%
\quad $ & $\quad \mathbf{\epsilon }_{1}+\mathbf{\epsilon }_{2}\quad $ & $%
\quad \mathbf{\epsilon }_{1}-\mathbf{\epsilon }_{2}\quad $ \\ \hline
\textbf{Fig \ref{3D}}-(b) & $\mathbf{\epsilon }_{1}-\mathbf{\epsilon }_{2}$
& $\mathbf{\delta -\mathbf{\epsilon }_{1}}$ & $\mathbf{\epsilon }_{1}+%
\mathbf{\epsilon }_{2}$ \\ \hline
\textbf{Fig \ref{3D}}-(c) & $\mathbf{\epsilon }_{1}+\mathbf{\epsilon }_{2}$
& $\mathbf{\epsilon }_{1}-\mathbf{\epsilon }_{2}$ & $\mathbf{\delta -\mathbf{%
\epsilon }_{1}}$ \\ \hline
\end{tabular}%
\end{equation}%
For the other non simple root, see eq(\ref{os2}) below. Regarding the closed
super Dynkin diagram, the three simple roots are ($\mathbf{\mathbf{\alpha }}%
_{1},\mathbf{\mathbf{\alpha }}_{2},\mathbf{\mathbf{\alpha }}_{3}$)
fermionic; they can be realised as follows%
\begin{equation}
\begin{tabular}{|c|c|c|c|}
\hline
simple root & $\mathbf{\mathbf{\alpha }}_{1}$ & $\mathbf{\mathbf{\alpha }}%
_{2}$ & $\mathbf{\mathbf{\alpha }}_{3}$ \\ \hline
\textbf{Fig \ref{3D}}-(d) & $\quad \mathbf{\delta }-\mathbf{\epsilon }%
_{2}\quad $ & $\quad \mathbf{\mathbf{\epsilon }_{1}-\delta }\quad $ & $\quad
\mathbf{\delta }+\mathbf{\epsilon }_{2}\quad $ \\ \hline
\end{tabular}%
\end{equation}%
together with%
\begin{equation}
\begin{tabular}{lll}
$\mathbf{\mathbf{\alpha }}_{1}+\mathbf{\alpha }_{2}$ & $=$ & $\mathbf{%
\mathbf{\epsilon }_{1}-\mathbf{\epsilon }_{2}}$ \\
$\mathbf{\alpha }_{2}+\mathbf{\alpha }_{3}$ & $=$ & $\mathbf{\mathbf{%
\epsilon }_{1}}+\mathbf{\epsilon }_{2}$%
\end{tabular}%
\quad ,\quad
\begin{tabular}{lll}
$\mathbf{\mathbf{\alpha }}_{1}+\mathbf{\alpha }_{2}+\mathbf{\alpha }_{3}$ & $%
=$ & $\mathbf{\delta }+\mathbf{\epsilon }_{1}$ \\
$\mathbf{\alpha }_{1}+2\mathbf{\alpha }_{2}+\mathbf{\alpha }_{3}$ & $=$ & $2%
\mathbf{\mathbf{\epsilon }_{1}}$%
\end{tabular}%
\end{equation}

$\bullet $ \emph{Distinguished basis set} $\left( \mathbf{\mathbf{\alpha }}%
_{1},\mathbf{\alpha }_{2},\mathbf{\alpha }_{3}\right) $ \emph{with} $\mathbf{%
\alpha }_{2}$\ \emph{odd.}\newline
For the distinguished super Dynkin diagram with one odd root taken as $%
\mathbf{\alpha }_{2}$ (i.e: $\mathbf{\alpha }_{2}^{2}=0$) and the other two $%
\left( \mathbf{\mathbf{\alpha }}_{1},\mathbf{\alpha }_{3}\right) $ even, we
have the following realisation of the roots of $\Phi _{{\small osp(4|2)}}$
in terms of three canonical weight vectors even $\mathbf{\epsilon }_{i}$ ($%
\mathbf{\epsilon }_{1}^{2}=\mathbf{\epsilon }_{2}^{2}=1$) and odd $\mathbf{%
\delta }$ ($\mathbf{\delta }^{2}=-1$) \cite{FS1,FS2}. First, the simple
roots are realised like $\mathbf{\mathbf{\alpha }}_{1}=\mathbf{\epsilon }%
_{1}-\mathbf{\epsilon }_{2}$ and$\ \mathbf{\mathbf{\alpha }}_{2}=\mathbf{%
\delta -\mathbf{\epsilon }_{1}}$ as well as $\mathbf{\mathbf{\alpha }}_{3}=%
\mathbf{\epsilon }_{1}+\mathbf{\epsilon }_{2}$ with the automorphism
symmetry $\mathbb{Z}_{2}$ acting like
\begin{equation}
\mathbb{Z}_{2}:\quad
\begin{tabular}{lll}
$\left( \mathbf{\epsilon }_{1},\mathbf{\delta },\mathbf{\epsilon }%
_{2}\right) $ & $\qquad \rightarrow \qquad $ & $\left( \mathbf{\epsilon }%
_{1},\mathbf{\delta },-\mathbf{\epsilon }_{2}\right) $ \\
$\left( \mathbf{\mathbf{\alpha }}_{1},\mathbf{\alpha }_{2},\mathbf{\alpha }%
_{3}\right) $ & $\qquad \rightarrow \qquad $ & $\left( \mathbf{\alpha }_{3},%
\mathbf{\alpha }_{2},\mathbf{\mathbf{\alpha }}_{1}\right) $%
\end{tabular}%
\end{equation}%
and \textrm{intersection }$\mathbf{\alpha }_{i}.\mathbf{\alpha }_{j}$ as
follows%
\begin{equation}
\mathbf{\alpha }_{i}.\mathbf{\alpha }_{j}=\left(
\begin{array}{ccc}
2 & -1 & 0 \\
-1 & 0 & -1 \\
0 & -1 & 2%
\end{array}%
\right) \qquad ,\qquad \left. \mathbf{\tilde{\alpha}}_{i}.\mathbf{\tilde{%
\alpha}}_{j}\right\vert _{\zeta =1}=\mathbf{\alpha }_{i}.\mathbf{\alpha }_{j}
\label{af}
\end{equation}%
corresponding to setting $\zeta =1$ in eq(\ref{kf}). The 4 other composite
positive roots are given by%
\begin{equation}
\begin{tabular}{lll}
$\mathbf{\mathbf{\alpha }}_{1}+\mathbf{\alpha }_{2}$ & $=$ & $\mathbf{\delta
-\mathbf{\epsilon }_{2}}$ \\
$\mathbf{\alpha }_{2}+\mathbf{\alpha }_{3}$ & $=$ & $\mathbf{\delta }+%
\mathbf{\epsilon }_{2}$%
\end{tabular}%
\quad ,\quad
\begin{tabular}{lll}
$\mathbf{\mathbf{\alpha }}_{1}+\mathbf{\alpha }_{2}+\mathbf{\alpha }_{3}$ & $%
=$ & $\mathbf{\delta }+\mathbf{\epsilon }_{1}$ \\
$\mathbf{\alpha }_{1}+2\mathbf{\alpha }_{2}+\mathbf{\alpha }_{3}$ & $=$ & $2%
\mathbf{\delta }$%
\end{tabular}
\label{os2}
\end{equation}%
from which $\left( \mathbf{i}\right) $ we read the three even roots $\mathbf{%
\mathbf{\beta }}_{i}$ of $sl(2)_{\mathbf{\mathbf{\beta }}_{1}}\oplus sl(2)_{%
\mathbf{\beta }_{2}}\oplus sl(2)_{\mathbf{\beta }_{3}}$ namely $\mathbf{%
\mathbf{\beta }}_{1}=\mathbf{\mathbf{\alpha }}_{1}$ and $\mathbf{\beta }_{2}=%
\mathbf{\mathbf{\alpha }}_{3}$ and $\mathbf{\beta }_{3}=2\mathbf{\delta }$
(three even 2-cycles)$;$ and $\left( \mathbf{ii}\right) $ the four odd roots
$\mathbf{\mathbf{\gamma }}_{1}=\mathbf{\mathbf{\alpha }}_{2}$ and $\mathbf{%
\gamma }_{2}=\mathbf{\mathbf{\alpha }}_{1}+\mathbf{\alpha }_{2}$ as well as $%
\mathbf{\gamma }_{3}=\mathbf{\alpha }_{2}+\mathbf{\alpha }_{3}$ and $\mathbf{%
\gamma }_{4}=\mathbf{\mathbf{\alpha }}_{1}+\mathbf{\alpha }_{2}+\mathbf{%
\alpha }_{3}\mathbf{.}$

\subsection*{Three pairs of oscillators for osp(4\TEXTsymbol{\vert}2)}

\qquad The super osp(4\TEXTsymbol{\vert}2) splits as $osp(4|2)_{\bar{0}%
}\oplus osp(4|2)_{\bar{1}}$ with $osp(4|2)_{\bar{0}}=so(4)\oplus sp(2)$. The
oscillator realisation of the $so(4,\mathbb{C})$ generators is achieved by
two\textrm{\ pairs} of fermionic oscillators $\left( f_{1},g_{1}\right) ,$ $%
\left( f_{2},g_{2}\right) $ satisfying the anticommutation relations $%
\{f_{i},g_{j}\}=\delta _{ij}$ and $\left\{ f_{i},f_{j}\right\} =\left\{
g_{i},g_{j}\right\} =0.$ For a unitary theory, the $g_{i}$'s are identified
with $f_{i}^{\dagger }.$ For simplicity, we consider the particular
oscillator realisation with $(f_{1}^{\dagger },f_{1}),$ $(f_{2}^{\dagger
},f_{2})$ satisfying the anticommutation relations $\{f_{i},f_{j}^{\dagger
}\}=\delta _{ij}$ in terms of which the 6 generators of $so(4,\mathbb{R}%
)\simeq su(2)\oplus su(2)$ are realised as follows
\begin{equation}
\begin{tabular}{lll}
$E_{+\mathbf{\beta }_{1}}$ & $=$ & $f_{1}^{\dagger }f_{2}$ \\
$E_{-\mathbf{\beta }_{1}}$ & $=$ & $f_{2}^{\dagger }f_{1}$ \\
$H_{\mathbf{\beta }_{1}}$ & $=$ & $f_{1}^{\dagger }f_{1}-f_{2}^{\dagger
}f_{2}$%
\end{tabular}%
\qquad ,\qquad
\begin{tabular}{lll}
$E_{+\mathbf{\beta }_{2}}$ & $=$ & $f_{1}^{\dagger }f_{2}^{\dagger }$ \\
$E_{-\mathbf{\beta }_{2}}$ & $=$ & $f_{2}f_{1}$ \\
$H_{\mathbf{\beta }_{2}}$ & $=$ & $f_{1}^{\dagger }f_{1}+f_{2}^{\dagger
}f_{2}-1$%
\end{tabular}
\label{sp1}
\end{equation}%
Notice that this oscillator realisation can be motivated by the following
expressions of the simple roots in terms of the canonical weight vectors
\begin{equation}
\pm \mathbf{\mathbf{\beta }}_{1}=\pm \left( \mathbf{\epsilon }_{1}-\mathbf{%
\epsilon }_{2}\right) \qquad ,\qquad \pm \mathbf{\beta }_{2}=\pm \left(
\mathbf{\epsilon }_{1}+\mathbf{\epsilon }_{2}\right)  \label{sp2}
\end{equation}%
Similarly, the realisation of the symplectic $sp(2,\mathbb{R})$ generators
is achieved by a bosonic oscillator $b,b^{\dagger }$ satisfying $\left[
b,b^{\dagger }\right] =1$ and $\left[ b,b\right] =0.$ Here, the three
generators are given by%
\begin{equation}
\begin{tabular}{lll}
$E_{+\mathbf{\beta }_{3}}$ & $=$ & $-\frac{1}{2}b^{\dagger }b^{\dagger }$ \\
$E_{-\mathbf{\beta }_{3}}$ & $=$ & $+\frac{1}{2}bb$ \\
$H_{\mathbf{\beta }_{3}}$ & $=$ & $b^{\dagger }b+\frac{1}{2}$%
\end{tabular}%
\end{equation}%
Here also, this realisation can be motivated from the realisation $\mathbf{%
\beta }_{3}=2\mathbf{\delta .}$

The 4+4 odd generators of $osp(4|2,\mathbb{R})_{\bar{1}}$ are given by
\begin{equation}
\begin{tabular}{lll}
$E_{+\mathbf{\gamma }_{1}}$ & $=$ & $b^{\dagger }f_{1}$ \\
$E_{+\mathbf{\gamma }_{2}}$ & $=$ & $b^{\dagger }f_{2}$ \\
$E_{-\mathbf{\gamma }_{1}}$ & $=$ & $f_{1}^{\dagger }b$ \\
$E_{-\mathbf{\gamma }_{2}}$ & $=$ & $f_{2}^{\dagger }b$%
\end{tabular}%
\qquad ,\qquad
\begin{tabular}{lll}
$E_{+\mathbf{\gamma }_{3}}$ & $=$ & $b^{\dagger }f_{2}^{\dagger }$ \\
$E_{+\mathbf{\gamma }_{4}}$ & $=$ & $b^{\dagger }f_{1}^{\dagger }$ \\
$E_{-\mathbf{\gamma }_{3}}$ & $=$ & $f_{2}b$ \\
$E_{-\mathbf{\gamma }_{4}}$ & $=$ & $f_{1}b$%
\end{tabular}
\label{sp3}
\end{equation}%
From these relations we calculate the anticommutator $\left\{ E_{+\mathbf{%
\gamma }_{1}},E_{-\mathbf{\gamma }_{1}}\right\} $ of the odd Chevalley
generator giving $H_{\mathbf{\gamma }_{1}}=b^{\dagger }b+f_{1}^{\dagger
}f_{1}.$

\section{Scheneurt realisation of $\mathfrak{d}$(2,1;$\protect\zeta $)}

\qquad \label{appC} In this appendix, we give the Scheneurt method for
realising the roots of the super Lie algebra $\mathfrak{d}$(2,1;$\zeta $) in
terms of weight vectors in $\mathbb{C}^{3}$.

\subsection*{Scheneurt construction}

\qquad Following \cite{FS1,Le}, the roots of the system $\tilde{\Phi}_{%
\mathfrak{d}{\small (2,1;\zeta )}}$ can be realised by using three complex
canonical weight vectors ($\mathbf{\varepsilon }_{1},\mathbf{\varepsilon }%
_{2},\mathbf{\varepsilon }_{3}$) generating $\mathbb{C}^{3}.$ These complex
weight vectors have (non hermitian) pairing products as $\mathbf{\varepsilon
}_{i}.\mathbf{\varepsilon }_{j}=$\textsc{s}$_{i}\delta _{ij}$ in which the
\textsc{s}$_{i}$'s are non vanishing complex numbers (\textsc{s}$_{i}\in
\mathbb{C}^{\ast }$). The properties of these weight vectors go beyond the
canonical values $\pm 1$ used in sl(2\TEXTsymbol{\vert}2). So, non zero
pairings are given by
\begin{equation}
\mathbf{\varepsilon }_{1}.\mathbf{\varepsilon }_{1}=\text{\textsc{s}}%
_{1}\qquad ,\qquad \mathbf{\varepsilon }_{2}.\mathbf{\varepsilon }_{2}=\text{%
\textsc{s}}_{2}\qquad ,\qquad \mathbf{\varepsilon }_{3}.\mathbf{\varepsilon }%
_{3}=\text{\textsc{s}}_{3}  \label{3e}
\end{equation}%
In the relations the \textsc{s}$_{i}$'s are not completely free; these are
three complex numbers \textsc{s}$_{l}=\left\vert \text{\textsc{s}}%
_{l}\right\vert e^{i\vartheta _{l}}$ such as \textsc{s}$_{1}+$\textsc{s}$%
_{2}+$\textsc{s}$_{3}=0$ and \textsc{s}$_{1}\times $\textsc{s}$_{2}\times $%
\textsc{s}$_{3}\neq 0.$ These conditions are required by the graded Jacobi
identity in $\mathfrak{d}$(2,1;$\zeta $) \textrm{\cite{6,1B,Le,Mu}; a
feature nicely exhibited on the \textbf{Figure} }\textbf{\ref{3D}}. In this
regard, recall that $\mathfrak{d}$(2,1;$\zeta $) is isomorphic to the simple
Lie superalgebra $\Gamma \left( \text{\textsc{s}}_{1},\text{\textsc{s}}_{2},%
\text{\textsc{s}}_{3}\right) $ \textrm{\cite{6,Le}}. Notice also that the
above constraint relations can be solved in several ways; an interesting way
is given by restricting to \textsc{s}$_{3}/$\textsc{s}$_{1}=\zeta $ and
\textsc{s}$_{2}/$\textsc{s}$_{1}=-1-\zeta $ with complex $\zeta \neq 0,-1$
due to \textsc{s}$_{1}$\textsc{s}$_{2}$\textsc{s}$_{3}\neq 0$. With this
choice, we have%
\begin{equation}
\frac{\text{\textsc{s}}_{1}\text{\textsc{s}}_{2}\text{\textsc{s}}_{3}}{\text{%
\textsc{s}}_{1}^{3}}=-\zeta \left( 1+\zeta \right)  \label{is}
\end{equation}%
having zeros at $\zeta =0,-1$. It breaks the $\mathcal{S}_{3}$ symmetry down
to its subgroup $\mathcal{Z}_{2}$ generated by the transposition $\zeta
\rightarrow -1-\zeta $; it has a fix point at $\zeta =-1/2$ (\textsc{s}$%
_{2}= $\textsc{s}$_{3}$) and exchanges the two singularities $\zeta =0,-1$
as depicted b\textrm{y the }\textbf{Figure} \textbf{\ref{area}}. Using the
weight vectors ($\mathbf{\varepsilon }_{1},\mathbf{\varepsilon }_{2},\mathbf{%
\varepsilon }_{3}$), the simple roots $\left( \mathbf{\tilde{\alpha}}_{1},%
\mathbf{\tilde{\alpha}}_{2},\mathbf{\tilde{\alpha}}_{3}\right) $ generating
the basis set $\Pi _{\mathfrak{d}{\small (2,1;\zeta )}}^{2}$ are realised as
follows%
\begin{equation}
\begin{array}{lll}
\mathbf{\tilde{\alpha}}_{1} & = & 2\mathbf{\varepsilon }_{1} \\
\mathbf{\tilde{\alpha}}_{2} & = & \mathbf{\varepsilon }_{2}-\mathbf{%
\varepsilon }_{1}-\mathbf{\varepsilon }_{3} \\
\mathbf{\tilde{\alpha}}_{3} & = & 2\mathbf{\varepsilon }_{3}%
\end{array}%
\qquad ,\qquad
\begin{array}{lll}
\mathbf{\tilde{\alpha}}_{1}.\mathbf{\tilde{\alpha}}_{1} & = & 4\text{\textsc{%
s}}_{1} \\
\mathbf{\tilde{\alpha}}_{2}.\mathbf{\tilde{\alpha}}_{2} & = & 0 \\
\mathbf{\tilde{\alpha}}_{3}.\mathbf{\tilde{\alpha}}_{3} & = & 4\text{\textsc{%
s}}_{3}%
\end{array}
\label{rea}
\end{equation}%
with the properties
\begin{equation}
\begin{tabular}{lllllll}
$\mathbf{\tilde{\alpha}}_{1}+\mathbf{\tilde{\alpha}}_{2}+\mathbf{\tilde{%
\alpha}}_{3}$ & $=$ & $\mathbf{\varepsilon }$ & $\quad ,\quad $ & $\mathbf{%
\varepsilon }$ & $=$ & $\mathbf{\varepsilon }_{1}+\mathbf{\varepsilon }_{2}+%
\mathbf{\varepsilon }_{3}$ \\
$\mathbf{\tilde{\alpha}}_{1}^{2}+\mathbf{\tilde{\alpha}}_{2}^{2}+\mathbf{%
\tilde{\alpha}}_{3}^{2}$ & $=$ & $-4$\textsc{s}$_{2}$ & $\quad ,\quad $ & $%
\mathbf{\varepsilon }^{2}$ & $=$ & $0$%
\end{tabular}%
\end{equation}%
The intersection matrix $\widetilde{\mathcal{I}}_{ij}$ associated with the
above realisation (\ref{rea}) reads as $\ $%
\begin{equation}
\widetilde{\mathcal{I}}_{ij}\left( \text{\textsc{s}}\right) =\left(
\begin{array}{ccc}
4\text{\textsc{s}}_{1} & {\small -2}\text{\textsc{s}}_{1} & {\small 0} \\
{\small -2}\text{\textsc{s}}_{1} & {\small 0} & {\small -2}\text{\textsc{s}}%
_{3} \\
{\small 0} & {\small -2}\text{\textsc{s}}_{3} & 4\text{\textsc{s}}_{3}%
\end{array}%
\right) \qquad ,\qquad \widetilde{\mathcal{I}}_{ij}\left( \zeta \right)
=\left(
\begin{array}{ccc}
{\small 2} & {\small -1} & {\small 0} \\
{\small -1} & {\small 0} & {\small -\zeta } \\
{\small 0} & {\small -\zeta } & {\small 2\zeta }%
\end{array}%
\right)  \label{cr}
\end{equation}%
\begin{equation*}
\end{equation*}%
with the property $\sum_{j=1}^{3}\widetilde{\mathcal{I}}_{ij}\left( \text{%
\textsc{s}}\right) =2$\textsc{s}$_{i}.$ By setting $2$\textsc{s}$_{1}=1$ and
$2$\textsc{s}$_{2}=-1-\zeta $ as well as $2$\textsc{s}$_{3}=\zeta ,$ the
intersection $\widetilde{\mathcal{I}}_{ij}\left( \zeta \right) $ with
determinant $\det \widetilde{\mathcal{I}}_{ij}=-2\zeta \left( \zeta
+1\right) .$ The super Cartan $\widetilde{\mathcal{K}}\left( \zeta \right) $
is given by $\widetilde{\mathcal{J}}.\widetilde{\mathcal{I}}\left( \zeta
\right) $ with \textrm{\cite{FS1,FS2}}%
\begin{equation}
\mathcal{\tilde{J}}=\frac{-1}{1+\zeta }\left(
\begin{array}{ccc}
\frac{4+\zeta }{2} & {\small 3} & \frac{{\small 3}}{2} \\
\frac{2+\zeta }{2} & 2+\zeta & \frac{{\small 1}}{2} \\
\frac{2+\zeta }{2} & {\small 1} & -\frac{2+\zeta }{2\zeta }%
\end{array}%
\right) \qquad ,\qquad \mathcal{\tilde{K}}_{ij}=\left(
\begin{array}{ccc}
{\small 2} & {\small -1} & {\small 0} \\
{\small 1} & {\small 0} & {\small \zeta } \\
{\small 0} & {\small -1} & {\small 2}%
\end{array}%
\right)
\end{equation}%
\begin{equation*}
\end{equation*}%
Comparing these matrices of $\mathfrak{d}$($2,1;\zeta $) with their sl(2%
\TEXTsymbol{\vert}2) homologue given by the relations (\ref{DR}), we see
that they can be recovered from the above (\ref{rea}) by setting $\zeta =-1.$
For the two singular points $\zeta =-1,$ we have critical matrices%
\begin{equation}
\zeta =-1:\quad \widetilde{\mathcal{I}}_{ij}=\left(
\begin{array}{ccc}
2 & -1 & 0 \\
-1 & 0 & 1 \\
0 & 1 & -2%
\end{array}%
\right)  \label{ci}
\end{equation}

\subsection*{Roots system in terms of weight vectors}

\qquad Here, we give the expression of the 7 positive roots $\mathbf{\tilde{%
\alpha}}$ in terms of the canonical weight vectors%
\begin{equation}
\begin{array}{lll}
\mathbf{\tilde{\alpha}}_{1} & = & 2\mathbf{\varepsilon }_{1} \\
\mathbf{\tilde{\alpha}}_{2} & = & \mathbf{\varepsilon }_{2}-\mathbf{%
\varepsilon }_{1}-\mathbf{\varepsilon }_{3} \\
\mathbf{\tilde{\alpha}}_{3} & = & 2\mathbf{\varepsilon }_{3} \\
\mathbf{\tilde{\alpha}}_{1}+\mathbf{\tilde{\alpha}_{2}} & = & \mathbf{%
\varepsilon }_{1}+\mathbf{\varepsilon }_{2}-\mathbf{\varepsilon }_{3} \\
\mathbf{\tilde{\alpha}_{2}+\tilde{\alpha}}_{3} & = & \mathbf{\varepsilon }%
_{2}-\mathbf{\varepsilon }_{1}+\mathbf{\varepsilon }_{3} \\
\mathbf{\tilde{\alpha}_{1}}+\mathbf{\tilde{\alpha}}_{2}+\mathbf{\tilde{\alpha%
}}_{3} & = & \mathbf{\varepsilon }_{1}+\mathbf{\varepsilon }_{2}+\mathbf{%
\varepsilon }_{3} \\
\mathbf{\tilde{\alpha}_{1}}+2\mathbf{\tilde{\alpha}}_{2}+\mathbf{\tilde{%
\alpha}}_{3} & = & 2\mathbf{\varepsilon }_{2}%
\end{array}
\label{RD}
\end{equation}%
Notice that using these weight vectors, the long root of $\mathfrak{d}$(2,1;$%
\zeta $) reads as%
\begin{equation}
\mathbf{\mathbf{\tilde{\psi}}}=\mathbf{\tilde{\alpha}_{1}}+2\mathbf{\tilde{%
\alpha}}_{2}+\mathbf{\tilde{\alpha}}_{3}  \label{psi}
\end{equation}%
it can be split in two ways like the sum of odd roots; either as $\mathbf{%
\tilde{\alpha}}_{2}+\left( \mathbf{\tilde{\alpha}}_{1}+\mathbf{\tilde{\alpha}%
}_{2}+\mathbf{\tilde{\alpha}}_{3}\right) ;$ or like $\left( \mathbf{\tilde{%
\alpha}}_{1}+\mathbf{\tilde{\alpha}}_{2}\right) +\left( \mathbf{\tilde{\alpha%
}}_{2}+\mathbf{\tilde{\alpha}}_{3}\right) .$ Using eq(\ref{psi}), we can
express the odd root $\mathbf{\tilde{\alpha}}_{2}$ in terms of halfs of the
bosonic $\mathbf{\tilde{\alpha}_{1}},$ $\mathbf{\tilde{\alpha}}_{3}$ and $%
\mathbf{\mathbf{\tilde{\psi}}}$ like%
\begin{equation}
\mathbf{\tilde{\alpha}}_{2}=\frac{1}{2}\mathbf{\mathbf{\tilde{\psi}}}-\frac{1%
}{2}\mathbf{\tilde{\alpha}_{1}}-\frac{1}{2}\mathbf{\tilde{\alpha}}_{3}
\end{equation}%
So, instead of (\ref{RD}), we can present the roots into representations of $%
\mathcal{S}_{3}$. For the 6 even roots, we have the triplet%
\begin{equation}
\begin{array}{lll}
\pm \mathbf{\tilde{\alpha}}_{1} & = & \pm 2\mathbf{\varepsilon }_{1} \\
\pm \mathbf{\mathbf{\tilde{\psi}}} & \mathbf{\mathbf{=}} & \pm 2\mathbf{%
\varepsilon }_{2} \\
\pm \mathbf{\tilde{\alpha}}_{3} & = & \pm 2\mathbf{\varepsilon }_{3}%
\end{array}%
\qquad ,\qquad
\begin{array}{lll}
\mathbf{\tilde{\alpha}}_{1}^{2} & = & 4\text{\textsc{s}}_{1} \\
\mathbf{\mathbf{\tilde{\psi}}}^{2} & \mathbf{\mathbf{=}} & 4\text{\textsc{s}}%
_{2} \\
\mathbf{\tilde{\alpha}}_{3}^{2} & = & 4\text{\textsc{s}}_{3}%
\end{array}
\label{re1}
\end{equation}%
For the 8 odd roots, they split into two singlets $\pm \mathbf{\tilde{\gamma}%
}_{0}$ with%
\begin{equation}
\begin{array}{lll}
\mathbf{\tilde{\gamma}}_{0} & \mathbf{=} & \mathbf{\varepsilon }_{2}+\mathbf{%
\varepsilon }_{1}+\mathbf{\varepsilon }_{3} \\
& \mathbf{=} & \frac{1}{2}\mathbf{\tilde{\alpha}_{1}}+\frac{1}{2}\mathbf{%
\mathbf{\tilde{\psi}}}+\frac{1}{2}\mathbf{\tilde{\alpha}}_{3}%
\end{array}%
\end{equation}%
and two triplets $\pm \mathbf{\tilde{\gamma}}_{i}$ with
\begin{equation}
\begin{array}{lllll}
\mathbf{\tilde{\gamma}}_{1} & \mathbf{=} & +\frac{1}{2}\mathbf{\tilde{\alpha}%
}_{1}-\frac{1}{2}\mathbf{\mathbf{\tilde{\psi}}}-\frac{1}{2}\mathbf{\tilde{%
\alpha}}_{3} & = & +\mathbf{\varepsilon }_{1}-\mathbf{\varepsilon }_{2}-%
\mathbf{\varepsilon }_{3} \\
\mathbf{\tilde{\gamma}}_{2} & \mathbf{=} & -\frac{1}{2}\mathbf{\tilde{\alpha}%
_{1}-}\frac{1}{2}\mathbf{\mathbf{\tilde{\psi}}}+\frac{1}{2}\mathbf{\tilde{%
\alpha}}_{3} & = & -\mathbf{\varepsilon }_{1}-\mathbf{\varepsilon }_{2}+%
\mathbf{\varepsilon }_{3} \\
\mathbf{\tilde{\gamma}}_{3} & \mathbf{=} & -\frac{1}{2}\mathbf{\tilde{\alpha}%
_{1}+}\frac{1}{2}\mathbf{\mathbf{\tilde{\psi}}}-\frac{1}{2}\mathbf{\tilde{%
\alpha}}_{3} & = & -\mathbf{\varepsilon }_{1}+\mathbf{\varepsilon }_{2}-%
\mathbf{\varepsilon }_{3}%
\end{array}
\label{re2}
\end{equation}

\section{Integrable superspin chains in CS theory}

\qquad \label{appD} In this appendix, we provide tools for the study of
integrable superspin chains \textrm{\cite{VAN1,VAN2}} by the help of CS
gauge potentials \textrm{\cite{1AA,2AA}}. For concreteness, we focus on the
integrable SL(2\TEXTsymbol{\vert}2) superspin chain as it is one of the
systems considered in this study. For more technical details and other
applications, we report to \textrm{\cite{3AA}} and refs therein.

\subsection*{SL(2\TEXTsymbol{\vert}2) potentials in 4D CS theory}

\qquad The Lie superalgebra sl(2\TEXTsymbol{\vert}2) of the Lie supergroup
SL(2\TEXTsymbol{\vert}2) over the complex field $\mathbb{C}$ has rank 3 and
15 dimensions. It belongs to the sub-family sl(n\TEXTsymbol{\vert}m) with $%
m=n\geq 2$ having a central charge generator. Apart from the very special
sl(1\TEXTsymbol{\vert}1) member, the super sl(2\TEXTsymbol{\vert}2)
constitutes the leading symmetry of superspin chains described by 4D
Chern-Simons 1-form potential $A\left( \mathrm{x}\right) $ with expansion
\begin{equation}
A\left( \mathrm{x}\right) =\mathfrak{A}_{0}\left( \mathrm{x}\right)
H_{0}+\dsum\limits_{even\text{ root }\mathbf{\beta }}\left[ A_{\mathbf{\beta
}}^{0}\left( \mathrm{x}\right) H_{\mathbf{\beta }}+A_{\mathbf{\beta }}^{\pm
}\left( \mathrm{x}\right) E_{\pm \mathbf{\beta }}\right] +\dsum\limits_{odd%
\text{ root }\mathbf{\gamma }}A_{\mathbf{\gamma }}^{\mp }\left( \mathrm{x}%
\right) F_{\pm \mathbf{\gamma }}  \label{AX}
\end{equation}%
where the $H_{0},$ $H_{\mathbf{\beta }},$ $E_{\pm \mathbf{\beta }}$ and $%
F_{\pm \mathbf{\gamma }}$ are matrix generators and where $\mathfrak{A}%
_{0}\left( \mathrm{x}\right) ,$ $A_{\mathbf{\beta }}^{0}\left( \mathrm{x}%
\right) ,$\ $A_{\mathbf{\beta }}^{\pm }\left( \mathrm{x}\right) $\ and $A_{%
\mathbf{\gamma }}^{\mp }\left( \mathrm{x}\right) $\ are (partial) 1-form
potentials along the 15 directions of sl(2\TEXTsymbol{\vert}2). The above 4D
CS gauge field $A\left( \mathrm{x}\right) $ has a central potential
\begin{equation*}
\mathfrak{A}_{0}\left( \mathrm{x}\right) H_{0}
\end{equation*}%
which is absent for the bigger family of sl(n\TEXTsymbol{\vert}m) superspin
chains with $n\neq m$. Recall that sl(n\TEXTsymbol{\vert}m) contains two
main sub-families labeled by $n\neq m$ and $n=m.$ The dynamics of $A\left(
\mathrm{x}\right) $ is described by the gauge field action%
\begin{equation}
S_{4dCS}=\int_{\Sigma \times \boldsymbol{C}}dz\wedge tr(A\wedge dA+\frac{2}{3%
}A\wedge A\wedge A)  \label{act}
\end{equation}%
where $\Sigma $ is a topological real plane (roughly $\mathbb{R}^{2}$)
parametrised by real variables (x,y); and a holomorphic curve $\boldsymbol{C}
$ (say $\mathbb{CP}^{1}$) labeled by the complex coordinate $z$ (rapidity)
\textrm{\cite{1AA,2AA,S}}. Notice that the gauge potential $A\left( \mathrm{x%
}\right) $ is a partial 1-form potential; it is given by $A_{x}dx+A_{y}dy+A_{%
\bar{z}}d\bar{z}$ where we have dropped the component $A_{z}dz$ because of
the $dz$ factor in the\textrm{\ }holomorphic volume form measure in\textrm{\
(\ref{act})}.

The field equation of motion following from the $S_{4dCS}$ is given by the
zero of the 2-form curvature; i.e:%
\begin{equation}
F=dA+A\wedge A=0
\end{equation}%
indicating that observables constructed out of it are trivial.

\subsection*{Line defects and RLL equation}

\begin{figure}[h]
\begin{center}
\includegraphics[width=13cm]{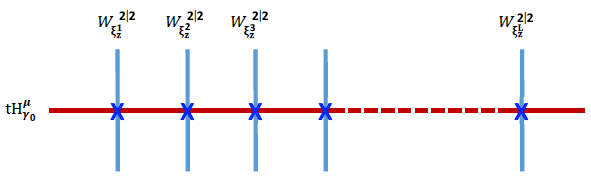}
\end{center}
\par
\vspace{-0.5cm}
\caption{Realization of an $sl(2|2)$ superspin chain of $L$ nodes in the
fundamental representation. In the 4D Chern-Simons theory, it is described
by L vertical super Wilson lines (in blue) crossing a horizontal 't Hooft
line (in red).}
\label{superchain}
\end{figure}

\begin{figure}[h]
\begin{center}
\includegraphics[width=11cm]{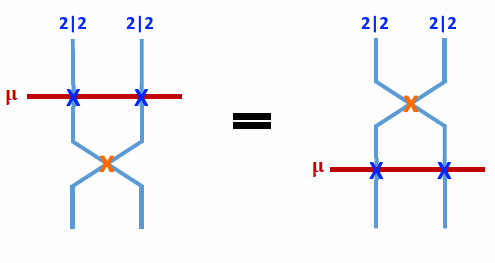}
\end{center}
\par
\vspace{-0.9cm}
\caption{Graphic representation of the RLL equation in terms of intersecting
line defects in $SL(2|2)$ 4D CS theory.}
\label{RLL}
\end{figure}

Despite that the 2-form field strength $F=0,$ there are still observables
describing physical entities. An interesting example is given by super
Wilson loops $W_{\xi _{z}}^{\boldsymbol{2|2}}$ defined by the supertrace of
the holonomy of the gauge field $A$\textrm{\ along a loop} $\xi _{z}$ in the
4D space $\mathbb{R}^{2}\times \mathbb{CP}^{1}$ as follows \textrm{\cite%
{BF,W3,W4}}
\begin{equation}
W_{\xi _{z}}^{\boldsymbol{2|2}}=str_{\boldsymbol{2|2}}\left[ P\exp \left(
\oint_{{\xi _{z}}}A\right) \right]
\end{equation}%
It plays a crucial role in dealing with (super) Yang Baxter equation of 2D
(super) integrable models. Another interesting example is given by the super
't Hooft lines $\mathcal{L}^{\mathbf{\mu }}(z)$ carrying a minuscule
coweight charge $\mathbf{\mu }$; it is defined as \textrm{\cite{TH1}-\cite%
{TH4}}%
\begin{equation}
\mathcal{L}^{\mathbf{\mu }}(z)=P\exp \left( \int_{y}A_{y}(z)\right)
\label{L1}
\end{equation}%
where the transport of the gauge fields $A_{y}(z)$ is measured from $y<0$ to
$y>0.$ The couplings of these topological super line defects is given by
line's crossings and are effectively described by the Yang Baxter R-matrix
and the so called Lax operator \textrm{\cite{S,BY,Lax1}}. A typical example
of such coupling is given by the sl(2\TEXTsymbol{\vert}2) superchain
depicted by the \textbf{Figure} \textbf{\ref{superchain}},
where the 't Hooft line plays\ the role of a transfer matrix modeling the
interactions between the electrically charged super atoms\ along the
superchain. In this realisation, each intersection of an electric $W_{\xi
_{z}}^{\boldsymbol{2|2}}$, \textrm{carrying a vector quantum space} $%
\boldsymbol{2|2}$, with the magnetic tH$_{\mathrm{\gamma }_{0}}^{\mathbf{\mu
}}$ \textrm{carrying an auxiliary (oscillator) space }{\large A}, yields the
super Lax operator for the corresponding node of the superspin chain. This
coupling operator acts on the tensor product of $End(\boldsymbol{2|2})$ of
the Wilson and the algebra {\large A} of functions in the phase space of the
't Hooft line\textrm{. As for the Yang-Baxter equation, the crossing of
Wilson and 't Hooft lines verifies the RLL equation of integrability namely}%
\begin{equation}
\boldsymbol{R}_{rs}^{ik}\left( z-w\right) \boldsymbol{L}_{j}^{r}\left(
z\right) \boldsymbol{L}_{l}^{s}\left( w\right) =\boldsymbol{L}_{r}^{i}\left(
w\right) \boldsymbol{L}_{s}^{k}\left( z\right) \boldsymbol{R}%
_{jl}^{rs}\left( z-w\right)  \label{lax}
\end{equation}%
In this relation, the $\boldsymbol{L}_{n}^{m}\left( z\right) $\textrm{\ is
the matrix realization of the L-operator and }$\boldsymbol{R}%
_{rs}^{ik}\left( z-w\right) $\textrm{\ is the usual R-matrix solving} the
Yang-Baxter equation. This RLL equation is represented by the graph of the
\textbf{Figure\ \ref{RLL}}.
For further details and applications, we refer to \textrm{\cite{1AA,2AA}}
for the general set up of the 4D Chern-Simons (CS) theory. Also, we refer to
\cite{3AA,IS1} for explicit field realisations, examples of superspin chains
and detailed calculations. For applications in strings, see for instance
\cite{Yama,IS1} and \textrm{\cite{t1,t2,t3,t4}}.

Concerning the algebraic structure of the 1-form graded potential $A\left(
\mathrm{x}\right) $ given by eq(\ref{AX}), it has two blocs $A^{{\small %
(even)}}\left( \mathrm{x}\right) $ and $A^{{\small (even)}}\left( \mathrm{x}%
\right) $. In this regard, recall that the even part sl(2\TEXTsymbol{\vert}2)%
$_{\bar{0}}$ of the superalgebra sl(2\TEXTsymbol{\vert}2) is given by $%
sl(2)\oplus sl(2)\oplus \mathbb{C}$I$_{id}$ where $sl(2)$ is the usual
complex sl(2$,\mathbb{C}$) and the block $\mathbb{C}$I$_{id}$ is a central
element spanned by the identity operator I$_{id}\mathbf{\ }$commuting with
all other generators; $\left[ I_{id},sl(2)\right] =0.$ The sl(2\TEXTsymbol{%
\vert}2)$_{\bar{0}}$ is 7 dimensional (7= 3+3+1). The odd part sl(2%
\TEXTsymbol{\vert}2)$_{\bar{1}}$ is 8 dimensional generated by fermionic
charges $F_{\pm \mathbf{\gamma }}$. The CS potentials associated with these
even and odd parts are given by%
\begin{equation}
A^{{\small (even)}}=\mathfrak{A}_{0}\left( \mathrm{x}\right)
H_{0}+\dsum\limits_{even\text{ root }\mathbf{\beta }}\left( A_{\mathbf{\beta
}}\left( \mathrm{x}\right) H_{\mathbf{\beta }}+A_{\mp \mathbf{\beta }}\left(
\mathrm{x}\right) E_{\pm \mathbf{\beta }}\right)
\end{equation}%
and%
\begin{equation}
A^{{\small (odd)}}=\dsum\limits_{odd\text{ root }\mathbf{\gamma }}A_{\mp
\mathbf{\gamma }}\left( \mathrm{x}\right) F_{\pm \mathbf{\gamma }}
\end{equation}%
Their topological 2-form field strengths are%
\begin{equation}
\begin{tabular}{lll}
$F^{{\small (even)}}$ & $=$ & $dA^{{\small (even)}}+A^{{\small (even)}%
}\wedge A^{{\small (even)}}+A^{{\small (odd)}}\wedge A^{{\small (odd)}}$ \\
$F^{{\small (odd)}}$ & $=$ & $dA^{{\small (odd)}}+A^{{\small (odd)}}\wedge
A^{{\small (even)}}+A^{{\small (even)}}\wedge A^{{\small (odd)}}$%
\end{tabular}%
\end{equation}%
they vanish on shell. The explicit expression of the Lax operator $%
\boldsymbol{L}_{j}^{r}\left( z\right) $ for the open super Dynkin diagram of
the superspin chain $\mathfrak{d}$(2,1:$\zeta $) solving (\ref{lax}) is
given by eq(\ref{ed}).


\begin{thebibliography}{99}
\bibitem{1A} V.G. Kac, \emph{Lie superalgebras}, Advances in Math. 26 (1977)
8-96 MR0486011.

\bibitem{2A} I. Heckenberger, F. Spill, A. Torrielli, H. Yamane, \emph{%
Drinfeld second realization of the quantum affine superalgebras of }$%
D(2,1;x)^{(1)}$\emph{\ via the Weyl groupoid,} Publ.Res.Inst.Math.Sci.Kyoto
B8:171,2008, arXiv:0705.1071 [math.QA].

\bibitem{3A} E. Poletaeva, \emph{Embedding of the Lie superalgebra} $%
D(2,1;\alpha )$ \emph{into the Lie superalgebra of pseudodifferential
symbols on} $S^{1|2}$, J. Math. Phys. 48 (2007) 103504, 17 pp. e-print
arXiv:0709.0083.

\bibitem{4A} J. de Boer, A. Pasquinucci and K. Skenderis, \textquotedblleft
\emph{AdS/CFT dualities involving large 2D N=4 superconformal symmetry,}%
\textquotedblright\ Adv. Theor. Math. Phys. 3 (1999) 577
[arXiv:hep-th/9904073].

\bibitem{5A} S. Gukov, E. Martinec, G.W. Moore and A. Strominger,
\textquotedblleft \emph{The Search for a holographic dual to} $AdS_{3}\times
S^{3}\times S^{3}\times S^{1},$\textquotedblright\ Adv. Theor. Math. Phys. 9
(2005) 435 [hep-th/0403090].

\bibitem{6A} Lorenz Eberhardta, Matthias R. Gaberdiela and Wei Li, \emph{A
holographic dual for string theory on} $AdS_{3}\times S^{3}\times
S^{3}\times S^{1},$ J. High Energ. Phys. 2017, 111 (2017), \emph{\ }%
arXiv:1707.02705 [hep-th].

\bibitem{BHH} \textrm{Niklas Beisert, Reimar Hecht, Ben Hoare}, \emph{%
Maximally extended} $sl(2|2),$ \emph{q-deforme}d $d(2,1;\varepsilon )$ \emph{%
and 3D kappa-Poincar\'{e}}, J. Phys. A. 50, 314003 (2017),
arXiv:1704.05093v2 [math-ph].

\bibitem{7A} \textrm{\ }N. Beisert and M. de Leeuw, \emph{The RTT
realization for the deformed} $gl(2|2)$ \emph{Yangian}, J. Phys.A: Math.
Theor. 47, 305201 (2014), doi:10.1088/1751-8113/47/30/305201.

\bibitem{8A} N. Beisert, M. de Leeuw and R. Hecht, \emph{Maximally extended}
$sl(2|2)$ \emph{as a quantum double}, J. Phys. A: Math. Theor. 49, 434005
(2016), doi:10.1088/1751-8113/49/43/434005.

\bibitem{relform} M. Parker, \textquotedblleft \emph{Classification of real
simple Lie superalgebras of classical type},\textquotedblright\ J. Math.
Phys., vol. 21, no. 4, pp. 689-697, 1980.

\bibitem{AdS1} Antonio Pittelli, \emph{Yangian Symmetry of String Theory on}
$AdS_{3}\times S^{3}\times S^{3}\times S^{1}$ \emph{with Mixed 3-form Flux},
Nuclear Physics B 935, 2018, 271-289, arXiv:1711.02468v2 [hep-th].

\bibitem{AdS2} I. Bars, C. Deliduman, D. Minic, \emph{String Theory on AdS}$%
_{3}$\emph{\ Revisited}, arXiv:hep-th/9907087v2 .

\bibitem{AdS3} \textrm{C. Bachas, E. D'Hoker, J. Estes, and D. Krym},
\textquotedblleft M-theory solutions invariant under $D(2,1;\gamma )\oplus
D(2,1;\gamma )$,\textquotedblright\ Fortschr. Phys. 62(3), 207
(2014).https://doi.org/10.1002/prop.201300039

\bibitem{1AB} Lorenz Eberhardt, Matthias R. Gaberdiel, \emph{Strings on} $%
AdS_{3}\times S^{3}\times S^{3}\times S^{1}$. J. High Energ. Phys. 2019, 35
(2019)., arXiv:1904.01585v1 [hep-th].

\bibitem{1ABA} G. Giribet, C. Hull, M. Kleban, M. Porrati, E. Rabinovici,
\emph{Superstrings on AdS}$_{3}$\emph{\ at k=1},\ JHEP08(2018)204,\
arXiv:1803.04420v3 [hep-th].

\bibitem{2ABB} Lorenz Eberhardt, Matthias R. Gaberdiel, \emph{String theory
on AdS3 and the symmetric orbifold of Liouville theory}, Nuclear Physics B
948, 2019, 114774, arXiv:1903.00421v1 [hep-th].

\bibitem{3AB} Gaston Giribet, \emph{String correlators in AdS3 from FZZ
duality}, JHEP12(2021)012, arXiv:2110.04197v2 [hep-th].

\bibitem{4AB} Lorenz Eberhardt, Matthias R. Gaberdiel, Rajesh Gopakumar,
\emph{Deriving the AdS3/CFT2 Correspondence}, J. High Energ. Phys. 2020, 136
(2020), arXiv:1911.00378v2 [hep-th].

\bibitem{Rajae1} Sammani, R., Boujakhrout, Y., Saidi, E. H., Laamara, R. A.,
\& Drissi, L. B. (2023). \emph{Higher spin AdS 3 gravity and Tits-Satake
diagrams}. Physical Review D, 108(10), 106019.

\bibitem{Rajae2} R. Sammani, E. H Saidi, \emph{Black Flowers and Real Forms
of Higher Spin Symmetries}, \textrm{JHEP (2024)}, arXiv:2406.01328v1
[hep-th].\newline
\emph{Finiteness of 3D higher spin gravity Landscape}, Classical and Quantum
Gravity 2024, DOI 10.1088/1361-6382/ad7cba

\bibitem{9A} Niklas Beisert, Egor Im, \emph{Classical Lie Bialgebras for
AdS/CFT Integrability by Contraction and Reduction},\textrm{\ SciPost Phys.
14, 157 (2023),} arXiv:2210.11150v2 [hep-th].

\bibitem{10A} Alessandro Torrielli, \emph{Review of AdS/CFT Integrability,
Chapter VI.2: Yangian Algebra}, Letters in Mathematical Physics 99(1),
arXiv:1012.4005v5 [hep-th].

\bibitem{10AT} Andrea Prinsloo, Vidas Regelskis, Alessandro Torrielli, \emph{%
Integrable open spin-chains in AdS3/CFT2 correspondences}, Phys. Rev. D 92,
106006 (2015), arXiv:1505.06767v3 [hep-th].

\bibitem{VAN1} N.I. Stoilova and J. Van der Jeugt, \emph{A classification of
generalized quantum statistics associated with basic classical Lie
superalgebras}, J. Math. Phys. 46 (2005) 113504[math-ph/0504013].

\bibitem{VAN2} N.I. Stoilova and J. Van der Jeugt, \emph{A classification of
generalized quantum statistics associated with the exceptional Lie
(super)algebras}, J. Math. Phys. 48 (2007) 043504 [math-ph/0611085].

\bibitem{10AA} Andrea Fontanella, Alessandro Torrielli, \emph{Massless
sector of AdS}$_{3}$\emph{\ superstrings: a geometric interpretation}, Phys.
Rev. D 94, 066008 (2016), arXiv:1608.01631v3 [hep-th].

\bibitem{fr} \textrm{R. Frassek and A. Tsymbaliuk}, \textquotedblleft
Rational Lax matrices from antidominantly shifted extended Yangians: BCD
types,\textquotedblright\ Commun. Math. Phys. 392, 545
(2022).https://doi.org/10.1007/s00220-022-04345-6

\bibitem{fr1} V.V. Bazhanov, R. Frassek, T.L. ukowski, C. Meneghelli, M.
Staudacher, Baxter Q-operators and representations of Yangians, Nucl. Phys.
B 850 (2011) 148, arXiv :1010 .3699.

\bibitem{fr2} R. Frassek, Oscillator realisations associated to the D-type
Yangian: towards the operatorial Q-system of orthogonal spin chains, Nuclear
Phys. B 956 (2020) 115063, 22 [2001.06825].

\bibitem{fr3} \textrm{G. Ferrando, R. Frassek, and V. Kazakov},
\textquotedblleft QQ-system and Weyl-type transfer matrices in integrable
SO(2r) spin chains,\textquotedblright\ J. High Energ. Phys. 2021, 193
(2021).https://doi.org/10.1007/JHEP02(2021)193.

\bibitem{ADE1} Y. Boujakhrout, E.H. Saidi, R.A. Laamara and L.B. Drissi,
\emph{'t Hooft lines of ADE-type and topological quivers}, SciPost Phys. 15
(2023) 078 [arXiv:2303.13879].

\bibitem{ADE2} Youssra Boujakhrout, El Hassan Saidi, \emph{Minuscule ABCDE
Lax Operators from 4D Chern-Simons Theory}, Nucl.Phys.B 981 (2022) 115859,
arXiv:2207.14777 [hep-th]

\bibitem{ODE} Katsushi Ito, Mingshuo Zhu, \emph{ODE/IM correspondence and
supersymmetric affine Toda field equations}, Nucl.Phys B985, December 2022,
116004, arXiv:2206.08024v2 [hep-th].

\bibitem{IS1} N. Ishtiaque, S.F. Moosavian, S. Raghavendran and J. Yagi,
Superspin chains from superstring theory, SciPost Phys. 13 (2022) 083
[arXiv:2110.15112].

\bibitem{SYM} Niklas Beisert, Matthias Staudacher, \emph{The N=4 SYM
Integrable Super Spin Chain}, Nuclear Physics B670, 2003, p 439-463,
arXiv:hep-th/0307042v3.

\bibitem{GR} Daniel Arnaudon, Jean Avan, Nicolas Cramp\'{e}, Anastasia
Doikou, Luc Frappat, Eric Ragoucy,\emph{\ Bethe Ansatz equations and exact S
matrices for the osp(M\TEXTsymbol{\vert}2n) open super spin chain}, \emph{%
bethe ansatz an algeb methods, }Nucl.Phys. B687 (2004) 257-278,
arXiv:math-ph/0310042v2.

\bibitem{1C} N. Beisert, The $su(2|2)$ dynamic S-matrix, Adv. Theor. Math.
Phys. 12 (2008) 945--979. e-print arXiv:hep-th/0511082.

\bibitem{1AA} K. Costello, E. Witten and M. Yamazaki, \emph{Gauge theory and
integrability}, I, ICCM Not. 6 (2018) 46, arXiv:1709.09993 [hep-th].

\bibitem{2AA} K. Costello, E. Witten and M. Yamazaki, \emph{Gauge theory and
integrability}, II, ICCM Not. 6 (2018) 120 arXiv:1802.01579 [hep-th].

\bibitem{CYW} K. Costello, M. Yamazaki, Gauge Theory And Integrability, III,
1908.02289.

\bibitem{S} E.H. Saidi, \emph{Quantum line operators from Lax pairs}, J.
Math. Phys. 61 (2020) 063501, arXiv :1812 .06701[hep -th].

\bibitem{2AB} Kevin Costello, Bogdan Stefa\'{n}ski jr, \emph{The
Chern-Simons Origin of Superstring Integrability}, Phys. Rev. Lett. 125,
121602 (2020), arXiv:2005.03064v3 [hep-th].

\bibitem{W0} \textrm{E. Witten}, \textquotedblleft Integrable lattice models
from gauge theory,\textquotedblright\ Adv. Theor. Math. Phys. 21(7), 1819
(2017).

\bibitem{FS1} L. Frappat, P. Sorba, \emph{Dictionary on Lie Superalgebras},
ENSLAPP-AL-600/96 and DSF-T-30/96, arXiv:hep-th/9607161.

\bibitem{FS2} L. Frappat, A. Sciarrino, and P. Sorba,\emph{\ Structure of
Basic Lie Superalgebras and of their Affine Extensions}, Commun. Math. Phys.
121, 457-500 (1989).

\bibitem{3AA} Youssra Boujakhrout, El Hassan Saidi, Rachid Ahl Laamara,
Lalla Btissam Drissi, \emph{Superspin Chains Solutions from 4D Chern-Simons
Theory}, J. High Energ. Phys. 2024, 43 (2024), arXiv:2309.04337v2 [hep-th].

\bibitem{Y1} Elena Poletaeva, \emph{On matrix realizations of the Lie
superalgebra} $D(2,1;\alpha ),$ Journal of Geometry and Physics 60 (2010)
1656--1664, arXiv:1008.2433 [math.RT].

\bibitem{GO} M. Gunaydin, Modern Physics Letters A, 6, 3239 (1991)

\bibitem{6} M. Scheunert, \emph{The Theory of Lie Superalgebras}, in:
Lecture Notes in Mathematics vol. 716, Springer, Berlin, 1979.

\bibitem{1B} K. Iohara , Y. Koga, \emph{Central extensions of Lie
superalgebras}, Comment. Math. Helv. 76 (2001) 110-154.

\bibitem{Le} Leyu Han, \emph{Centres of centralizers of nilpotent elements
in exceptional lie superalgebras}, Journal of Algebra and Its Applications
(2021): 2250053, arXiv:2203.04423v1 [math.RT].

\bibitem{2B} V\textrm{.K. Dobrev}, \emph{Note on Centrally Extended} $%
su(2|2) $ \emph{and Serre Relations}, Fortsch.Phys.57:542-545,2009,
arXiv:0903.0511v3 [hep-th].

\bibitem{SV} N.I. Stoilova, J. Van der Jeugt, \emph{Orthosymplectic} $%
Z_{2}\times Z_{2}$-\emph{graded Lie superalgebras and parastatistics}, J.
Phys. A: Math. Theor. 57 095202 (2024), arXiv:2402.11952 [math-ph].

\bibitem{BY} Y Boujakhrout et al 2022, \emph{Lax operator and superspin
chains from 4D CS gauge theory}, J. Phys. A: Math. Theor. 55 415402,

\bibitem{SE} K. Intriligator, N. Seiberg,\ \emph{Mirror Symmetry in Three
Dimensional Gauge Theories}, Phys.Lett.B387:513-519,1996,
arXiv:hep-th/9607207.

\bibitem{SU2S} Saidi, E.H. \emph{Mutation symmetries in BPS quiver theories:
building the BPS spectra}. J. High Energ. Phys. 2012, 18 (2012),
arXiv:1204.0395v2 [hep-th].

\bibitem{W1} Edward Witten, Fermion Path Integrals And Topological Phases,
Rev. Mod. Phys. 88, 35001 (2016), arXiv:1508.04715.

\bibitem{W2} Edward Witten, Three Lectures On Topological Phases Of Matter,
La Rivista del Nuovo Cimento, 39 (2016) 313-370.

\bibitem{FK} Liang Fu, Charles Kane, Eugene Mele, Topological Insulators in
Three Dimensions, Phys. Rev. Lett. 98 (2007) 106803

\bibitem{Z} Xiao-Liang Qi, Shou-Cheng Zhang, Topological insulators and
superconductors, Rev. Mod. Phys. 83 (2011) 1057-1110.

\bibitem{TM1} L. B. Drissi, S. Lounis, E. H. Saidi, \emph{Higher order
topological matter and fractional chiral states}, Eur. Phys. J. Plus 137,
796 (2022), arXiv:2211.02362 [cond-mat.mes-hall].

\bibitem{TM2} L. B Drissi, E. H Saidi, \emph{A signature index for third
order topological insulators}, Journal Phys Condensed Matter, 32 (36)
(2020), arXiv:2207.02901v1 [cond-mat.mes-hall]

\bibitem{TM3} Lalla Btissam Drissi, El Hassan Saidi, \emph{Domain Walls in
Topological Tri-hinge Matter}, European Physical Journal Plus 136, (68)
(2021), arXiv:2206.11984 [cond-mat.mtrl-sci].

\bibitem{Conifold} Rhiannon Gwyn, Anke Knauf, \emph{Conifolds and Geometric
Transitions}, Rev.Mod.Phys.8012:1419-1453,2008, arXiv:hep-th/0703289v3.

\bibitem{cone2} \textrm{B. S. Acharya, L. Foscolo, M. Najjar, and E. E.
Svanes}, \textquotedblleft New G2-conifolds in M-theory and their field
theory interpretation,\textquotedblright\ High Energ. Phys. 2021, 250
(2021).https://doi.org/10.1007/JHEP05(2021)250

\bibitem{cone3} El Hassan Saidi, \emph{Topological SL(2) Gauge Theory on
Conifold}, arXiv:hep-th/0601020.

\bibitem{LEUNG} N.C. Leung, C. Vafa, Branes and Toric Geometry,
Adv.Theor.Math.Phys.2:91-118,1998.

\bibitem{SEBB} A. Belhaj, L. B. Drissi, J. Rasmussen, E. H. Saidi, A.
Sebbar, \emph{Toric Calabi-Yau supermanifolds and mirror symmetry},
J.Phys.A38:6405-6418,2005, arXiv:hep-th/0410291v2.

\bibitem{AbouN} R.Abounasr, M.Ait Ben Haddou, A.El Rhalami, E.H.Saidi, \emph{%
Algebraic Geometry Realization of Quantum Hall Soliton}, J.Math.Phys. 46
(2005) 022302, arXiv:hep-th/0406036.

\bibitem{SUP} \textrm{S. Barbier and S. Claerebout,} \textquotedblleft A
superunitary Fock model of the exceptional Lie supergroup $D(2,1;\alpha )$%
,\textquotedblright\ Commun. Math. Phys. 403(1), 451--472
(2023).https://doi.org/10.1007/s00220-023-04793-8

\bibitem{IV} Sergey Fedoruk, Evgeny Ivanov, Olaf Lechtenfeld, \emph{New
D(2,1;}$\alpha $\emph{) Mechanics with Spin Variables, }JHEP 1004:129,2010,
arXiv:0912.3508v2 [hep-th].

\bibitem{IVA} Sergey Fedoruk, Evgeny Ivanov, Olaf Lechtenfeld, \emph{OSp(4%
\TEXTsymbol{\vert}2) Superconformal Mechanics}, JHEP
0908:081,2009arXiv:0905.4951v2 [hep-th].

\bibitem{bakas} \textrm{S. Barbier and S. Claerebout}, \textquotedblleft A
Schrodinger model, Fock model and intertwining Segal, Bargmann transform for
the exceptional Lie superalgebra $D(2,1;\alpha )$,\textquotedblright\ J. Lie
Theory 31(4), 1153--1188 (2021).

\bibitem{DS} L.B Drissi, E.H Saidi, \emph{From orthosymplectic structure to
super topological matter}, Nuclear Physics B, Volume 989, April 2023,
116128, arXiv:2303.12584v1 [hep-th].

\bibitem{DSS} E. H Saidi, L. B Drissi, \emph{5D N=1 super QFT: symplectic
quivers}, Nuclear Physics B 974, January 2022, 115632, arXiv:2112.04695
[hep-th].

\bibitem{BE} Youssra Boujakhrout, El Hassan Saidi, \emph{On Exceptional 't
Hooft Lines in 4D-Chern-Simons Theory}, Nuclear Physics B 980, 2022, 115795,
arXiv:2204.12424v1 [hep-th].

\bibitem{BF} \textrm{R. Sammani et al.}, Class. Quantum Grav. 41, 215012
(2024).https://doi.org/10.1088/1361-6382/ad7cb

\bibitem{Yama} \textrm{K. Costello, D. Gaiotto, and J. Yagi},
\textquotedblleft Q-operators are 't Hooft lines,\textquotedblright\ J. High
Energ. Phys. 2024, 3 (2024).https://doi.org/10.1007/JHEP11(2024)003

\bibitem{Lax2} \textrm{R. Frassek}, \textquotedblleft Algebraic bethe ansatz
for Q-operators: The Heisenberg spin chain,\textquotedblright\ J. Phys. A:
Math. Theor. 48, 294002 (2015);
arXiv:1504.04501.https://doi.org/10.1088/1751-8113/48/29/294002

\bibitem{FRC} \textrm{R. Frassek, T. Lukowski, C. Meneghelli, and M.
Staudacher}, \textquotedblleft Oscillator construction of $su(n|m)$
Q-operators,\textquotedblright\ Nucl. Phys. B 850(1), 175--198
(2011).https://doi.org/10.1016/j.nuclphysb.2011.04.008

\bibitem{Fra} \textrm{R. Frassek and A. Tsymbaliuk}, \textquotedblleft
Orthosymplectic superoscillator Lax matrices,\textquotedblright\ Lett. Math.
Phys. 114, 49 (2024).https://doi.org/10.1007/s11005-024-01789-w

\bibitem{1AC} \textrm{S. Gerigk}, \textquotedblleft String states on $%
AdS_{3}\times S^{3}$ from the supergroup,\textquotedblright\ J. High Energy
Phys. 2012, 84; arXiv:1208.0345v3
[hep-th].https://doi.org/10.1007/jhep10(2012)084

\bibitem{VB1} Gerhard Gotz, Thomas Quella, Volker Schomerus, \emph{The WZNW
model on PSU(1,1\TEXTsymbol{\vert}2)}, JHEP 0703:003,2007,
arXiv:hep-th/0610070v2.

\bibitem{VB2} \textrm{M. R. Gaberdiel, B. Guo, and S. D. Mathur},
\textquotedblleft Tensionless strings on AdS3 orbifolds,\textquotedblright\
J. High Energ. Phys. 2024, 57 (2024).https://doi.org/10.1007/JHEP04(2024)057

\bibitem{Mu} I.M. Musson, Lie superalgebras and enveloping algebras,
Graduate Studies in Mathematics 131, Amer. Math. Soc. (2012).

\bibitem{W3} \textrm{F. Delduc, S. Lacroix, M. Magro, and B. Vicedo},
\textquotedblleft A unifying 2D action for integrable $\sigma $-models from
4D Chern--Simons theory,\textquotedblright\ Lett. Math. Phys. 110(4),
1645--1687 (2020).https://doi.org/10.1007/s11005-020-01268-y

\bibitem{W4} T. Okuda, Line operators in supersymmetric gauge theories and
the 2d-4d relation, in New Dualities of Supersymmetric Gauge Theories, J.
Teschner ed., Springer, Cham (2016), p. 195-222
[DOI:10.1007/978-3-319-18769-3 7] [arXiv:1412.7126].

\bibitem{TH1} H. Hayashi, T. Okuda and Y. Yoshida, ABCD of 't Hooft
operators, JHEP 04 (2021) 241 [arXiv:2012.12275].

\bibitem{TH2} K. Maruyoshi, T. Ota and J. Yagi, Wilson-'t Hooft lines as
transfer matrices, JHEP 01 (2021) 072 [arXiv:2009.12391].

\bibitem{TH3} A. Kapustin and E. Witten, Electric-Magnetic Duality And The
Geometric Langlands Program, Commun. Num. Theor. Phys. 1 (2007) 1
[hep-th/0604151].

\bibitem{TH4} A. Kapustin, Wilson-'t Hooft operators in four-dimensional
gauge theories and S-duality, Phys. Rev. D 74 (2006) 025005 [hep-th/0501015].

\bibitem{Lax1} V.V. Bazhanov, T. Lukowski, C. Meneghelli and M. Staudacher,
A Shortcut to the Q-Operator, J. Stat. Mech. 1011 (2010) P11002
[arXiv:1005.3261].

\bibitem{t1} \textrm{E. H. Saidi and M. B. Sedra}, \textquotedblleft
HyperKhaler metrics building and integrable models,\textquotedblright\ Mod.
Phys. Lett. 9, 3163--3174 (1994).https://doi.org/10.1142/S0217732394002987

\bibitem{t2} \textrm{L. B. Drissi, E. H. Saidi, O. Fassi-Fehri, and M.
Bousmina}, Eur. Phys. J. Plus 138, 1105
(2023).https://doi.org/10.1140/epjp/s13360-023-04590-1

\bibitem{t3} A. Kapustin, Wilson-'t Hooft operators in four-dimensional
gauge theories and S-duality, Phys.Rev. D74 (2006) 025005
(arXiv:hep-th/0501015).

\bibitem{t4} \textrm{B. Charbonneau and J. Hurtubise}, \textquotedblleft
Singular Hermitian-Einstein monopoles on the product of a circle and a
Riemann surface,\textquotedblright\ Int. Mat. Res. Not. 2011, 175--216
(2011).https://doi.org/10.1093/imrn/rnq059

\bibitem{t5} K. Maruyoshi, T. Ota, J. Yagi, \emph{Wilson- 't Hooft lines as
transfer matrices}. J. High Energ. Phys. 2021, 72 (2021).
https://doi.org/10.1007/JHEP01(2021)072.
\end{thebibliography}
\end{document}